\documentclass[
10pt,
]{iopart}
\usepackage{iopams}  
\usepackage{graphicx}
\usepackage{bm}
\usepackage{soul}
\usepackage{pifont}
\usepackage{color} 
\usepackage{subfiles}
\usepackage{adjustbox}
\usepackage{fancyhdr}
\usepackage[T1]{fontenc}
\usepackage[utf8]{inputenc}
\usepackage{textcomp}

\newcommand {\tn}{T_{\rm{N}}}

\begin{document}

\title[Metal-insulator transitions in pyrochlore oxides]{Metal-insulator transitions in pyrochlore oxides}

\author{Yoshinori Tokura$^{1,2,3}$, Yukitoshi Motome$^{1}$, and Kentaro Ueda$^{1}$}
\address{$^{1}$Department of Applied Physics, University of Tokyo, Tokyo 113-8656, Japan}
\address{$^{2}$RIKEN Center for Emergent Matter Science (CEMS), Wako 351-0198, Japan}
\address{$^{3}$Tokyo College, University of Tokyo, Tokyo 113-8656, Japan}
\ead{ueda@ap.t.u-tokyo.ac.jp}





\noindent\textbf{Note:} This manuscript is under consideration for publication in \textit{Reports on Progress in Physics}.

\begin{abstract}
Pyrochlore oxides with chemical formula of $A_2B_2$O$_7$ exhibit a diverse range of electronic properties as a representative family of quantum materials.
These properties mostly stem from strong electron correlations at the transition metal $B$ site and typical geometrical frustration effects on the pyrochlore lattice.
Furthermore, the coupling between the magnetic moments of the rare-earth $A$ site and the conduction electrons at the $B$ site, along with the relativistic spin-orbit coupling particularly affecting the $4d$/$5d$ electrons at the $B$ site, gives rise to the topological characteristics of the correlated electrons.
This review paper focuses on the metal-insulator transitions in pyrochlore oxides as evidence of the strong electron correlation, which is highlighted as a rich source of intriguing charge dynamics coupled with frustrated spin-orbital entangled magnetism.
\end{abstract}

%
%
%
%
\ioptwocol

\tableofcontents

\clearpage

\pagestyle{fancy}
\fancyhf{} 

\fancyhead[L]{\leftmark}

\fancyfoot[C]{\thepage} 

\section{Introduction}
\label{sec:1_Introduction}

Oxide pyrochlores, characterized by the general formula $A_2B_2$O$_7$ (where $A$ and $B$ represent metallic elements), constitute a vast family of complex oxides. This family is comparable in breadth to the family of perovskite-type oxides. Pyrochlore compounds feature two types of $A_4$- or $B_4$-tetrahedral networks, known as the pyrochlore lattice. These networks are offset from each other by half a unit cell, as depicted in Fig.~\ref{1-1_lattice_structure_all}(a). The pyrochlore lattice can be perceived as a three-dimensional (3D) extension of the kagome lattice, which is renowned as a typical two-dimensional (2D) frustrated lattice.
Among these, the pyrochlore oxides, where the $B$ site is filled by a transition metal with $3d$, $4d$, and $5d$ electrons, exhibit a range of fascinating physical properties. These properties stem from the inherent strong electron correlation on the $B$ site, the geometrical frustration of the $A$/$B$-site magnetism, and the interaction between the local moments on $A$ sites and the conduction electrons on $B$ sites.

Pyrochlore compounds with $3d$ electrons are predominantly correlated insulators (Mott insulators), while certain pyrochlore compounds with $4d$/$5d$ electrons undergo metal-insulator transitions under physical and/or chemical stimuli due to the intermediate or critical strength of electron correlation.
When the $A$ site is occupied by a rare-earth element with $4f$ electrons, the coupling of the correlated $4d$/$5d$ electrons with the $4f$ local moments leads to a host of intriguing phenomena and provides a potent means of controlling electronic properties.
In particular, the well-known spin ice physics in the geometrically frustrated pyrochlore lattice of the $A$ site ($4f$ moment) can be combined with the $B$-site magnetism or electron correlated state to give rise to the formation of scalar spin chirality (SSC) and resultant topological Hall effect, as exemplified by Mo-oxide pyrochlores.
Moreover, the relativistic spin-orbit coupling in the $B$-site pyrochlore lattice further introduces vibrant features of spin-coupled charge transport in magnetic topological states, such as the realization of Weyl semimetals as exemplified by Ir-oxide pyrochlores.

In terms of oxide pyrochlores, a seminal review paper by Subramanian, Aravamudan, and Subba Rao~\cite{1983Subramanian} offers a comprehensive overview of the material aspects spanning the entire family. The issues of magnetic frustration on the $A$-site $4f$-electron moments, such as classical/quantum spin ice liquid and exotic spin excitations, are thoroughly reviewed in the literatures~\cite{2010RMPGardner,2014RPPGingras,Udagawa2021}.

In this review article, we focus on the metal-insulator phenomena of oxide pyrochlore compounds with frustrated magnetism, which are stimulated by the critical electron correlation effect on the $B$-site transition metal elements with $4d$/$5d$ electrons. In Sect.~\ref{sec:2_GeneralFeatures}, we initially outline some fundamental concepts of crystal chemistry and physics characteristic of the pyrochlore, such as the lattice structure, the metal-insulator transition, the electron (spin and orbital) configurations of the transition-metal ions, the versatile frustrated magnetism, and the interaction of the local $4f$ moments and $4d$/$5d$ conduction electrons. Subsequently, we proceed to the specific topics of pyrochlore Mo-oxides (Sect.~\ref{sec:3_Molybdates}), Ru-oxides (Sect.~\ref{sec:4_Ruthenates}), Re/Os-oxides (Sect.~\ref{sec:5_5dsystems}), and Ir-oxides (Sect.~\ref{sec:6_Iridates}). Perspective/outlook on the study of pyrochlore is provided in Sect.~\ref{sec:7_Perspective}.

\section{General features}
\label{sec:2_GeneralFeatures}

\subsection{Lattice structure $A_2B_2$O$_7$}
\label{sec:2_LatticeStructure}

\begin{figure}[tb]
\centering
\includegraphics[width=0.95\columnwidth]{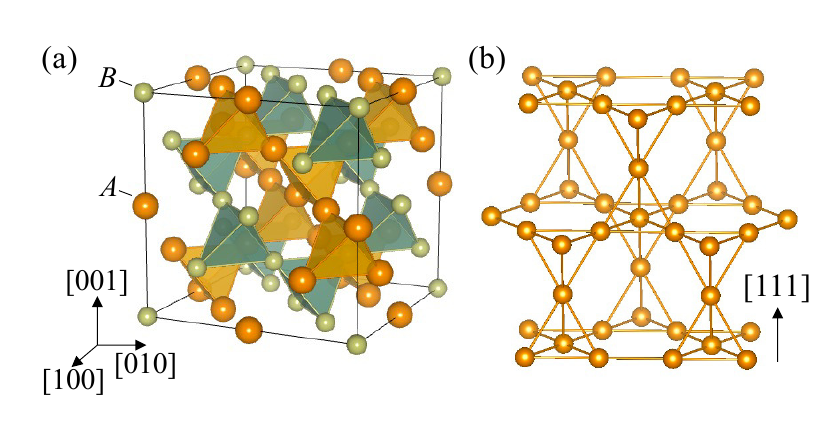}
\caption{\label{1-1_lattice_structure_all}
(a) Crystal structure of pyrochlore lattice $A_2B_2X_7$. Orange (green) balls denote $A$ ($B$) atoms.
(b) Sublattice of $A$ atoms parallel to [111] direction. Kagome and triangular layers are stacked alternatively.
}
\end{figure}

\begin{figure}[tb]
\centering
\includegraphics[width=0.95\columnwidth]{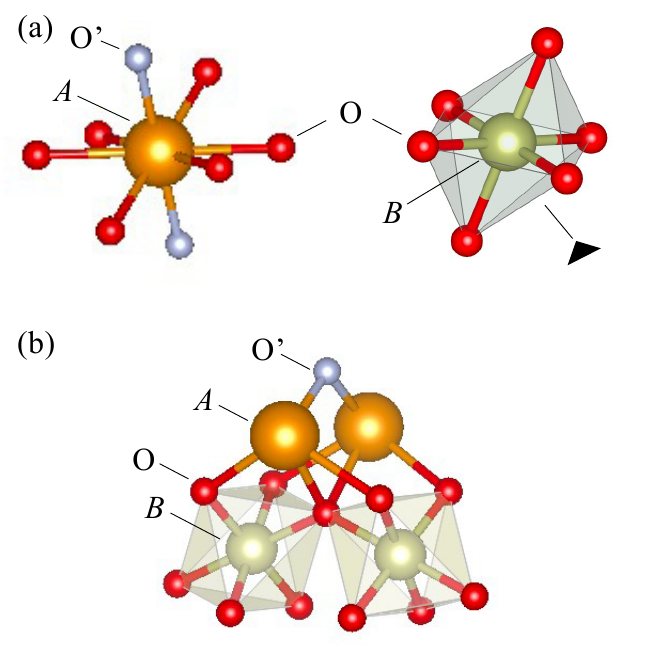}
\caption{\label{1-1_lattice_structure_detail}
(a) Geometry of the $A$ atom (surrounded by O and O' atoms) and $B$ atom (surrounded by O atoms), respectively.
The orange balls denote $A$ atoms, the greens $B$ atoms, the reds O atoms, and the whites O' atoms, respectively. The bond $A$-O' is shorter than $A$-O, forming a distorted AO$_6$O'$_2$ cube. $B$O$_6$ octahedron is trigonally distorted as shown by a black triangle mark.
(b) Corner-sharing octahedra of $B$O$_6$ connected to $A$ atoms.
}
\end{figure}

The pyrochlore materials $A_2B_2X_7$ crystallize in the space group $Fd\overline{3}m$ (No. 227).
$X$ is an anion such as oxygen or fluorine. We will focus on pyrochlore oxides ($X$ = O) in this review.
A standard formula is $A_2B_2$O$_6$O' where the $B$ ion is located at $16c$ as the origin of the second setting for $Fd\overline{3}m$ in the International Tables of Crystallography, $A$ at $16d$, O at $48f$, and O' at $8b$ (see Table~\ref{1-1table}).
Here, there is one structural parameter $x$ for the position of O at $48f$.
The six nearest neighbour O ions surrounding the $B$ ion forms a perfect octagon at $x=0.3125$, whereas the nearest O ions of the $A$ ion are cubic at $x=0.375$.
In reality, since $x$ takes the range of $0.32$-$0.35$ for the most pyrochlore oxides, both geometries are distorted from the ideal polyhedra.
For instance, the eight oxygens surrounding the $A$ site deviate significantly from the cube as shown in the left panel of Fig.~\ref{1-1_lattice_structure_detail}(a).
The bond length of $A$-O is 2.4-2.5 \AA , whereas that of $A$-O' is about $\sim 2.2$ \AA .
Reflecting $D_{3d}$ symmetry, the $16d$ site hosts an axial symmetry along the local $\langle 111\rangle $ direction, which is responsible for the strong anisotropy of rare-earth magnetic moments~\cite{2010RMPGardner}.
On the other hand, the $B$-O bond lengths at the $16c$ site are all equal. However, as shown in the right panel of Fig.~\ref{1-1_lattice_structure_detail}(a), the $B$O$_6$ octahedron is distorted along the local $\langle 111\rangle $ direction.
In such a trigonal crystal field, the triply degenerate $t_{2g}$ orbitals are split into $e_g'$ and $a_{1g}$ orbitals (see Sect.~\ref{sec:2_ElectronConfiguration} for details).

\begin{table*}
    \centering
    \begin{tabular}{cccc}
        Atom & Wyckoff position & Point symmetry & Minimal coordinates \\ \hline
        $A$ & $16d$ & $\overline{3}m$ ($D_{3d}$) & 1/2,1/2,1/2 \\
        $B$ & $16c$ & $\overline{3}m$ ($D_{3d}$) & 0,0,0 \\
        O & $16d$ & $mm$ ($C_{2v}$) & $x$,1/8,1/8 \\
        O' & $16c$ & $\overline{4}3m$ ($T_{d}$) & 3/8,3/8,3/8 \\ \hline
    \end{tabular}
    \caption{The crystallographic positions for the space group $Fd\overline{3}m$ (No. 227) in pyrochlore $A_2B_2$O$_6$O'.}
    \label{1-1table}
\end{table*}


As shown in Fig.~\ref{1-1_lattice_structure_detail}(b), the pyrochlore lattice can be regarded as being formed by interpenetrating networks composed of zigzag chains of $A_2$O' and corner-sharing octahedra of $B_2$O$_6$ in terms of chemical bonding.
It is noteworthy that the latter network is similar to that of perovskite oxides, although the $B$-O-$B$ bond angle is around 125-135$^{\circ }$, which is small and rigid compared to the 140-180$^{\circ }$ for perovskite-type oxides.
Therefore, the transfer integral of $d$-electrons over $B$-O-$B$ chains in pyrochlore oxides tends to be small compared to that of perovskites.
In fact, most of the pyrochlore $3d$-electron systems are insulators while some of the $4d$ and $5d$-electron systems are metals because of the spatially-spread wavefunctions.
Furthermore, the bond angle is modulated by changing the $A$ ionic radius in pyrochlore oxides, although it is rigid compared to perovskites.
In the case of perovskites, there is a general relation between the $A$ ionic radii and the $B$-O-$B$ bond angle (so-called tolerance factor)~\cite{1992PRBTorrance}.
Since the $d$-electron transfer integral via the O $2p$ state is sensitive to the bond angle, the one-electron bandwidth is effectively controlled by the variation of tolerance factor (chemical pressure)~\cite{1998RMPImada}.
Similarly, in pyrochlore molybdates, where the insulator-to-metal transition is realized by changing $A$ ions, the bond angle systematically broadens as the rare-earth ionic radius at $A$ site increases~\cite{2001PRBMoritomo}.
Thus, pyrochlore oxides offer an ideal platform to explore the Mott physics especially for $4d$ and $5d$ electron systems.

Figure~\ref{1-1_lattice_structure_all}(a) shows the schematic distribution of $A$ and $B$ ions in the unit cell of a pyrochlore oxide.
The $A$ and $B$ ions form corner-linked tetrahedra, respectively, resulting in the geometrical frustration.
Therefore, it has been studied for a long time from the viewpoint of magnetism, such as spin ice~\cite{2010RMPGardner}.
The pyrochlore structure can also be regarded as an alternating stacking of kagome and triangular lattices when viewed from the [111] crystalline direction [Fig.~\ref{1-1_lattice_structure_all}(b)].
Some materials such as La$_3$Sb$_3$Mg$_2$O$_{14}$ have been reported in which nonmagnetic (Mg) ions are substituted on the triangular lattice and hence rare-earth ions materialize pseudo-2D kagome networks, similar to the Herbertsmithite structure~\cite{2016PSSSanders}.

There are a wide variety of pyrochlore $A_{2}B_{2}$O$_7$ which basically form $A_{2}^{3+}B_{2}^{4+}$O$_7$ or $A_{2}^{2+}B_{2}^{5+}$O$_7$.
Rare-earth ions $R^{3+}$ frequently occupy the $A$ site in the former while divalent ions such as Ca, Cd, and Hg in the latter.
Some materials have complex formulae, such as $A_{2}^{3+}B^{3+}B'^{5+}$O$_7$~\cite{1983Subramanian}, which are not discussed in detail in this review.
There is also an oxygen-deficient $A_2B_2$O$_{6.5}$ and $\beta $-type $AB_2$O$_6$.
The former crystallizes in the space group $F\overline{4}3m$ as a result of the loss of oxygen, in which the inversion symmetry is broken.
The latter is the $Fd\overline{3}m$ structure in which alkali metal $A$ ions are replaced at the $8b$ site, resulting in the large size mismatch~\cite{2009JPSJNagao}.
Therefore, the $A$ ions rattle in the $B$-O cage like skutterudites~\cite{1998NatureKeppens}, leading to the unusual lattice dynamics.


\subsection{Control of metal-insulator transition in strongly correlated electron systems}
\label{sec:2_SCES}

\begin{figure}[tb]
\centering
\includegraphics[width=0.5\columnwidth]{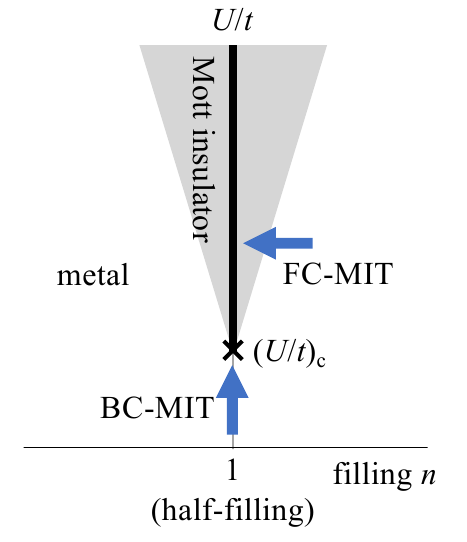}
\caption{\label{1-2_MIT_review}
Schematics of the metal-insulator transition near the simple half-filled case in strongly correlated electron systems.
BC stands for bandwidth control, FC stands for filling control, MIT stands for metal-insulator transition.
The abscissa stands for the band filling $n=1\pm x$ where $x$ is the electron/hole doping.
$(U/t)_{\mathrm{c}}$ stands for critical electron correlation strength ($U/t$) at $n=1$ corresponding to the metal-insulator boundary.
}
\end{figure}

The strong electron correlation for $d$-electron systems produces the electron localization, as referred to as a Mott insulator~\cite{1998RMPImada}, in particular when the compound shows the integer-number band filling ($n$) in the $d$-electron related conduction band. (In the case of the simple band only with the spin degeneracy, the $n=1$ case is called the half-filling.)
Figure \ref{1-2_MIT_review} shows a schematic diagram in the band filling ($n$) vs. the electron correlation strength ($U/t$) for the insulator-metal transition near the simple half-filled case; here $U$ and $t$ represent the on-site coulomb repulsion (Hubbard $U$) and transfer integral, respectively.
In general, there are two pathways to cause the insulator-metal transition starting from the Mott insulator with $(U/t)>(U/t)_{\mathrm{c}}$ and $n=1$ (or $n=\mathrm{integer}$): One is to decrease the effective correlation $U/t$, and the other is to change the band filling from $n=1$ (or $n=\mathrm{integer}$), termed the bandwidth control and the band filling control, respectively. The increasing (decreasing) procedure of the band filling, {\it i.e.}, $n=1+x$ ($n=1-x$), is conventionally termed the electron (hole) doping $x$. Near the metal-insulator boundary, the Drude weight measured by $n_{\mathrm{e}}e^2/m^{*}$ ($m^{*}$ and $n_{\mathrm{e}}$ standing for the effective mass and effective concentration of the conduction electron, respectively) tends to decrease to zero, either by the divergence of $m^{*}$ or by reduction of $n_{\mathrm{e}}$ to zero. The most simple canonical case shows the divergence of $m^{*}$, as typically observed in LaTiO$_3$~\cite{1993PRLTokura}, while there is a variety of metal-insulator transition depending on the correlations of other degrees of freedom of correlated electron, such as spin, orbital, as well as the electron-phonon interactions and random potential effect, as typically  observed in layered-cuprate superconductors~\cite{1998RMPImada}. In reality, against hole or electron doing, the Mott insulating state remains robustly up to some critical doping level $x_{\mathrm{c}}$, which is also critically dependent on the electron correlation such as $x_{\mathrm{c}}\rightarrow 0$ with ($U/t$)$\rightarrow (U/t)_{\mathrm{c}}$. This is in most cases due to the electron-lattice interaction and the quenched disorder effect in conjunction of the dominant effect of electron correlation, all of which collectively decrease the Drude weight to lead to the electron localization or to the insulator~\cite{1995PRLKatsufuji}.

\begin{figure}[tb]
\centering
\includegraphics[width=0.95\columnwidth]{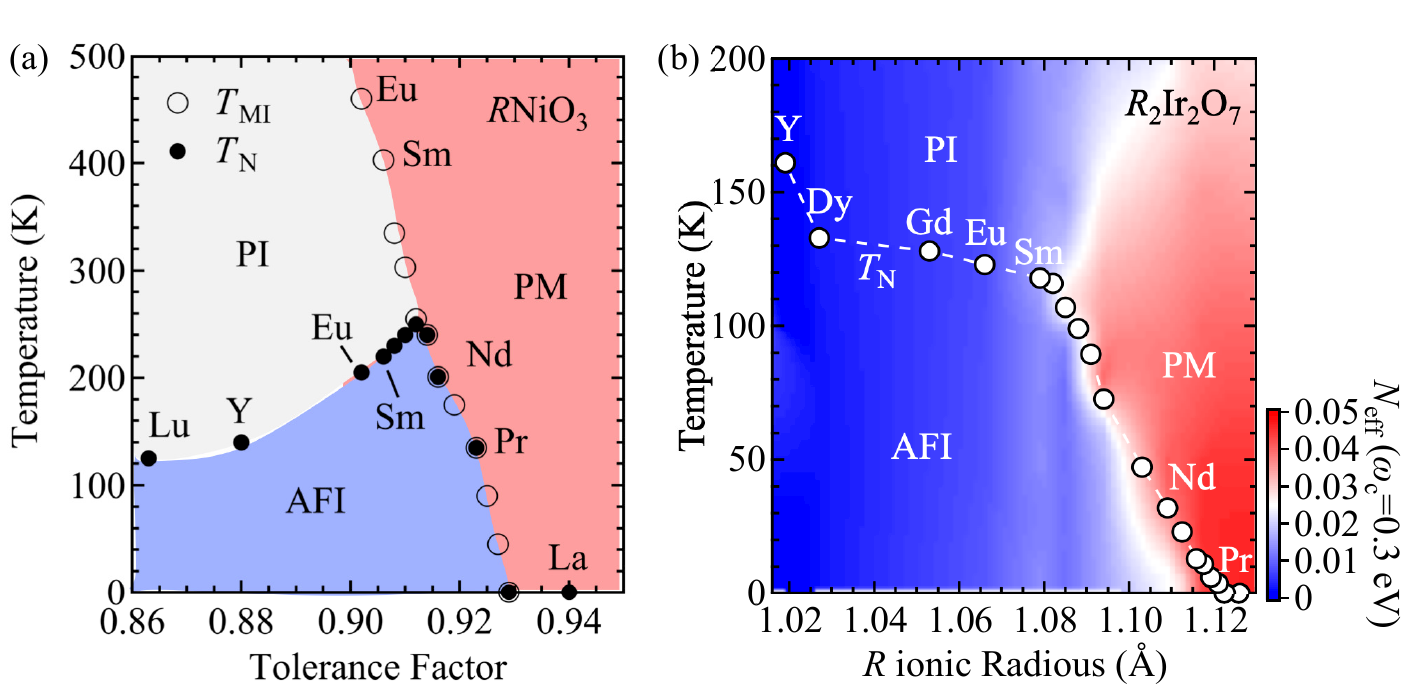}
\caption{\label{1-2_Nickelates}
Metal-insulator phase diagram in the $r_R$-temperature plane: (a) perovskite $R$NiO$_3$ and (b) pyrochlore $R_2$Ir$_2$O$_7$. The $R$-site ionic radius $r_R$ determines the Ni-O-Ni and Ir-O-Ir bond angles ($\theta $): the $\theta $ variation between 140-180$^{\circ }$ for the perovskite and 125-135$^{\circ }$ for the pyrochlore. Thus the change in $r_R$ ($\Delta r_R$ ) represents the variation of the effective electron correlation $\Delta (U/t)$; $\Delta r_{R}\propto -\Delta (U/t)$. PI, AFI, and PM stand for paramagnetic insulator, antiferromagnetic insulator, and paramagnetic metal, respectively.
Panel (a) is adapted from \cite{1992PRBTorrance}.
Reproduced with permission from \cite{2016PRBUeda} for panel (b), Copyright and (2016) by the American Physical Society.
}
\end{figure}

In the actual $d$-electron systems, the spin and orbital degrees of freedom affect the critical behaviors of the insulator-metal transitions.
The crystal field or the related Jahn-Teller coupling, partially lifts the degeneracy of the $d$-electron orbital state.
In the case of the pyrochlore oxides, the originally triply-degenerate $t_{\mathrm{2g}}$ electron level in the $B$O$_6$ octahedron is further split into the doubly degenerate $e_{\mathrm{g}}'$ state and the lower-lying $a_{\mathrm{1g}}$ state under the trigonal crystal field characteristic of the $3d$ and $4d$-electron pyrochlore systems, while the strong relativistic spin-orbit coupling leads to the formation of the $j_{\mathrm{eff}}=1/2$ and $3/2$ states in the $5d$-electron ({\it e.g.}, Ir-oxide) pyrochlore systems, as described in detail in the next section.
The Hund's-rule coupling of the spins on the respective orbital levels define not only the antiferromagnetic state of the ground state of the insulator but also the ferromagnetic or antiferromagnetic magnetic states in the metallic state, {\it e.g.}, via the double-exchange and super-exchange interactions.
The variation of the magnetic state with the change of the band filling or bandwidth, occasionally accompanying the metal-insulator phenomena, is one of the major topics in this article.

Here we briefly show the examples of the metal-insulator transitions with the bandwidth and band-filling control in the pyrochlore system, comparing with the well-known cases for the 3$d$-electron perovskite system. 
Figure~\ref{1-2_Nickelates} shows the electronic phase diagram for (a) perovskite $R$NiO$_3$~\cite{1992PRBTorrance} and (b) pyrochlore $R_2$Ir$_2$O$_7$~\cite{2016PRBUeda} plotted as a function of the ionic radius of rare-earth $R^{3+}$.
As described in Sect.~\ref{sec:2_LatticeStructure}, the $R$ ionic radius ($r_R$) determines the Ni-O-Ni or Ir-O-Ir bond angles; a deviation below 180$^{\circ }$ and around 130$^{\circ }$ for the perovskite and pyrochlore, respectively. The larger (smaller) bond angle tends to increase (decrease) the $d$-electron transfer integral via the supertransfer process via the O $2p$ state and hence the one-electron bandwidth, {\it i.e.}, the effective electron correlation $U/t\propto r_R$ in a small variation range of $r_R$.
For both compounds, roughly a $\sim 10$ \% change of $r_R$ causes the metal-insulator phenomena in accord with the idea of the Mott transition via the bandwidth control. Note here the Ni$^{3+}$ ion in the Mott insulating state (more precisely a charge-transfer insulator according to the Zaanen-Sawatzky-Allen scheme~\cite{1985PRLZaanen}) bears $S=1/2$ spin, while Ir$^{4+}$ ion in the Mott-insulating pyrochlore bears $j_{\mathrm{eff}}=1/2$ spin. Occasionally and as seen even in the quantum spin system in these compounds, the antiferromagnetic order (AFI) shows up in the ground state of the Mott insulating state, although the phase boundary between the AFI and Pauli-paramagnetic metal (PM) states shows generally a slanted line in the $r_R$ vs. temperature $T$ plane. We will come back to the phase diagram of $R_2$Ir$_2$O$_7$ [Fig. \ref{1-2_Nickelates}(b)] in Sect.~\ref{sec:6_Iridates}.

\begin{figure}[tb]
\centering
\includegraphics[width=0.95\columnwidth]{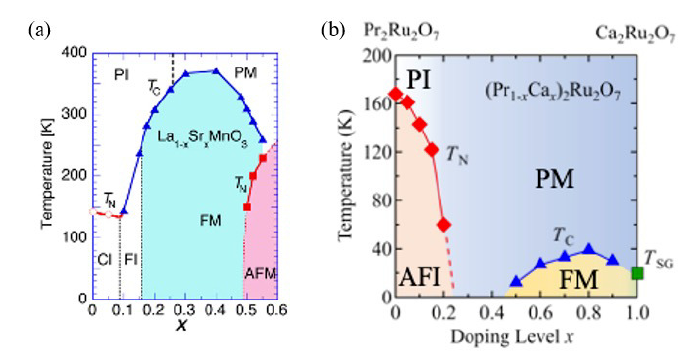}
\caption{\label{1-2_Manganites}
Electronic phase diagram in the hole doping ($x$)-temperature plane; (a) perovskite La$_{1-x}$Sr$_x$MnO$_3$ and (b) pyrochlore (Pr$_{1-x}$Ca$_x$)$_2$Ru$_2$O$_7$.
PI, CI, FI, AFI, FM, PM, and AFM stand for paramagnetic insulator, canted-spin insulator, ferromagnetic insulator, antiferromagnetic insulator, ferromagnetic metal, paramagnetic metal, and antiferromagnetic metal, respectively.
Reproduced from \cite{2006RPPTokura} for panel (a). $\copyright $ IOP Publishing Ltd. All rights reserved.
}
\end{figure}

Figure \ref{1-2_Manganites} shows the examples of the electronic phase diagram in the case of the band filling control, for (a) the perovskite Mn-oxides La$_{1-x}$Sr$_x$MnO$_3$ and (b) the pyrochlore Ru-oxides (Pr$_{1-x}$Ca$_x$)$_2$Ru$_2$O$_7$. The band filling can be changed systematically by forming the solid solution or the mixed crystal with partially substituting the $A$-site trivalent ion ({\it e.g.}, La and Pr) with the divalent ions ({\it e.g.}, Sr or Ca) in the present case, in analogy to the well-known high-temperature superconductors, such as La$_{2-x}$Sr$_x$CuO$_4$. Note here that the electronic configuration of Mn$^{3+}$ and Ru$^{4+}$ is $3d^4$ and $4d^3$ in which the Hund's rule works on the spin alignment between the respective $d$ orbitals, as described in Sect.~\ref{sec:2_ElectronConfiguration}, characteristic of the multi-orbital systems. Reflecting this, the doping induced metallic phase in these compounds ($x>0.17$ for La$_{1-x}$Sr$_x$MnO$_3$~\cite{2006RPPTokura}, and $x>0.2$ for (Pr$_{1-x}$Ca$_x$)$_2$Ru$_2$O$_7$~\cite{2020PRBKaneko}) show various magnetic states, including paramagnetic-metal (PM), ferromagnetic-metal (FM), and antiferromagnetic-metal (AFM) phases.  We will come back to the electronic phase diagram in the pyrochlore (Pr$_{1-x}$Ca$_x$)$_2$Ru$_2$O$_7$ with the detailed discussion in Sect.~\ref{sec:4_Ruthenates}.

Finally, we note some differences in the metal-insulator phenomena that stem from the electron correlation between oxide perovskites and oxide pyrochlores. In the perovskite $AB$O$_3$, the $4d$-electron $B$ site compounds with the integer band filling are mostly metallic, reflecting smaller electron correlation $U/t$. For the $3d$-electron $B$ site case, it is either Mott insulator or metal, depending on the valence of the integer filling state. In contrast, the pyrochlore case, the $4d$/$5d$-electron related conduction band becomes narrower due to the more distorted $B$-O-$B$ bond than in the corresponding perovskite, sometimes placing the $4d$/$5d$ electron pyrocholore compounds near the Mott transition even at the integer band filling. The examples are pyrochlore $R_2M_2$O$_7$ compounds ($M$ being transition metals such as Mo, Ru, and Ir), which undergo the bandwidth control and/or filling control metal-insulator transitions (or incipient transitions) depending on the $R^{3+}$ ionic radius $r_R$ which governs the $B$-O-$B$ bond angle and hence the transfer integral $t$. This is a source of rich intriguing metal-insulator phenomena in the pyrochlores as highlighted in this article.

\subsection{Crystal field, electron correlation, and spin-orbit coupling in $d$ electrons at $B$ sites}
\label{sec:2_ElectronConfiguration}

\begin{figure}[tb]
\centering
\includegraphics[width=0.95\columnwidth]{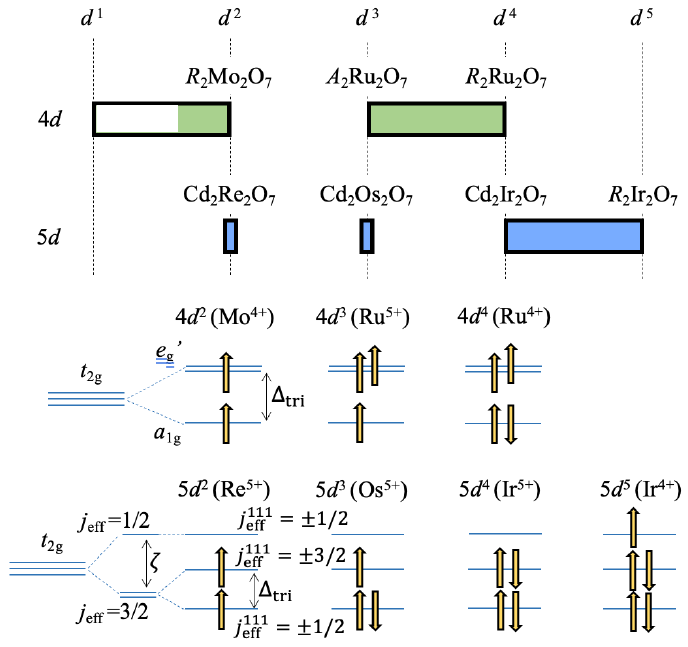}
\caption{\label{fig:dele}
Electron configurations of $B$ ions in the pyrochlore compounds $A_2B_2$O$_7$.
(a) The range of $d$ electron numbers in the $4d$ and $5d$ pyrochlore compounds.
Green and blue colors indicate the compounds obtained thus far.
(b) The electron configurations in each nominal case. 
}
\end{figure}

The pyrochlore oxides $A_2B_2$O$_7$ exhibit a variety of electronic and magnetic properties depending on the number of $d$ electrons and the occupied orbitals of $B$ ions. 
The typical nominal valence of $B$ ions is $4+$ or $5+$, while it also takes the values in between depending on chemical substitution of $A$ ions. 
Figure~\ref{fig:dele}(a) summarizes the range of $d$-electron numbers for the $4d$ and $5d$ pyrochlore compounds obtained thus far. 
As each $B$ ion is surrounded by six oxygens, forming a distorted $B$O$_6$ octahedron [Fig.~\ref{1-1_lattice_structure_detail}(a)], the dominant cubic crystal field splits the fivefold $d$-orbital manifold into the threefold $t_{2g}$ and twofold $e_g$ manifolds. 
In these $4d$ and $5d$ compounds, the Coulomb interactions are relatively weak compared to the $3d$ cases, and all the $d$ electrons occupy the low-energy $t_{2g}$ manifold even in the $d^4$ and $d^5$ cases, resulting in the low-spin electron configurations.

\subsubsection{$4d$ case.}
\label{sec:2_4dcase}

We first discuss the fundamental electronic structure of the $4d$ pyrochlores. 
In the $4d$ cases, the additional crystal field splitting $\Delta_{\rm tri}$ due to a trigonal distortion of the $B$O$_6$ octahedron is relatively larger than the spin-orbit coupling. 
The trigonal crystal field splits the $t_{2g}$ manifold into the $a_{1g}$ singlet and the $e_g^\prime$ doublet, and the $4d$ electrons occupy these manifolds obeying the Hund's rule, as shown in Fig.~\ref{fig:dele}(b). 
Below, we discuss two cases with $B$ = Mo and Ru. 

In the Mo pyrochlore compounds $A_2$Mo$_2$O$_7$, each Mo ion takes the nominal valence of $4+$ with $4d^2$ electron configuration. 
In this case, one of the two $4d$ electrons occupies the low-energy $a_{1g}$ orbital, and the rest occupies one of the high-energy $e_g^\prime$ orbitals, with aligning their spins in parallel, as shown in Fig.~\ref{fig:dele}.
In this electron configuration, the $a_{1g}$ orbital is half filled, resulting in the Mott insulating state under electron correlations, while the $e_g^\prime$ orbitals are quarter filled and generally prone to form a metallic band with active orbital degree of freedom. 
On the pyrochlore lattice, neighboring localized spins in the $a_{1g}$ orbital tend to interact with each other antiferromagnetically by the superexchange mechanism, but the itinerant electrons in the $e_g^\prime$ orbitals induce effective ferromagnetic interactions due to the double-exchange mechanism~\cite{Zener1951}. 
The latter can be enhanced when additional carriers are doped by chemical substitution of $A$ ions. 
However, first-principles calculations indicate that the effective magnetic interactions can be more intricate because of the keen competition between the trigonal distortion, the spin-orbit coupling, and the orbital degrees of freedom~\cite{Shinaoka2013}.
Such competition in the $t_{2g}$ manifold contributes to intriguing magnetic properties in $A_2$Mo$_2$O$_7$, as discussed in Sect.~\ref{sec:3_Molybdates}.

In the Ru pyrochlore compounds $A_2$Ru$_2$O$_7$, the nominal valence of Ru ions takes $5+$ or $4+$ depending on the $A$ ion. 
In the $5+$ case, each Ru ion has three $4d$ electrons, and the additional electron occupies the $e_g^\prime$ orbital that is unoccupied in the Mo$^{4+}$ case. 
This makes the $e_g^\prime$ manifold half filled as well as $a_{1g}$, resulting in the Mott insulating state as a whole of the $t_{2g}$ manifold. 
This state has spin $S=3/2$ in total and no orbital degree of freedom, leading to dominant antiferromagnetic interactions between neighboring localized magnetic moments on the pyrochlore lattice in general. 

Meanwhile, in the Ru$^{4+}$ case with $4d^4$ electron configuration, one electron is further added to the $a_{1g}$ orbital, which realizes a band insulating state in the full-filled $a_{1g}$ orbital. 
The $e_{1g}^\prime$ manifold is still half filled, which is expected to form a localized moment with total spin $S=1$.
In this case, however, the resultant electronic and magnetic states appear to be affected by strong correlation effects near the Mott transition, as discussed in Sect.~\ref{sec:4_Ruthenates}.

In addition, interestingly, the Ru pyrochlores realize intermediate electronic states between $4d^3$ and $4d^4$ due to chemical substitution of the $A$ ions, as shown in Fig.~\ref{fig:dele}(a). 
It provides a platform to study intriguing physics by carrier doping that bridges between the half-filled $t_{2g}$ state with $S=3/2$ and the half-filled $e_g^\prime$ state with $S=1$.

\subsubsection{$5d$ case.}
\label{sec:2_5dcase}

Next, we turn to the $5d$ cases. 
In contrast to the $4d$ cases, the spin-orbit coupling $\zeta$ is rather larger than the trigonal crystal field splitting $\Delta_{\rm tri}$. 
The strong spin-orbit coupling mixes the six orbitals including spins in the $t_{2g}$ manifold and splits them into a doublet and a quartet, respectively characterized by the effective spin-orbital entangled pseudospins $j_{\rm eff}=1/2$ and $3/2$, as shown in Fig.~\ref{fig:dele}(b). 
We discuss three cases with $B$ = Re, Os, and Ir below.

In the Re case, typically realized in Cd$_2$Re$_2$O$_7$, each Re ion takes the nominal valence of $5+$ with $5d^2$ electron configuration. 
In this case, the two $5d$ electrons occupy two out of the $j_{\rm eff}=3/2$ quartet. 
This results in the spin $S=1$ and orbital $L=1$ state, with the total angular momentum $J=2$~\cite{Chen2011,Khaliullin2021}. 
In reality, however, the quartet is further split by a weak trigonal crystal field into two doublet. 
In this case, $j_{\rm eff}$ is no longer a good quantum number, but its projection onto the local $[111]$ axis, $j_{\rm eff}^{111}$ remains quantized. 
The two electrons occupy one each of the two doublets with their spins aligned, as shown in Fig.~\ref{fig:dele}(b).
The electronic properties of Cd$_2$Re$_2$O$_7$ will be discussed in Sect.~\ref{sec:5_Cd2Re2O7}.

In the Os case whose canonical example is Cd$_2$Os$_2$O$_7$, each Os ion is nominally in the $5d^3$ electron configuration. 
In this case, the additional electron is apt to occupy the lower-energy doublet with $j_{\rm eff}^{111}=1/2$ and makes it fully occupied. 
However, first-principles calculations indicate that the splitting between the $j_{\rm eff}=1/2$ and $j_{\rm eff}=3/2$ by the spin-orbit coupling is not large, and they overlap and form metallic bands when the electron correlation is less important~\cite{Shinaoka2012}. 
In addition, quantum-chemical calculations showed that a large easy-axis magnetic anisotropy arises from the trigonal distortion in the presence of the spin-orbit coupling~\cite{Bogdanov2013}.
The electronic and magnetic properties of Cd$_2$Os$_2$O$_7$ will be discussed in Sect.~\ref{sec:5_Cd2Os2O7}.

Finally, in the Ir pyrochlores, the nominal valence of Ir ions takes $5+$ or $4+$ depending on the $A$ ion, as in the Ru case. 
In the Ir$^{5+}$ case with $5d^4$ electron configuration, one electron is further added to the quartet compared to the Os$^{5+}$ case. 
This results in the fully-occupied $j_{\rm eff}=3/2$ state, having $S=1$, $L=1$, and $J=0$. 
This state is apparently a simple band insulator, but has attracted attention as the van Vleck-type Mott insulator due to the peculiar properties, such as excitonic magnetism~\cite{Khaliullin2013}.

Meanwhile, in the Ir$^{4+}$ case with $5d^5$ electron configuration, one further additional electron occupies one of the $j_{\rm eff}=1/2$ doublet. 
This state, which can be regarded as a single hole state in the $t_{2g}$ manifold, results in the half-filled $j_{\rm eff}=1/2$ doublet.
This is a source of exotic magnetism by the spin-orbital entangled $j_{\rm eff}=1/2$ moments, such as possible quantum spin liquid behavior in quasi-two-dimensional magnets $A_2$IrO$_3$ ($A$ = Na and Li)~\cite{Jackeli2009}. 
Moreover, under the consideration of electron correlations, it becomes a Mott insulating state, dubbed the spin-orbit coupled Mott insulator, as firstly discovered in Sr$_2$IrO$_4$~\cite{Kim2008,Kim2009}. 
Likewise, the Ir$^{4+}$ pyrochlores provide an intriguing platform for three-dimensional frustrated magnetism with $j_{\rm eff}=1/2$ moments. 

As shown in Fig.~\ref{fig:dele}(a), the Ir pyrochlores $A_2$Ir$_2$O$_7$ possess intermediate electronic configurations between the $5d^{4+}$ and $5d^{5+}$ cases depending on the chemical substitution of the $A$ ions. 
Thus, this bridges two intriguing situations: the $J=0$ van-Vleck type Mott insulator and the $j_{\rm eff}=1/2$ spin-orbit coupled Mott insulator. 
It was pointed out that carrier doping to the former leads to a spin-triplet superconductivity~\cite{Chaloupka2016}.
The intriguing physics in the Ir pyrochlores $A_2$Ir$_2$O$_7$ will be discussed in Sect.~\ref{sec:6_Iridates}.

\subsection{Magnetism in $f$ electrons at $A$ sites and $d$-$f$ interaction}
\label{sec:2_c-f interaction}

The pyrochlore structure composed of corner-sharing network of tetrahedra is a geometrically frustrated lattice in three dimensions, as shown in Fig.~\ref{1-1_lattice_structure_all}. 
Even at the classical level, the strong frustration is capable of preventing the system from the formation of magnetic orderings. 
For instance, the antiferromagnetic Ising model on the pyrochlore lattice does not show any ordering down to absolute zero temperature, forming macroscopically-degenerate ground-state manifold~\cite{Anderson1956}. 
The problem is closely related with that for proton configurations in water ice~\cite{Bernal1933, Pauling1935}, and hence, called the spin ice. 
In this section, after introducing some fundamentals of the spin ice problem in Sect.~\ref{sec:spin_ice}, we discuss its implications to the $4d$ and $5d$ pyrochlore oxides in Sects.~\ref{sec:Scalar spin chirality: 2-in_2-out}, \ref{sec:3-in_1-out}, and \ref{sec:all-in_all-out}.

\subsubsection{Spin ice.}
\label{sec:spin_ice}

\begin{figure}[tb]
\centering
\includegraphics[width=0.95\columnwidth]{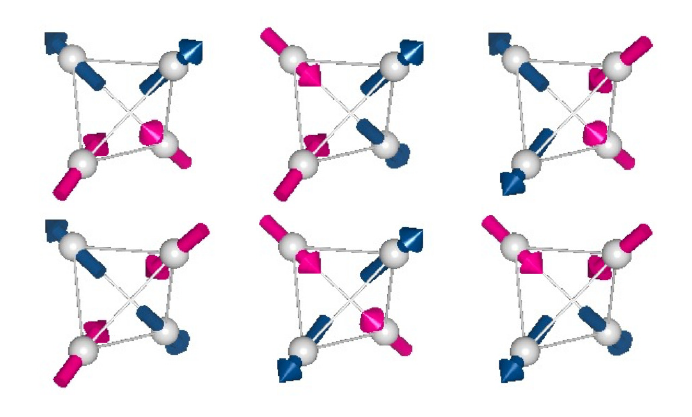}
\caption{\label{fig:six2i2o}
Six patterns of 2-in 2-out states on a single tetrahedron. Magenta arrows point inwards and blue outwards.
}
\end{figure}

Spin ice is a peculiar magnetically-disordered state realized on the pyrochlore lattice. 
In this state, each magnetic moment, which locates at a corner of tetrahedron, is subject to strong uniaxial anisotropy in the direction connecting the tetrahedron center and the corner. 
Due to strong ferromagnetic interactions between the anisotropic Ising-type moments, the lowest-energy state in each tetrahedron is realized by two out of four moments pointing inward and the other two pointing outward, which is dubbed 2-in 2-out state. 
The 2-in 2-out state has sixfold degeneracy with different spin configurations (Fig.~\ref{fig:six2i2o}). 
Due to this degeneracy, the local energy minimization in each tetrahedron does not enforce any long-range ordering throughout the pyrochlore lattice, resulting in macroscopically-degenerate ground states at the classical level. 
This is called the spin ice. 
Since there are a number of reviews for this intriguing state~\cite{Bramwell2001,Melko2004,Bramwell2005,Gingras2011,Castelnovo2012,Bramwell2020,Udagawa2021}, we briefly summarize the fundamental nature below. 

The typical compounds of the spin ice are the $f$-electron pyrochlores Ho$_2$Ti$_2$O$_7$~\cite{Harris1997} and Dy$_2$Ti$_2$O$_7$~\cite{Ramirez1999}. 
In these compounds, Ho$^{3+}$ and Dy$^{3+}$ ions provide the anisotropic Ising-type magnetic moments under the strong spin-orbit coupling. 
These moments interact with each other predominantly via short-range ferromagnetic exchange interactions, in addition to subdominant long-range dipolar interactions. 
When focusing on the ferromagnetic interactions and neglecting long-range interactions beyond the nearest neighbors, the system can be effectively regarded as an antiferromagnetic Ising model with a single common anisotropy axis, which was originally studied for possible charge ordering in the so-called Verwey transition in magnetite Fe$_3$O$_4$~\cite{Anderson1956}. 
This model has strong frustration and resulting in the macroscopically-degenerate ground state with 2-up 2-down configurations in all tetrahedra. 
The ground state is equivalent to the so-called Pauling state for water ice~\cite{Bernal1933,Pauling1935}, which is the reason why it is called spin ice. 
Associated with the macroscopic number of degenerate states, the ground state has residual entropy, whose value was estimated by the Pauling approximation~\cite{Pauling1935} as well as the series expansion~\cite{Nagle1966}. 
It is noteworthy that the experimental estimates of the residual entropy from the specific heat measurements are very close to the theoretical ones~\cite{Ramirez1999}. 

The local 2-in 2-out constraint in the spin ice manifold can be regarded as the divergence free condition of an effective magnetic field. 
This gives rise to peculiar spatial correlations between the Ising-type moments, which manifest as the so-called pinch point in the spin structure factor in momentum space~\cite{Isakov2004}. 
The pinch point structures are indeed observed in inelastic neutron scattering experiments~\cite{Fennell2009,Kadowaki2009}. 

In the spin ice manifold, it is not allowed to flip a single spin, since it breaks the local 2-in 2-out configuration. 
However, flipping all the spins along a closed loop is allowed at no energy cost, and changes the spin configuration to another in the 2-in 2-out manifold. 
The minimum closed loop is a hexagon on the pyrochlore lattice, and hence, the loop flip on a hexagon forms a zero-energy excitation in the spin ice systems. 
Such loop flips are also important for efficient sampling in Monte Carlo simulations for the spin ice models~\cite{Melko2001}. 

The lowest-energy excitation in the spin ice manifold, except for the loop flips, is a flip of single moment, resulting in 3-in 1-out and 1-in 3-out configuration in two tetrahedra sharing the flipped moment. 
Interestingly, they can be regarded as a pair of monopole and antimonopole in terms of the effective magnetic field~\cite{Castelnovo2008}. 
By successively flipping neighboring spins, it is possible to move the monopole and antimonopole with no energy cost. 
The line of flipped spins between the monopole and antimonopole is regarded as the Dirac string.


The macroscopic degeneracy in the spin ice manifold is lifted by long-range dipolar interactions. 
The stable magnetic order depends on the magnitudes and signs of the further-neighbor exchange interactions and the dipolar interactions~\cite{Melko2001}. 
When the nearest-neighbor exchange interactions are predominantly ferromagnetic, the system undergoes a crossover from high-temperature paramagnet to the spin ice state with local 2-in 2-out constraint, and at a lower temperature it exhibits a magnetic long-range order due to the subdominant long-range dipolar interactions. 
Meanwhile, when the nearest-neighbor exchange interactions are antiferromagnetic, the system selects out a long-range ordered state consisting of all-in and all-out spin states (see Sec.~\ref{sec:all-in_all-out}).

\begin{figure}[tb]
\centering
\includegraphics[width=0.95\columnwidth]{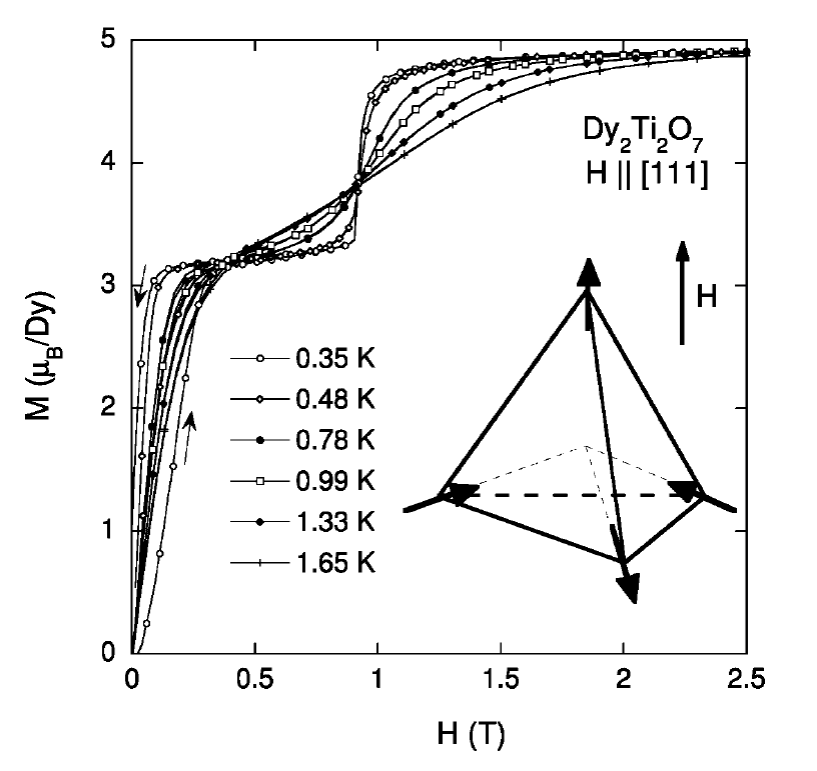}
\caption{\label{fig:MH_Dy2Ti2O7}
Magnetic field dependence of magnetization in Dy$_2$Ti$_2$O$_7$. The direction of the magnetic field is along [111] crystalline direction. The inset shows the spin configuration of the kagome ice state. 
Reproduced with permission from \cite{2003PRLSakakibara}, Copyright (2003) by the American Physical Society.
}
\end{figure}

When one applies an external magnetic field, the spin ice materials exhibit a variety of interesting behaviors depending on the field direction. 
For instance, for the magnetic field along the [111] direction, the spins having the anisotropic axis in that direction are all aligned, and the rest three spins in each tetrahedron form the 2-in 1-out or 1-in 2-out state on the triangle.
This leaves again macroscopic degeneracy in each kagome plane on the [111] cut of the pyrochlore lattice, called kagome ice. 
Correspondingly, the magnetization curve shows a plateau, as observed in experiments (Fig. \ref{fig:MH_Dy2Ti2O7})~\cite{Cornelius2001,Matsuhira2002,Sakakibara2003}. 
With further increasing the magnetic field, the local constraint in the kagome ice is broken, and the all-in (3-in) or all-out (3-out) spin configurations are chosen on the kagome lattices, leading to the full saturation of the magnetization (Fig. \ref{fig:MH_Dy2Ti2O7}).
For the [001] and [110] fields, the system shows different types of crossovers and phase transitions; see the details in the reviews~\cite{Bramwell2005,Gingras2011,Udagawa2021}. 


Thus far we discussed the spin ice at the classical level assuming the magnetic moments are Ising type. 
However, when allowing them to have transverse components, quantum tunneling can take place between different spin configurations in the 2-in 2-out manifold, possibly generating the so-called quantum spin ice~\cite{Hermele2004,Molavian2007,Banerjee2008}. 
The quantum spin ice is a $U(1)$ quantum spin liquid, where dynamical emergent electromagnetism gives rise to exotic fractional excitations, such as artificial photons, visons, and monopoles. 
The promising candidates include Pr, Tb, and Yb-based pyrochlores. 
For the details of quantum spin ice, see the reviews, {\it e.g.}, Refs.~\cite{Gingras2014,Udagawa2021}.

\subsubsection{Hall effect and scalar spin chirality.}
\label{sec:Scalar spin chirality: 2-in_2-out}
Conductors exhibit the Hall effect where a voltage is generated perpendicular to both an electric current and a magnetic field.
Phenomenologically, the Hall resistivity $\rho _{\mathrm{H}}$ is classified into three types, expressed as; 
\[
\eqalign{
\rho _{\mathrm{H}}=R_{\mathrm{0}}B+R_{\mathrm{S}}M+\rho_{\mathrm{H}}^{\mathrm{T}}.
\label{eq:Hall}
}
\]
The first term is the normal Hall effect, which arises from the circular motion of electrons due to the Lorentz force and is generally proportional to the magnetic field.
The second term is the anomalous Hall effect, which is proportional to magnetization.
The third term is called the topological Hall effect, frequently observed in itinerant magnets with noncoplanar spin textures; see below.
Among them, the anomalous Hall effect can be classified into two mechanisms~\cite{2010RMPNagaosa}. One is the extrinsic mechanism, caused by asymmetric scattering due to magnetic impurities, and the other is the intrinsic mechanism, which originates from the geometrical feature of the band structure. 
It is known that there is a scaling law between $R_{\mathrm{S}}$ and the longitudinal resistivity $\rho _{xx}$ for each mechanism~\cite{2008PRBOnoda}.
Recently, particular attention has been given to the intrinsic anomalous Hall effect.
For instance, since the Weyl points act as sources and sinks of Berry curvature, a giant Hall response even with minimal magnetization has been observed in topological Weyl semimetals such as Mn$_3$Sn~\cite{2015NatureNakatsuji}, Co$_3$Sn$_2$S$_2$~\cite{2018NPLiu}, and pyrochlore-type iridates (see Sect.~\ref{sec:6_Iridates magnetotransport}).

\begin{figure}[tb]
\centering
\includegraphics[width=0.7\columnwidth]{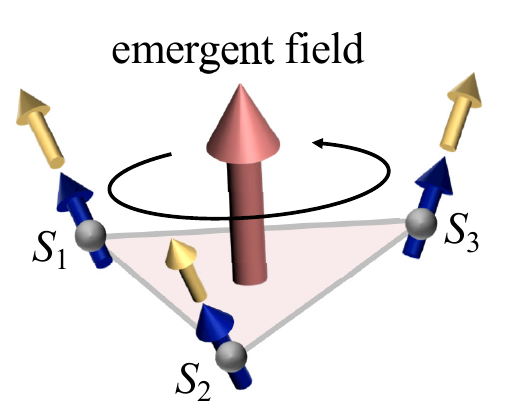}
\caption{\label{fig:spinchirality}
Schematic picture of emergent magnetic field in a noncoplanar spin configuration given by scalar spin chirality ${\bf S}_1 \cdot ({\bf S}_2 \times {\bf S}_3)$.
Yellow arrows indicate the spins of conduction electrons coupled to the localized spins denoted by blue arrows.
}
\end{figure}

When the spin configurations have a noncoplanar structure as in the spin ice, electrons traversing the system feel an effective magnetic field arising from the Berry phase effect. 
This is called the emergent magnetic field. 
It is proportional to the solid angle spanned by three spins, which is approximately given by the triple product ${\bf S}_1 \cdot ({\bf S}_2 \times {\bf S}_3)$ called SSC (Fig.~\ref{fig:spinchirality}).
Such an emergent magnetic field plays an important role in quantum transport and optical phenomena, such as the topological Hall effect~\cite{Bruno2004, Nagaosa2010}.

The topological Hall effect by the emergent magnetic field was firstly proposed for the double-exchange model on the kagome lattice~\cite{Ohgushi2000}. 
It was shown that the electronic band structure modulated by noncoplanar spin configurations acquires the Berry curvature, resulting in nonzero Hall conductivity. 
When the chemical potential locates in the band gap, the Hall conductivity can be quantized corresponding to the Chern number of the occupied bands. 
Experimentally, the topological Hall effect was first observed in the pyrochlore compound Nd$_2$Mo$_2$O$_7$~\cite{Taguchi2001}. 
In this material, the spin-ice like noncoplanar configuration of Nd moments affects the conduction electrons in Mo ions through the interaction between conduction electron and $4f$ moment ($d$-$f$ interaction), and results in the topological Hall effect. 
Subsequently, this intriguing effect has been observed in the wide range of  materials, including magnets hosting topological spin textures such as skyrmions~\cite{Nagaosa2013,Tokura2021}.

\subsubsection{2-in 2-out / 3-in 1-out configurations.}
\label{sec:3-in_1-out}
As seen in the topological Hall effect mentioned above, the coupling between rare-earth magnetism and conduction electrons often influences transport properties.
This effect is particularly pronounced in iridates, which possess topologically nontrivial electronic states due to the strong spin-orbit coupling.
As discussed in detail in Sect.~\ref{sec:6_Iridates theory}, the paramagnetic metallic phase exhibits a unique electronic structure known as a zero-gap semiconductor with inverted bands~\cite{2013PRLMoon,2015NCommKondo}.
Similar to HgTe and $R$PtBi, this electronic structure serves as an ingredient which generates topological electronic states such as topological insulators due to the symmetry breaking.
Pyrochlore iridates provide an ideal platform to explore physical phenomena arising from the interplay between topological electronic states and magnetism.
For instance, the 2-in 2-out magnetic structure at the $B$ sites breaks the time-reversal symmetry, transforming the system into a semimetal with a line node appearing on the mirror plane perpendicular to the magnetic field.
On the other hand, in a 3-in 1-out magnetic structure, three of the four three-fold rotation axes disappear except for the one parallel to the magnetic field, leading to the three pairs of Weyl points.
The orders-of-magnitude magnetoresistance effect and the large Hall effect observed in Nd$_2$Ir$_2$O$_7$ suggest the possible realization of these topological semimetal states (see Sect.~\ref{sec:6_Iridates magnetotransport} in detail).

The $d$-$f$ interaction also induces unconventional magnetic states that are not realized in compounds with a single type of magnetic ion.
When the Coulomb-like interaction between monopoles are large enough, a monopole crystal, which consists of the staggered 3-in 1-out and 1-in 3-out configuration, can be stabilized.
This state can be regarded as the fragmented state composed of an ordered antiferromagnetic phase and a disordered ferromagnetic phase. 
Such a state is reported in Ho$_2$Ir$_2$O$_7$ where, without considering the $d$-$f$ interaction, the Ho magnetic moments favor the 2-in 2-out configuration due to the dipolar interaction while the Ir ones the all-in all-out state.
When the $d$-$f$ interaction is taken into account, the Ir all-in all-out configuration generates the molecular field along the local $\langle 111 \rangle $ directions on the Ho sites, leading to the monopole crystal phase which shows magnetic fragmentation~\cite{2017NCLefrancois}.

\subsubsection{All-in all-out configuration.}
\label{sec:all-in_all-out}

Some of $5d$ pyrochlore compounds realize a magnetically-ordered state composed of all-in and all-out tetrahedra at the $B$ sites; for instance, Cd$_2$Os$_2$O$_7$~\cite{Yamaura2012}, Nd$_2$Ir$_2$O$_7$~\cite{Tomiyasu2012}, and Eu$_2$Ir$_2$O$_7$~\cite{Sagayama2013}. 
The pyrochlore lattice is an alternating array of two types of tetrahedra connected by inversion symmetry, and this magnetically-ordered state has all-in spin configurations in one type and all-out in the other. 
In contrast to the spin ice state with 2-in 2-out spin configurations, this state does not have degeneracy, except that related by time reversal operation. 
Note that the all-in all-out ordered state is stabilized in the spin ice model when the antiferromagnetic interactions are strong, as mentioned in Sect.~\ref{sec:spin_ice}. 

Interestingly, the all-in all-out state can be regarded as a ferroic order of magnetic octupoles, and hence exhibit peculiar responses to external fields~\cite{Arima2013}. 
In this state, each [111] kagome plane realizes an alternating arrangement of all-in all-out spin configurations in each triangular unit, which gives rise to nonzero SSC, as discussed in Ref.~\cite{Ohgushi2000}. 
Besides, for the Ir pyrochlores, it was theoretically pointed out that the all-in all-out order is stabilized by the synergy between electron correlations and the spin-orbit coupling~\cite{Wan2011,Go2012,Witczak-Krempa2013,Shinaoka2015}. 
Remarkably, in the vicinity of correlation-driven metal-insulator transition, the electronic band structure realizes a Weyl semimetallic state and an axion insulator~\cite{Wan2011}. 
These interesting possibilities will be discussed in Sect.~\ref{sec:6_Iridates}.

\subsection{Minimal theoretical framework}

On general grounds, the electronic and magnetic properties of pyrochlore oxides $A_2B_2$O$_7$ can be captured by itinerant $d$ electrons at $B$-site transition metals, localized $f$-electrons at $A$-site rare-earth elements, and their coupling on the frustrated pyrochlore lattice structure. 
We describe these three contributions in the following.

First, the fundamental physics of the $d$ electrons is described by a multi-orbital Hubbard model, whose Hamiltonian is given by
\[
\eqalign{
\mathcal{H}_d = \mathcal{H}_{\rm hop} + \mathcal{H}_{\rm int} + \mathcal{H}_{\rm CF} + \mathcal{H}_{\rm SOC}.
\label{eq:Hd}
}\]
Here, $\mathcal{H}_{\rm hop}$ represents electron hopping on the $B$-site pyrochlore lattice, $\mathcal{H}_{\rm int}$ represents electron-electron interactions, $\mathcal{H}_{\rm CF}$ represents the effects of the crystal field primarily from surrounding ligand oxygens, and $\mathcal{H}_{\rm SOC}$ represents the relativistic spin-orbit coupling. 
The explicit form of $\mathcal{H}_{\rm int}$ for the local contributions can be written by using the Slater-Kanamori interaction parameters as
\[
\eqalign{
\mathcal{H}_{\rm int} = \frac12 \sum_i \sum_{\alpha\beta\alpha'\beta'} \sum_{\sigma\sigma'} U_{\alpha\beta\alpha'\beta'} c_{i\alpha\sigma}^\dagger c_{i\beta\sigma'}^\dagger c_{i\beta'\sigma'} c_{i\alpha'\sigma},
}\] 
where $i$ denotes the lattice site; $\alpha$, $\beta$, $\alpha'$, and $\beta'$ are the orbital indices; $\sigma$ and $\sigma'$ denotes the spin degrees of freedom. 
This form includes the intra- and inter-orbital Coulomb repulsion, the Hund's-rule coupling, and the pair-hopping between different orbitals.

In the transition metals, Coulomb interactions are of the same order or larger compared to the crystal field splitting between the $t_{2g}$ and $e_g$ manifolds, while they become weaker when moving from 3$d$ to 5$d$. 
On the contrary, the spin-orbit coupling becomes stronger from 3$d$ to 5$d$. 
The typical energy scales for Coulomb interaction, crystal field splitting, and spin-orbit coupling are approximately $O(1)$~eV, $\sim 1$~eV, and $O(0.01)$~eV, respectively, for 3$d$, while $O(0.1)$~eV, $\sim 1$~eV, and $O(0.1)$~eV, respectively, for 5$d$; with 4$d$ falling in between these values. 
This energy scheme in the local components of Eq.~\ref{eq:Hd}, $\mathcal{H}_{\rm int} + \mathcal{H}_{\rm CF} + \mathcal{H}_{\rm SOC}$, determines the atomic electronic states discussed in Sect.~\ref{sec:2_ElectronConfiguration}, based on the number of $d$ electrons for each $4d$ and $5d$ case. 
The interplay between the local energies and the kinetic energy from electron hopping $\mathcal{H}_{\rm hop}$ drives rich physics in $d$-electron systems, such as metal-insulator transitions, unconventional superconductivity, and the formation of topological insulators and semimetals~\cite{1998RMPImada,Hasan2010,2014ARCMPKrempa}.

Next, the $f$ electrons of rare-earth elements are well localized and form local magnetic moments on the $A$-site pyrochlore lattice.
The local magnetic state is primarily determined by strong Coulomb interaction and spin-orbit coupling, based on the Russell-Saunders scheme, with additional influence from weak crystal fields. 
The localized moments often experience strong anisotropy due to the spin-orbit coupling. 
An example is the Ising-type anisotropy along the [$111]$ axes, leading to the spin ice state discussed in Sec.~\ref{sec:spin_ice}. 
These moments interact with each other, with primary contributions described by
\[
\eqalign{
\mathcal{H}_f = \mathcal{H}_{\rm exch} + \mathcal{H}_{\rm dipolar},
}\]
where the first term represents short-range exchange interactions and the second term represents long-range dipolar interactions, respectively (see reviews listed in Sec.~\ref{sec:spin_ice}).

Finally, on the interpenetrating pyrochlore networks of $A$ and $B$ sites [Fig.~\ref{1-1_lattice_structure_all}(a)], the itinerant $d$ electrons and the localized $f$ moments interact and couple with each other through the $d$-$f$ interaction. 
The simplest form of this interaction is given by the Kondo coupling as
\[
\eqalign{
\mathcal{H}_{d{\mbox{-}}f} = J_{d{\mbox{-}}f} \sum_{i\alpha} \sum_{\sigma\sigma'} c_{i\alpha\sigma} \boldsymbol{\sigma}_{\sigma\sigma'} c_{i\alpha\sigma'} \cdot {\bf{J}}_i,
}\]
where $J_{d{\mbox{-}}f}$ denotes the coupling strength, $\boldsymbol{\sigma}$ represents the vector of Pauli matrices, and ${\bf{J}}_i$ describes the localized $f$ moment at site $i$. 
Through this $d$-$f$ interaction, the magnetism in the $f$ moments may significantly influence the electronic states of $d$ electrons, as mentioned in Sects.~\ref{sec:Scalar spin chirality: 2-in_2-out} and \ref{sec:all-in_all-out}. 
In addition, the local $f$ moments also experience the effects of itinerant $d$ electrons via the Ruderman-Kittel-Kasuya-Yosida (RKKY) interaction, which is effectively mediated by the itinerant $d$ electrons~\cite{Ruderman1954,Kasuya1956,Yosida1957}.
This complex interplay between $d$ and $f$ electrons plays a crucial role in understanding the electronic and magnetic properties of the pyrochlore oxides, as discussed in the following sections.

\section{Molybdates ($R$,$A$)$_2$Mo$_2$O$_7$}
\label{sec:3_Molybdates}

\subsection{Metal-insulator transitions}
\label{sec:3_MIT}

\begin{figure}[tb]
\centering
\includegraphics[width=0.95\columnwidth]{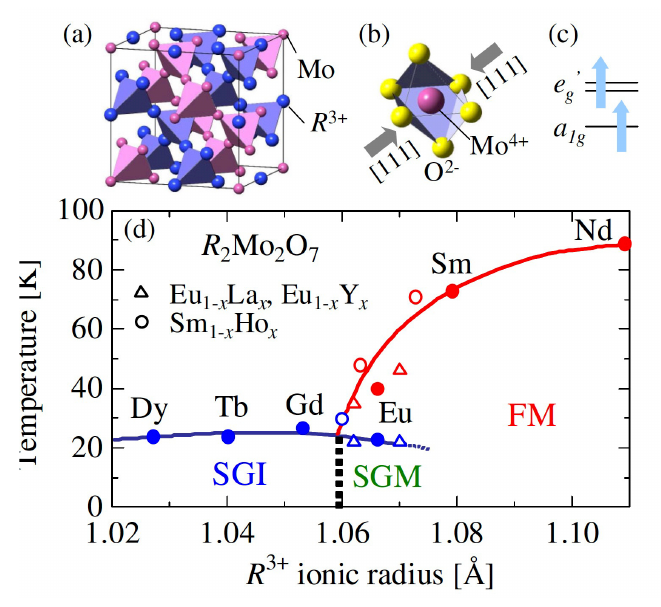}
\caption{\label{2-1_phasediagram}
(a) Pyrocholore structure of $R_2$Mo$_2$O$_7$. (b) Trigonal crystal field on the Mo site. (c) The crystal field splitting of Mo $4d$ orbitals. (d) Magnetic phase diagram in the $R^{3+}$ ionic radius ($r_R$) vs. temperature in $R_2$Mo$_2$O$_7$ with $R$ site ranging from Nd to Dy and their solid solution. The abscissa stands for the change of the electron correlation strength, inversely proportional to $r_R$. FM, SGM, and SGI indicate ferromagnetic metal, spin-glass like metal, and spin glass insulator, respectively.
Reproduced with permission from \cite{2009PRLIguchi}, Copyright (2009) by the American Physical Society.
}
\end{figure}

Pyrochlore molybdates in form of $R_2$Mo$_2$O$_7$ [Fig.~\ref{2-1_phasediagram}(a)] position around the Mott criticality to undergo metal-insulator transitions.
This is in contrast to the perovskite molybdate SrMoO$_3$~\cite{1979MRBHayashi,2005APLNagai} that is known as a good metal characterized by Pauli paramagnetism.
This difference mostly stems from the difference in the one-electron bandwidth of Mo $4d$ $t_{\mathrm{2g}}$ electrons, which is relatively small in the pyrochlore due to the bent Mo-O-Mo bond angle ($\sim 130^{\circ }$~\cite{2001PRBMoritomo}) and relatively large in the perovskite due to the nearly straight bond ($\sim 180^{\circ }$).
Furthermore, the Mo $t_{\mathrm{2g}}$ electron level under the trigonal crystal field in the pyrochlore lattice [Fig.~\ref{2-1_phasediagram}(b)] splits into $a_{\mathrm{1g}}$ and $e_{\mathrm{g}}'$ levels, as shown in Fig.~\ref{2-1_phasediagram}(c).
One electron occupies in $a_{\mathrm{1g}}$, playing a role in a local spin, while the other resides in $e_{\mathrm{g}}'$ as a conduction electron, which is consistent with the result of the cutting-edge x-ray techniques~\cite{2023PRBKitou}.
In some cases, this leads to a ferromagnetic metal state with a double-exchange interaction due to the Hund's-rule coupling between the $a_{\mathrm{1g}}$ and $e_{\mathrm{g}}'$ spins like the case of the hole doped perovskite manganites with Hund's-rule coupled $t_{\mathrm{2g}}$ (local spin) and $e_{\mathrm{g}}$ (conduction electrons)~\cite{2006RPPTokura}.
Figure~\ref{2-1_phasediagram}(d) shows the magnetic phase diagram of $R_2$Mo$_2$O$_7$ with $R$ site ranging from Nd to Dy~\cite{2009PRLIguchi}. 
The abscissa stands for the ionic radius  $r_R$ of $R$ ion, which is known to determine the Mo-O-Mo bond angle and hence the relative magnitude of the one-electron bandwidth $W$($\propto t$) or the inverse of the electron correlation as represented by $U/W$, as described in Sect.~\ref{sec:2_SCES}.
Judging from the change of the Mo-O-Mo bond angle $\theta $, the relative change of the electron correlation from Nd to Dy is around 5\%, viewed almost linear in scale with the $R$ ionic radius. As often observed in the strongly correlated $3d$ electron oxides~\cite{1998RMPImada}, even such a small variation brings about the critical change of electronic state, as exemplified by the metal-insulator (Mott) transition. 
In a large $r_R$ (low $U/W$) region from $R$ = Eu - Nd, the ferromagnetic (FM) metal ground state shows up with the ferromagnetic transition (Curie) temperature $T_{\mathrm{C}}$ (red line) increases with $r_R$ (or $W$), in accord with the feature of the double-exchange ferromagnetic metal~\cite{2006RPPTokura}.
In a smaller $r_R$ region ($r_R<1.06$ \AA ), by contrast, the spin glass insulator state shows up with the nearly constant spin-glass transition temperature $T_{\mathrm{g}}\sim 20$ K.
In this insulating region, the antiferromagnetic interaction takes over the ferromagntism characteristic of the metallic state. In the present pyrochlore case with magnetic frustration, however, no long-range magnetic order shows up in the insulating side, but the spin glass state appears.
The origin of the emergence of the spin glass in the neat pyrochlore lattice has been discussed in literature~\cite{1999PRLGardner}; the glassiness appears to come from the combined effect of frozen spin and $e_{\mathrm{g}}'$-orbital degrees of freedom~\cite{2013PRBShinaoka} or the effect of magnetoelastic coupling~\cite{2014PRBShinaoka,2019PRLSmerald} on the pyrochlore lattice.

\begin{figure}[tb]
\centering
\includegraphics[width=0.95\columnwidth]{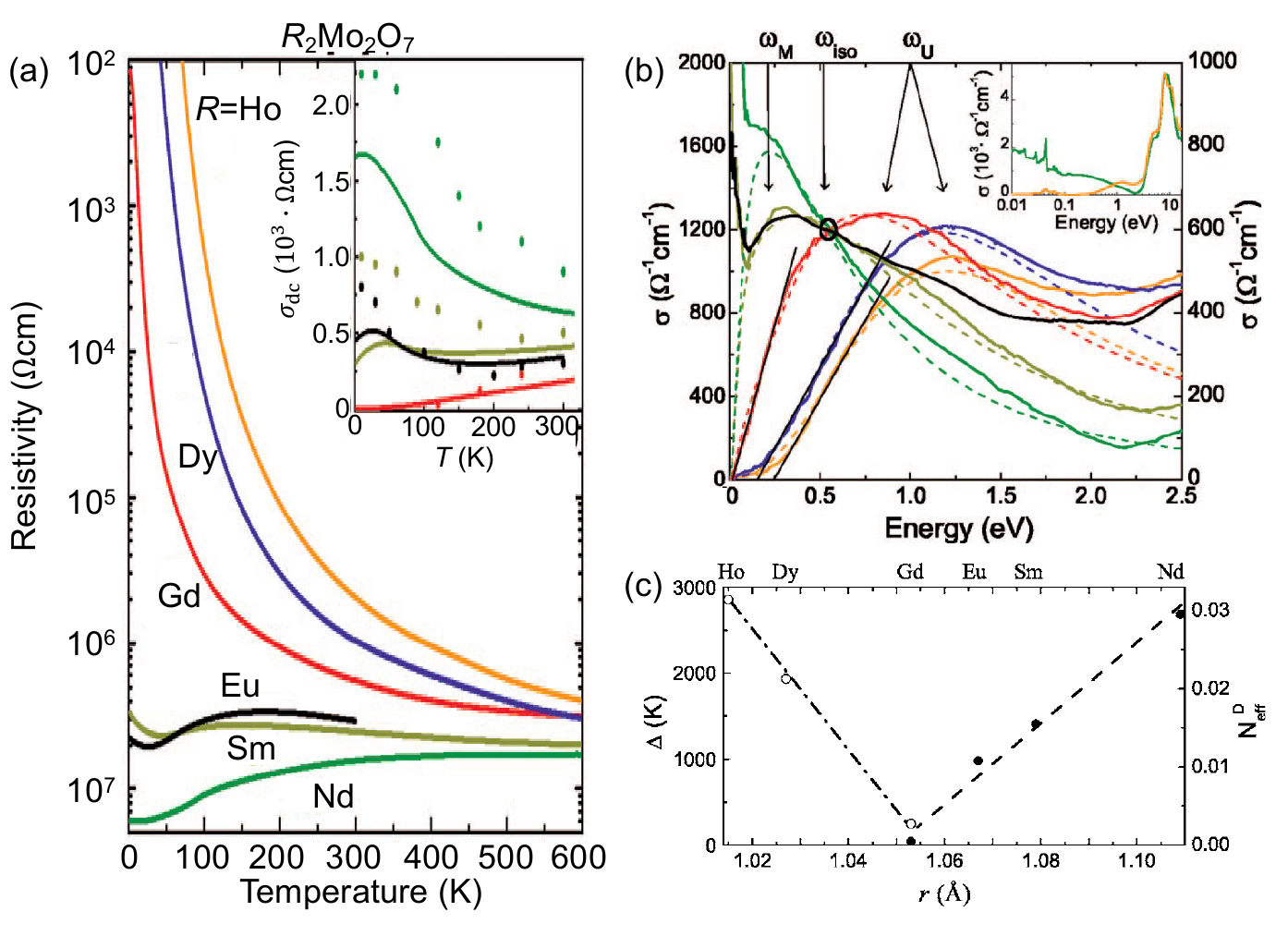}
\caption{\label{2-1_optics}
(a) Temperature dependence of resistivity ($\rho $) or dc conductivity ($\sigma _{\mathrm{dc}}\sim \rho ^{-1}$, inset) for a series of $R_2$Mo$_2$O$_7$. (b) The optical conductivity spectra $\sigma (\omega )$. The inset shows the spectra in the extended photon energy range for the typical metal ($R$ = Nd) and insulator ($R$ = Ho) compounds. (c) The optical conductivity gap energy (open circles) for the insulating compounds and the optical spectral weight (closed circles) below the isosbectic energy $\omega  _{\mathrm{iso}}$ for the metallic compounds.
Reproduced with permission from \cite{2006PRBKezsmarki}, Copyright (2006) by the American Physical Society.
}
\end{figure}

The variation of temperature dependence of resistivity with $r_R$ is shown in Fig.~\ref{2-1_optics}(a). Near $R$ = Gd, $R_2$Mo$_2$O$_7$ shows an insulator to metal transition.
The dramatic change of the low-energy electronic structure upon the spin-orbital coupled metal-insulator transition manifests itself in the optical conductivity $\sigma (\omega )$ spectra related to the charge gap excitations, as shown in Fig.~\ref{2-1_optics}(b) \cite{2006PRBKezsmarki}. For metallic $R$ = Nd and Sm compounds, $\sigma (\omega )$ is composed of the Drude component ($\omega <0.05$ eV) sharply increasing with $\omega  \rightarrow 0$, which is followed by a broad hump with the maximum in the mid-infrared energy (indicated by $\omega _{\mathrm{M}}$ in the panel).
The broad hump structure is ascribed to the remnant Mott-Hubbard gap, representing the incoherent part of the charge dynamics. Such a broad hump or a central energy of the Mott gap excitation ($\omega _{\mathrm{U}}$) moves to higher energy with the decrease (increase) of $r_R$ (electron correlation), while the Drude component completely disappears and the gapped feature, {\it i.e.}, $\sigma (\omega ) \sim 0$ emerges in the lower energy.
The charge-gap energy ($\Delta $) is approximately estimated by the linear extrapolation of $\sigma (\omega )$ tail to zero value and plotted as a function of $r_R$, {\it i.e.}, $W$, in Fig.~\ref{2-1_optics}(c) together with the Drude weight $N_{\mathrm{eff}}^{\mathrm{D}}$ in the metallic region~\cite{2006PRBKezsmarki}.
The Mott gap $\Delta $ is critically decreased with increase of $W/U$ toward the Mott transition point $(W/U)_{\mathrm{c}}$, whereas the Drude weight representing the coherent charge dynamics critically increases above $(W/U)_{\mathrm{c}}$.
The strong correlation effect in the metallic state near the Mott criticality is thus manifested by a tiny Drude weight compared to the large spectral intensity of the remnant Mott-Hubbard gap transition even in the metallic state. Such a critical feature appears to be generic for the bandwidth control Mott transition systems~\cite{1998RMPImada}. 

\begin{figure}[tb]
\centering
\includegraphics[width=0.5\columnwidth]{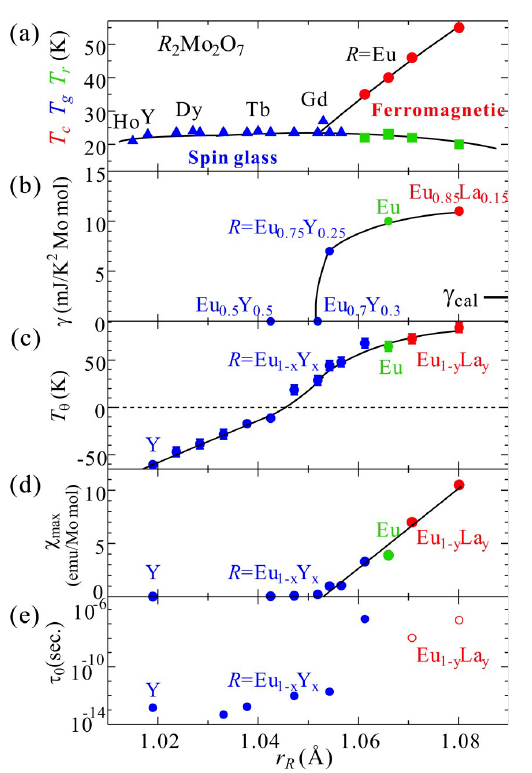}
\caption{\label{2-1_Hanasaki}
(a) Magnetic transition temperatures, $T_{\mathrm{C}}$ (ferromagnetic), $Tg$ (spin glass) and $Tr$ (reentrant spin glass), (b) electronic specific heat coefficient $\gamma $, (c) Curie-Weiss temperature $\Theta _{\mathrm{CW}}$, and (d) maximum value $\chi _{\mathrm{max}}$ of the ac magnetic susceptibility with variation of temperature for $R_2$Mo$_2$O$_7$. (e)spin flipping time $\tau _{0}$ estimated by the dynamical scaling of the ac susceptibility.
Reproduced with permission from \cite{2007PRLHanasaki}, Copyright (2007) by the American Physical Society.
}
\end{figure}

Figure~\ref{2-1_Hanasaki} shows the variation of various transport and magnetic characteristics in the course of metal-insulator transition in $R_2$Mo$_2$O$_7$ as a function of $r_R$~\cite{2007PRLHanasaki}.
First, the correlation effect in the metallic state is also manifested by the relatively large electronic specific heat coefficient $\gamma $ ($\sim 10$ mJ/K$^2$mol) satisfying that $C_{\mathrm{el}}=\gamma T$, as compared with the band calculation result ($\gamma _{\mathrm{cal}} \sim 2.5$ mJ/K$^2$mol~\cite{2003PRBSolovyev}) in Fig.~\ref{2-1_Hanasaki}(b).
The metal vs. insulator state is globally characterized by the switching of magnetism of ferromagnetic vs. antiferromagnetic interaction, as seen in the sign change of Curie-Weiss temperature $\theta _{\mathrm{CW}}$ [Fig.~\ref{2-1_Hanasaki}(c)].

As remarked above, the low temperature state in the antiferromagnetic and Mott-insulating regime is dominated by the spin glass state.
Typically, Y$_2$Mo$_2$O$_7$ without magnetic $R$-site has been examined in detail by measurements of neutron scattering to reveal that at temperatures below $T_{\mathrm{g}}$ low energy spin fluctuations are replaced by static antiferromagnetic short range order with a frozen staggered magnetization showing the four sublattice ordering with zero total magnetization~\cite{1999PRLGardner}.
The spin glass state can be analyzed with the scaling relation for the ac magnetic susceptibility; $\tau /\tau _{0}=[(T_{\mathrm{f}}-T_{\mathrm{g}})/T_{\mathrm{g}}]^{-z\nu }$~\cite{2001PRBMathieu}.
Here, $T_{\mathrm{f}}$ and $T_{\mathrm{g}}$ denote the peak temperature measured at the frequency $f=\tau ^{-1}$ and the spin glass transition temperature in the low frequency limit, respectively, while $z$ and $\nu $ denote the dynamical critical exponent and the critical exponent of the spin correlation length, respectively.
The spin glass state for Y$_2$Mo$_2$O$_7$ analyzed by the ac susceptibility shows the values of $z\nu =7.7$ with $T_{\mathrm{g}}=20.8$ K and the spin flipping time $\tau _{0}\sim 1\times 10^{-13}$ sec. This is typical of the atomic-scale spin glass caused by the geometrical frustration~\cite{2007PRLHanasaki}.

Such a frequency dependent anomaly in ac magnetic susceptibility is also observed in the metallic regime below the ferromagnetic transition temperature $T_{\mathrm{C}}$.
This is marked as the reentrant magnetic phase below $T_{\mathrm{r}}$. The similar scaling analysis for (Eu$_{0.85}$La$_{0.15}$)$_2$Mo$_2$O$_7$ at $r_R$(averaged)=1.06 \AA ~gives the long spin flipping time $\tau _{0}\sim 2\times 10^{-7}$ sec, suggesting that the reentrant phase is a spin glass state where the ferromagnetic cluster is randomly oriented~\cite{2007PRLHanasaki}.
Figure~\ref{2-1_Hanasaki}(e) shows the variation of the spin flipping time; the critical increase in $\tau _{0}$ from the atomic spin glass to the cluster spin glass occurs upon the insulator to metal transition accompanying the concomitant switching from antiferromagnetic to ferromagnetic interaction.  
Note the $R$ = Gd compound is situated in close vicinity of the Mott transition (as shown in Figs.~\ref{2-1_phasediagram} and \ref{2-1_Hanasaki}).
In the pyrochlore compounds, Mo conduction $4d$ electrons couple with $R$-ion localized $4f$ moments, as described in Sect.~\ref{sec:2_c-f interaction}.
This coupling plays a crucial role in the topological Hall effect observed in the metallic $R_2$Mo$_2$O$_7$, which will be discussed in the following subsections.
A unique situation for $R$ = Gd is that the Gd$^{3+}$($4f^7$) exhibits the maximum moment value ($\sim 7$ $\mu _{\mathrm{B}}$), while retaining the nearly isotropic character of a Heisenberg spin.
Thus, the application of an external magnetic field rather easily polarizes the Gd moments, which tends to polarize the Mo moments as well via the $f$-$d$ interaction and hence energetically favors the ferromagnetic metallic state.
In reality, even when the Gd and Mo moments in Gd$_2$Mo$_2$O$_7$ are almost fully polarized to the forcedly ferromagnetic state, the compound still remains Mott-insulating at the resistivity value several orders of magnitude higher than the Ioffe-Regel limit ($\rho \sim 5\times 10^{-4}$ $\Omega $cm).
However, under moderate hydrostatic pressure ({\it e.g.}, 1.8 GPa) the compound is observed to undergo the rather continuous insulator-metal transition with the increase of magnetic field (up to a few Tesla) from a localized regime ($\sigma =\rho ^{-1} \rightarrow 0$ as $T\rightarrow 0$) to a weak-localization regime ($\sigma =\mathrm{const.}+T^{1/2}$).
According to the finite-temperature scaling analysis, the obtained critical exponent $\mu =1.04\pm 0.04$ [$\sigma \sim (x-x_{c})^{\mu }$, where $x$ is a control parameter] is in between the values for the genuine Mott transition ($\mu =0.33$) and the Anderson transition of the noninteracting electron system ($\mu $ = 1.4-1.6), exemplifying the mixed character of Mott-Anderson transition~\cite{2006PRLHanasaki}. 

\subsection{High pressure effect and hole doping effect on the electronic phase diagram}
\label{sec:3_highpressure_holedoping}

\begin{figure}[tb]
\centering
\includegraphics[width=0.95\columnwidth]{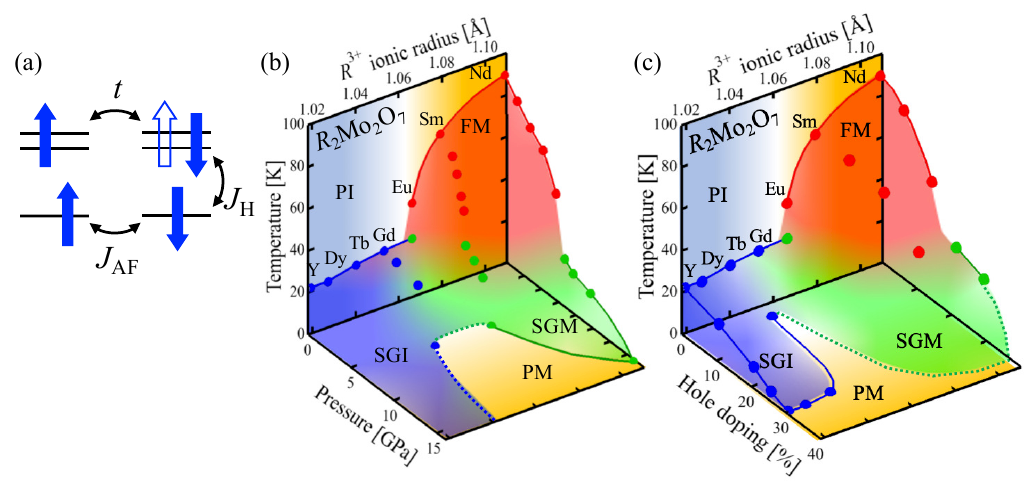}
\caption{\label{2-2_hole_phasediagram}
(a) Scheme of the transfer integral $t$ of $e_{\mathrm{g}}'$ conduction electron and the antiferromagnetic superexchange interaction $J_{\mathrm{AF}}$ under the onsite Coulomb interaction $U$ and Hund's-rule coupling $J_{\mathrm{H}}$ on the pyrochlore Mo lattice. The schematic electronic phase diagrams with varying (b) the hydrostatic pressure and (c) the hole doping for $R_2$Mo$_2$O$_7$.
}
\end{figure}


Application of high pressure is believed to be the most effective method to promote the insulator-metal transition near the Mott critical region. This is partly correct for the present case of $R_2$Mo$_2$O$_7$. However, the simple one-to-one relation between the ferromagnetism and the metallicity does not hold for $R_2$Mo$_2$O$_7$. Namely, the applied pressure always tends to drive the system toward metallic but not necessarily ferromagnetic. This is likely due to the trigonal crystal-field splitting of $t_{\mathrm{2g}}$ orbital and resultant complex competition of the double-exchange ferromagnetism and the super-exchange antiferromagnetism, as depicted in Fig.~\ref{2-2_hole_phasediagram}(a). Let us begin with a comprehensive review of the phase diagram in the $r_R$($W$)-pressure ($P$)-temperature ($T$) space, as depicted in Fig.~\ref{2-2_hole_phasediagram}(b)~\cite{2012PRLUedaMo}.
There, the application of pressure tends to extinguish the ferromagnetic (FM) state toward the paramagnetic (PM) state in the metallic regime at ambient pressure with large $r_R$. In the Mott insulating regime with smaller $r_R$, on the other hand, pressure tends to extinguish the spin glass (SG) phase and also stabilize the paramagnetic metallic (PM) state.

\begin{figure}[tb]
\centering
\includegraphics[width=0.95\columnwidth]{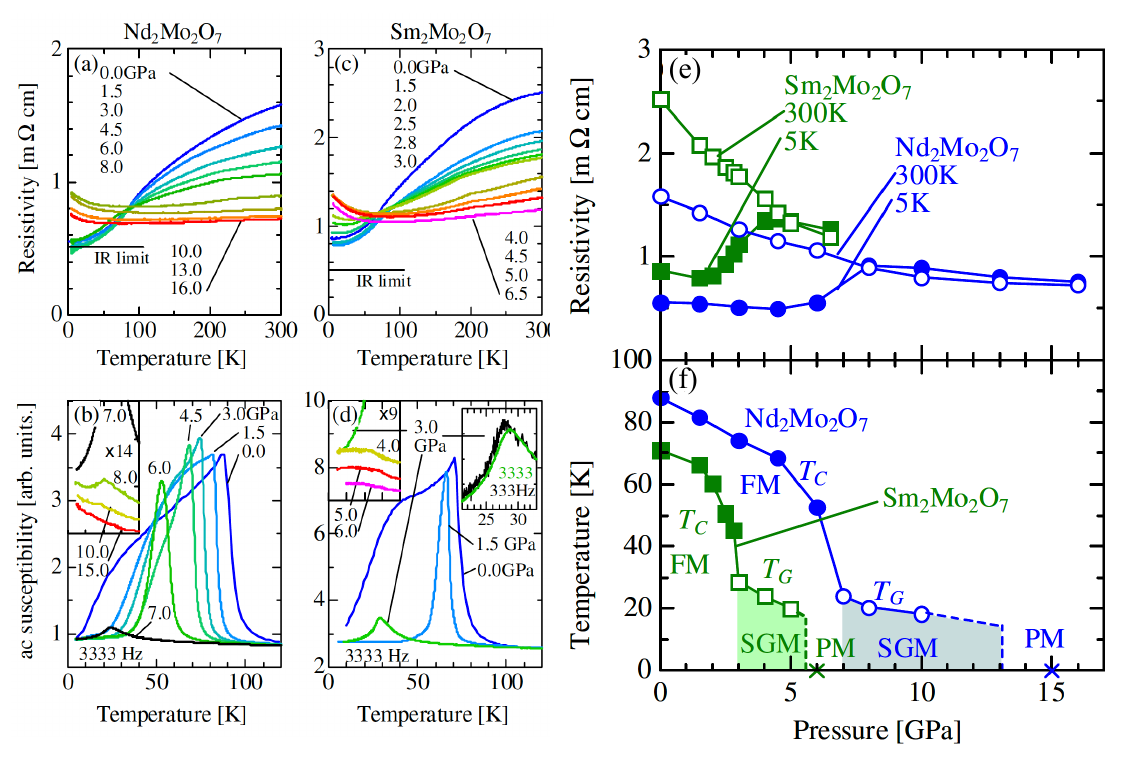}
\caption{\label{2-2_pressure}
Temperature ($T$) dependence of resistivity and ac magnetic susceptibility ($\chi _{\mathrm{ac}}$) for (a),(b) Nd$_2$Mo$_2$O$_7$ (NMO) and (c),(d) Sm$_2$Mo$_2$O$_7$ (SMO) under high pressure. Insets to (b) and (d) show the magnified view of ($\chi _{\mathrm{ac}}$) in the spin-glass or paramagnetic phase. The frequency dependence of the cusps [right inset of (d)] is observed for SGM phase. Pressure dependence of (e) the resistivity at 300 and 5 K and (f) the magnetic transition temperature for NMO and SMO. The abbreviations FM, SGM and PM mean ferromagnetic meal, spin-glass like metal, and paramagnetic metal, respectively. The cross symbols on the pressure axis show the disappearance of any magnetic transition.
Reproduced with permission from \cite{2009PRLIguchi}, Copyright (2009) by the American Physical Society.
}
\end{figure}

We show the examples of the effect of high-pressure application on the ferromagnetic metals, Nd$_2$Mo$_2$O$_7$ and Sm$_2$Mo$_2$O$_7$, in Fig.~\ref{2-2_pressure}~\cite{2009PRLIguchi}. Both compounds exhibit a qualitatively similar behavior of resistivity change with the increase of pressure. Temperature dependence of resistivity $\rho $ [Figs.~\ref{2-2_pressure}(a) and (c)] shows a rather sharp change above 7 GPa (3 GPa) for $R$ = Nd ($R$ = Sm) compounds from a well metallic behavior ($d\rho /dT<0$) to much less temperature-dependent behavior; in particular, the residual resistivity is rather sharply increased in a counter-intuitive manner at these critical pressures. With further application of pressure, however, the rather temperature-independent resistivity is again decreased, or partly recovering coherency in charge transport, while accompanying the up-turn in the low temperature region.
Corresponding to these critical pressures, the ac susceptibility data [Figs.~\ref{2-2_pressure}(b) and (d)] indicates the change of magnetic order from the ferromagnetic to spin-glass-like behavior. With further increase of pressure the ac susceptibility peak indicative of spin-glass state is extinguished at $P_{\mathrm{QCP}}$, 10-15 GPa (5-6 GPa) for $R$ = Nd ($R$ = Sm) compounds, indicating the emergence of the PM state.
The pressure-induced change of the electronic and magnetic states is more visible in Figs.~\ref{2-2_pressure}(e) and (f); the resistivity at 5 K is enhanced with pressure upon the transition to the spin-glass metal (SGM) phase, while that at 300 K decreases monotonously as in the conventional pressure effect. As evidenced by the further pressure-induced change (disappearance) of the ac susceptibility cusp, the SGM phase is turned to the paramagnetic metal (PM) phase. These complex behaviors may be attributed to the competition between the double-exchange interaction mediated by the $e_{\mathrm{g}}'$ electron Hund's-coupled to the localized $a_{\mathrm{1g}}$ electron spin and the super-exchange interaction of the $a_{\mathrm{1g}}$ electron spins. Above the possible quantum critical point $P_{\mathrm{QCP}}$ at the SGM-PM transition, the charge transport remains diffusive, indicating the strong quantum fluctuation of $a_{\mathrm{1g}}$ electron spins which are Hund's-coupled to the $e_{\mathrm{g}}'$ conduction electrons.
This may be interpreted in terms of the non-Fermi-liquid behavior in orbital-selective Mott systems with Hund's-rule coupling as predicted by dynamical mean-field (DMFT) theory~\cite{2005PRLLiebsch,2005PRLBiermann}.
Considering the local Dzyalonshinskii-Moriya interaction on the Mo tetrahedron in the pyrochlore lattice~\cite{2005PRBElhajal}, the short-range order in the SG phase and the spin correlation in the PM phase are likely to be composed of the noncoplanar spin correlation with SSC, $\textbf{S}_i\cdot (\textbf{S}_j\times \textbf{S}_k)$ with $\textbf{S}_i$, $\textbf{S}_j$, and  $\textbf{S}_k$ being the adjacent spins on the Mo tetrahedron (see Sect. \ref{sec:Scalar spin chirality: 2-in_2-out}). This is also the subject related to the topological Hall effect observed in the broad phase region of pyrochlore Mo-oxides, that is the issues reviewed in the following subsections.

One other means to control the metal-insulator phenomena in such a correlated system is to change the band filling. The band filling control in pyrochlore molybdates is performed by doping holes (reducing the band filling), {\it e.g.}, in a form of ($R_{1-x}$Ca$_{x}$)$_{2}$Mo$_{2}$O$_{7}$.
In spite of the limited examples of the crystal-synthetic study~\cite{2012PRLUedaMo,2021PRBHirschberger,2022PRBFukuda}, the magnetic orders, both ferromagnetic state and spin glass state, are observed to be converted with hole doping to the paramagnetic state even at low temperatures.
In particular, the spin glass (SG) insulator affected by the antiferromagnetic interaction is turned to the paramagnetic metal (PM), in a parallel behavior with the case of pressure application. Again, the PM state thus derived by hole doping procedure is of great interest as a possible candidate of non-Fermi liquid or chiral spin liquid state. 

\subsection{Scalar spin chirality and geometrical Hall effect}
\label{sec3:Spin chirality and geometrical Hall effect}

\subsubsection{Strong coupling case: spontaneous scalar spin chilarity.}

\begin{figure*}[tb]
\begin{center}
\includegraphics[width=1.6\columnwidth]{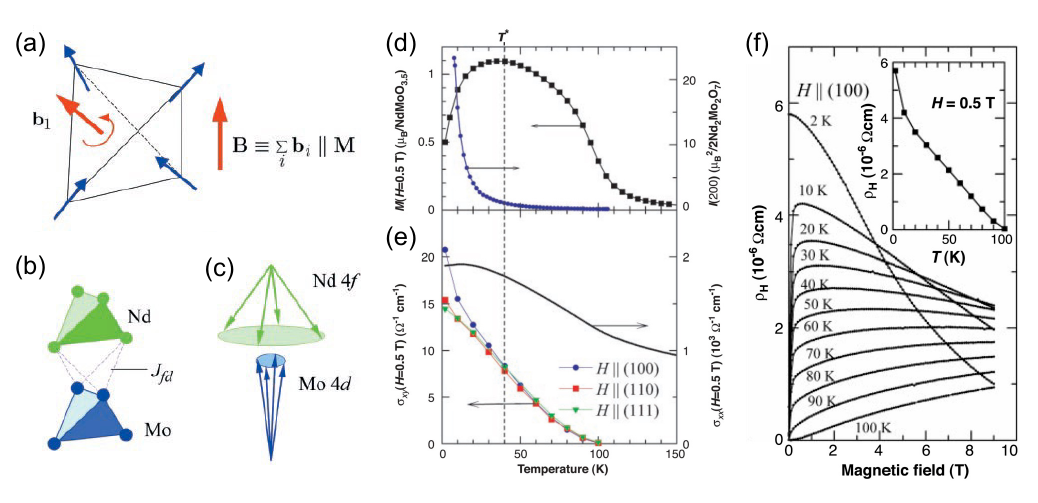}
\caption{\label{2-3_Mo_Hall1}
Schematic magnetic and crystal structures of pyrochlore. (a) Spin chirality in the  2-in 2-out spin structure, in which each spin points along the line that connects the center of the tetrahedron and the vertex. The total emergent magnetic field is the vector sum of each emergent magnetic flux that penetrates each plaquette. (b) Relative position of Nd tetrahedron (green circles) and Mo tetrahedron (blue circles) and $f$-$d$ coupling $J_{fd}$ in Nd$_2$Mo$_2$O$_7$ pyrochlore. (c) The local umbrella structure observed for Nd$_2$Mo$_2$O$_7$ by a neutron diffraction study. Magnetic unit cell contains four inequivalent Nd $4f$ moments $n_i$ and four Mo $4d$ moments $m_i$. (d) Temperature dependence of magnetization at $H=0.5$ T and the background-subtracted and properly normalized neutron scattering intensity $I$ of the (200) reflection, which represents the ordering of the transverse
component. (e) Temperature dependence of the longitudinal conductivity $\sigma _{xx}$ at $H=0.5$ T and the Hall conductivity $\sigma _{xy}$ at $H=0.5$ T. (f) Magnetic field dependence of Hall resistivity $\rho _{\mathrm{H}}$ with $H$// [100] at several temperatures. The inset shows the temperature dependence of $\rho _{\mathrm{H}}$ at 0.5 T, which is a measure of the spontaneous Hall component.  
From \cite{Taguchi2001}. Reprinted with permission from American Association for the Advancement of Science.
}
\end{center}
\end{figure*}

A most typical ferromagnet Nd$_2$Mo$_2$O$_7$ in the pyrochlore molybdate family shows a slightly canted Mo spin configuration at the zero-field ground state. This spin canting occurs due to the antiferromagnetic coupling between Nd $4f$ moment and Mo $4d$ spin.
The Nd $4f$ moment shows the strong Ising anisotropy whose easy axis is along $\langle 111\rangle $ axis, directing the respective $4f$ moments inward or outward between each apex and the center of mass of $R$ tetrahedron [Fig.~\ref{2-3_Mo_Hall1}(a)]. Then the $f$-$d$ interaction cants the originally collinear Mo spins, as shown in Figs.~\ref{2-3_Mo_Hall1}(b) and (c). The canted angle estimated by neutron scattering studies~\cite{Taguchi2001,2003JPSJYasui} is as small as 6-10$^{\circ }$. Nevertheless, this canting, which shows the temperature- and field-variation, can produce the nonvanishing sum of SSC, $\langle \textbf{S}_i\cdot (\textbf{S}_j \times \textbf{S}_k)\rangle $, or the real-space Berry curvature, and hence cause the large topological (or sometimes termed geometrical) Hall effect.

\begin{figure}[tb]
\centering
\includegraphics[width=0.95\columnwidth]{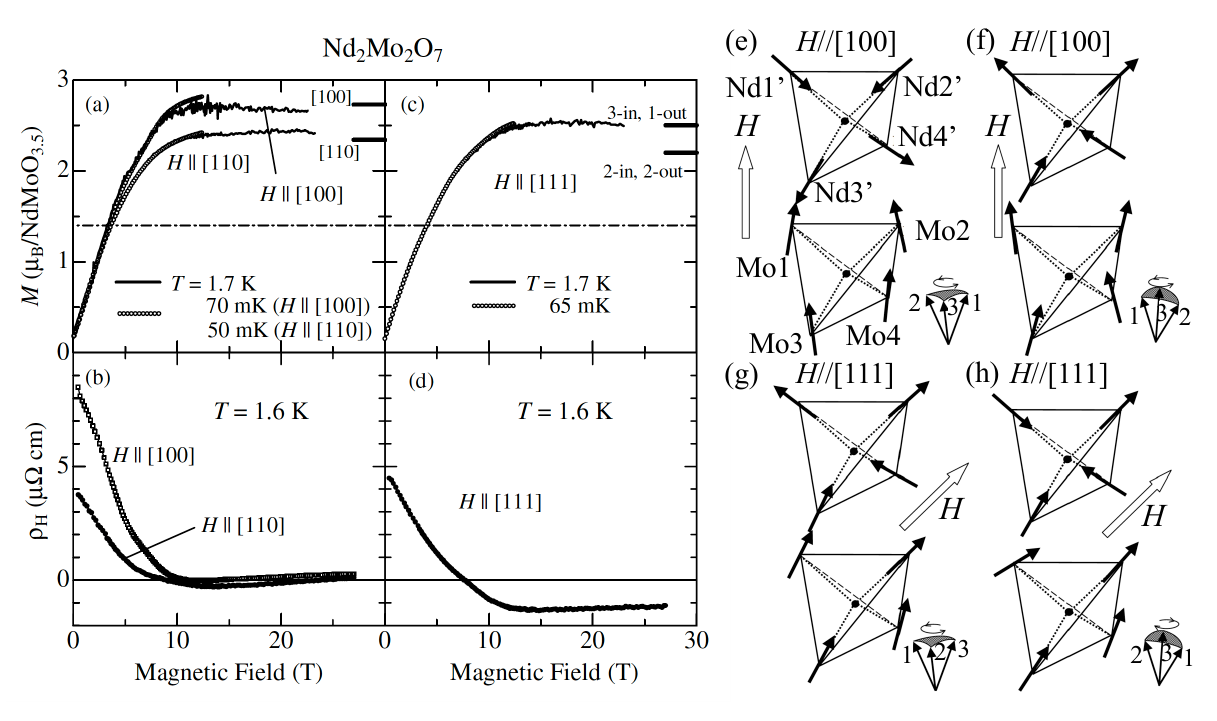}
\caption{\label{2-3_Mo_Hall2}
Magnetization curves for the applied field (a) along the [100] and [110] directions, and (c) along the [111] direction. $\rho _{\mathrm{H}}$ is plotted as a function of the magnetic field which is applied (b) along the [100] and [110] directions, and (d) along the [111] direction. In (a) and (c), the horizontal dot-dashed line represents the contribution from the Mo spin moment.
Thick horizontal short bars in (a) indicate the expected values of the saturation moment for each field direction under the assumption of strong Ising anisotropy of Nd moments. In (c), thick horizontal short bars indicate the expected values of magnetization when the magnetic configurations of all the Nd tetrahedra are ``3-in 1-out" and ``2-in 2-out", respectively.
(e)-(h): Schematic figures of spin configurations for Nd and Mo tetrahedra. In each panel, upper and lower tetrahedra correspond to Nd and Mo tetrahedra, respectively. The relationship among the Mo spin 1, 2, and 3 is schematically shown at the right-hand side of each Mo tetrahedron. Open arrows depict directions of applied magnetic field.
In (e) and (f) [(g) and (h)], the magnetic field is applied along the [100] ([111]) directions. For weak magnetic fields [(e) and (g)], Nd magnetic moments form the 2-in 2-out configuration for both magnetic field directions. As the magnetic field becomes sufficiently strong [(f) and (h)] and the Zeeman energy overcomes the $d$-$f$ interaction, the Nd magnetic configuration changes and hence affects the scalar spin chirality of Mo electrons.
Reproduced with permission from \cite{2003PRLTaguchi}, Copyright (2003) by the American Physical Society.
}
\end{figure}

\begin{figure}[tb]
\centering
\includegraphics[width=0.95\columnwidth]{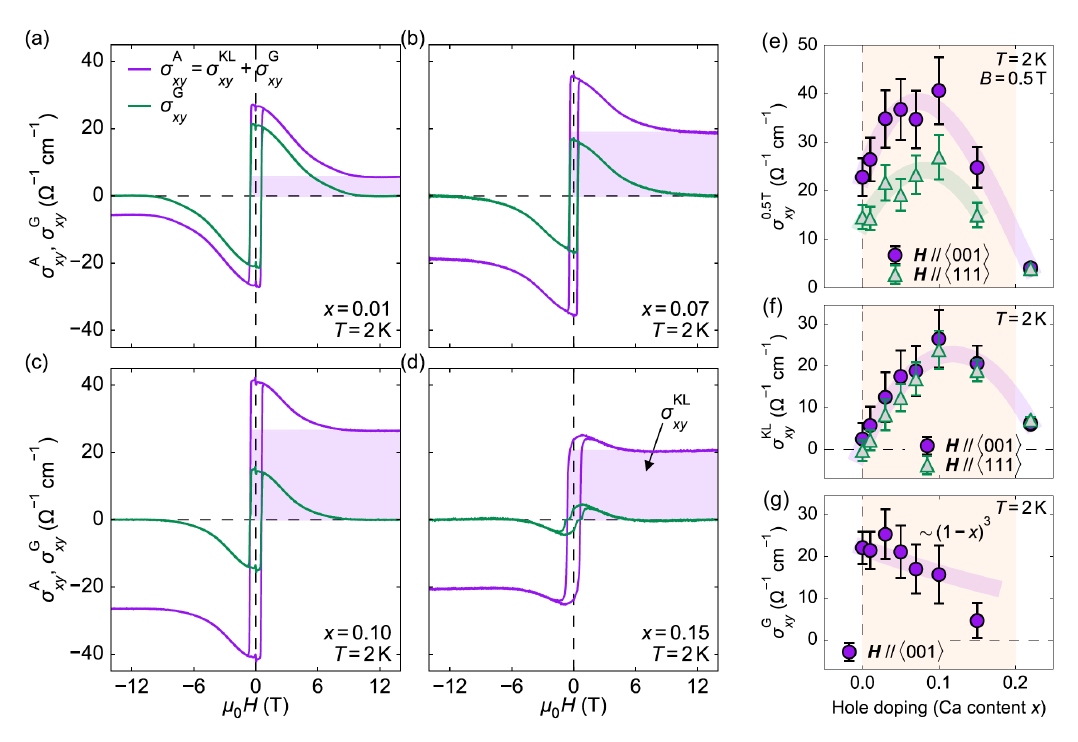}
\caption{\label{2-3_MaxHirschberger}
(a)-(d): Separation of Hall conductivities in hole-doped Nd$_2$Mo$_2$O$_7$, (Nd$_{1-x}$Ca$_x$)$_2$Mo$_2$O$_7$ at $T=2$ K and $H$//[001].
We show the anomalous (in the broad sense, including the topological component) Hall conductivity $\sigma _{xy}^{\mathrm{A}}$ (violet curves) after subtraction of the field-linear normal Hall signal and the geometrical (or topological) Hall signal $\sigma _{xy}^{\mathrm{G}}$ generated by SSC (green curves). The Hall signal originating from spin-orbit coupling $\sigma _{xy}^{\mathrm{KL}}$ is marked by a violet shaded box in each plot. Note that $\sigma _{xy}^{\mathrm{A}}=\sigma _{xy}^{\mathrm{KL}}+\sigma _{xy}^{\mathrm{G}}$. Hole doping dependence of (e) low-field Hall conductivity $\sigma _{xy}$(0.5 T) in (Nd$_{1-x}$Ca$_x$)$_2$Mo$_2$O$_7$, a proxy for the anomalous Hall signal $\sigma _{xy}^{\mathrm{A}}=\sigma _{xy}^{\mathrm{G}}+\sigma _{xy}^{\mathrm{KL}}$. (f) Karplus-Luttinger type Hall conductivity $\sigma _{xy}^{\mathrm{KL}}$ and (g) SSC-driven geometrical Hall conductivity $\sigma _{xy}^{\mathrm{G}}$. In (g), the thick purple line indicates the $(1-x)^3$ power law expected from dilution of rare-earth magnetic moments.
Reproduced with permission from \cite{2021PRBHirschberger}, Copyright (2021) by the American Physical Society.
}
\end{figure}

Figure~\ref{2-3_Mo_Hall1}(d) shows the temperature and magnetic-field dependences of the Hall effect, which are mostly dominated by the topological Hall effect overwhelming the anomalous Hall effect of spin-orbit-coupling origin (in the narrow sense) in proportion with the magnetization $M$ or the normal Hall effect in proportion to the applied magnetic field $B$, as argued in the following (Figs.~\ref{2-3_Mo_Hall2} and \ref{2-3_MaxHirschberger}).
At 2 K and zero field, the Mo spin moments ($\sim 1.4$ $\mu _{\mathrm{B}}$/Mo) direct nearly along the [100] axis with the umbrella form of the canting which is coupled antiferromagnetically to the 2-in 2-out Nd moments [Fig.~\ref{2-3_Mo_Hall1}(c)]. This is clearly evidenced by the gradual decrease of $M$ toward zero temperature from $T^*\sim 40$ K, where the antiferromagnetic $f$-$d$ coupling is switched on. The difference of the Hall resistivity $\rho _{\mathrm{H}}$ between $B$//[100] and $B$//[111] can be ascribed to the difference of the projection of 2-in 2-out like spin canting induced SSC onto the respective $B$ directions. As the temperature $T$ is increased, the field anisotropy in $\rho _{\mathrm{H}}$ tends to decrease, while the total magnitude of $\rho _{\mathrm{H}}$ is decreased but still survive above $T^*$ up to the ferromagnetic transition temperature. The characteristic thermal change of $\rho _{\mathrm{H}}$ near and above $T^*$ may be ascribed to the remnant SSC due to the thermal fluctuation of the Nd moment SSC and/or to the persistent Mo SSC due to the thermal spin fluctuation under the influence of Dzyaloshinskii-Moriya interaction~\cite{2021PNASKolincio,2023PRLKolincio}.
The $\rho _{\mathrm{H}}$ vs. $B$ curve shown in Fig.~\ref{2-3_Mo_Hall1}(d) provides the additional evidence for the averaged SSC as affected by the $B$-induced magnetic structure change and by the thermal fluctuation. At the lowest temperature 2 K, $\rho _{\mathrm{H}}$ is monotonously decreased with $B$ as opposed to the case of the simple anomalous Hall effect being proportional to $M$. This indicates that the Mo spin umbrella causing SSC [Fig.~\ref{2-3_Mo_Hall1}(c)] is successively closed with increasing $B$ to the forced spin-collinear ferromagnetic state, leading to the disappearance of topological Hall effect. On the other hand, at relatively higher temperature above $T^*$, $\rho _{\mathrm{H}}$ is rather field enhanced due to the alignment of the averaged SSC component along $B$.

Figure~\ref{2-3_Mo_Hall2} shows the magnetic field dependence of the field-anisotropic topological Hall effect and the magnetization at low temperatures, 1.6 K or below (down to 50 mK)~\cite{2003PRLTaguchi}. With the magnetic field $H$ along [100], $\rho _{\mathrm{H}}$ decreases to nearly zero around 10 T where the magnetization tends to nearly saturate to the value expected for the collinearly-aligned Mo spin moments plus the 2-in 2-out Nd moments.
This is in accord with the expectation of reduction in Mo SSC at high fields, corresponding to the magnetic structural change as depicted in Figs.~\ref{2-3_Mo_Hall2}(e) and (f). (The $H$//[110] case is similar except the difference of the projection of 2-in 2-out related configuration on to the [110] direction in the low field region.) In the case of $H$//[111] [Figs.~\ref{2-3_Mo_Hall2}(c) and (d)], on the other hand, $M$//[111] is also saturated above 12 T to the expected value of the nearly collinearly-aligned Mo spin moments plus the 3-in 1-out Nd moments. 
$\rho _{\mathrm{H}}$ tends to decrease towards a negative value with $H$, corresponding to the spin configuration change from the counter-clockwise Mo SSC antiferromagnetically coupled with the 2-in 2-out Nd moments near zero field [Fig.~\ref{2-3_Mo_Hall2}(g)] to the clockwise Mo SSC ferromagnetically coupled with the 3-in 1-out Nd moments at moderate field along [111] [Fig.~\ref{2-3_Mo_Hall2}(h)].
Such a sign change of Mo SSC appears to be manifested by the sign change of $\rho _{\mathrm{H}}$ in the course of increasing the $H$//[111] field across $\sim 7$ T. Incidentally, the high-field limit of the perfectly collinearly-aligned Mo spins would show no topological Hall effect for both cases of $H$//[100] and [111]; the discerned small values of $\rho _{\mathrm{H}}$ above 20 T may be attributed to the other origins of $\rho _{\mathrm{H}}$, such as the normal and anomalous Hall effects.

It is important to note the distinction between the topological Hall effect of the SSC origin and the intrinsic (Karplus-Luttinger type) anomalous Hall effect of the relativistic spin-orbit coupling origin~\cite{Nagaosa2010} in the pyrochlore Mo oxides.
Figure~\ref{2-3_MaxHirschberger} shows the respective variations of the topological Hall and anomalous Hall conductivities in the hole-doped Nd$_2$Mo$_2$O$_7$, (Nd$_{1-x}$Ca$_x$)$_2$Mo$_2$O$_7$ [see also the phase diagram for the holed-doped compounds in Fig.~\ref{2-2_hole_phasediagram}(c)]~\cite{2021PRBHirschberger}.
The whole Hall conductivity denoted as $\sigma _{xy}^{\mathrm{A}}$ shown in purple curves in Figs.~\ref{2-3_MaxHirschberger}(a)-(d) are composed of the sharply field-dependent part, representing the topological Hall component $\sigma _{xy}^{\mathrm{G}}$ (green curves) and the nearly field-independent anomalous Hall component $\sigma _{xy}^{\mathrm{KL}}$; the latter intrinsic (Karplus-Luttinger type) anomalous Hall conductivity (shaded in purple) is proportional with the Mo magnetization value along the field direction $M_z$ which is viewed nearly constant as a function of $H$//[001] apart from the tiny contribution of the closing umbrella of the Mo spins. While the noncoplanar spin texture is robust against the hole-doping concentration ($x$)~\cite{2021PRBHirschberger2}, the respective Hall components show the characteristic $x$ dependences; the topological component is rather insensitive to $x$ after the correction [$\sim (1-x)^3$] considering the effect of the Nd site dilution on the SSC formation [Fig.~\ref{2-3_MaxHirschberger}(g)], whereas the intrinsic anomalous Hall component is critically dependent on $x$, forming a dome-like curve [Fig.~\ref{2-3_MaxHirschberger}(f)]. This large difference was qualitatively explained in terms of the difference of the mechanism for Berry curvature generation in momentum space~\cite{2021PRBHirschberger}: The Karplus-Luttinger type intrinsic anomalous Hall effect possibly stems from all the band crossings of the spin-up and spin-down $4d$ (Mo) electron bands affected by the spin-orbit coupling. In the topological Hall effect of the SSC origin, by contrast, the Berry curvature is generated around the band crossing between the spin-up and spin-down bands without action of the spin-orbit coupling. An elaborate calculation could semi-quantitatively reproduce the observed features of the respective Hall effects~\cite{2021PRBHirschberger}, although the accurate prediction remains challenging, partly because of the strong electron correlation near the Mott criticality in the Mo $4d$ state in the pyrochlore oxides.

\begin{figure}[tb]
\centering
\includegraphics[width=0.95\columnwidth]{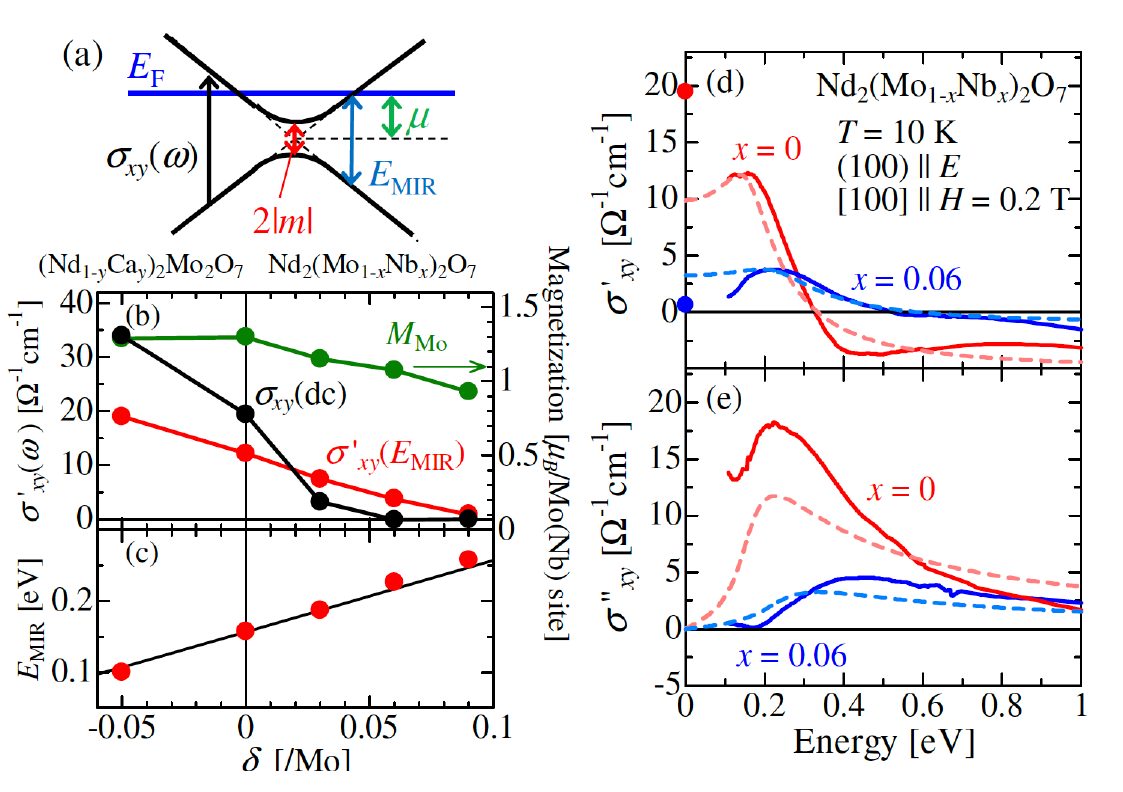}
\caption{\label{2-3_Kerr}
(a) The schematic view of the two-band model and its parameters $E_{\mathrm{F}}$, $\mu $, and $m$ representing the Fermi energy, the distance of the band-anticrossing point from $E_{\mathrm{F}}$, and half of the energy gap, respectively. (b) Hole ($y$) and electron ($x$) doping (band-filling change $\delta =x-y$) dependence of the dc anomalous Hall conductivity $\sigma _{xy}(\mathrm{dc})$, the peak magnitude of the real part of the optical Hall conductivity in the midinfrared region $\sigma _{xy}(E_{\mathrm{MIR}})$, and the estimated magnetization of the Mo spin ($M_{\mathrm{Mo}}$) at 10 K. (c) Doping dependence of the peak position $E_{\mathrm{MIR}}$ in $\sigma _{xy}(\omega )$ spectra located in the midinfrared region. The line shows the peak position shift corresponding to the slope $E_{\mathrm{MIR}}/\delta $ of 1 eV$\cdot $Mo. Comparison between the experimental spectra (solid lines) and the two-band model calculations (broken lines) for (d) the real and (e) the imaginary part of $\sigma _{xy}(\omega  )$ for $x=0$ and 0.06.
Reproduced with permission from \cite{2009PRLIguchi_Kerr}, Copyright (2009) by the American Physical Society.
}
\end{figure}

It is important to reveal the band crossing or the Weyl node generation in the band structure as the source of the Berry curvature of SSC origin. One such means to detect the Berry curvature signature is to examine the magneto-optical spectra in the infrared photon energy region.  We show the optical transverse (Hall) conductivity spectra $\sigma _{xy}(\omega  )=\sigma '_{xy}(\omega  )+i\sigma ''_{xy}(\omega )$ which are deduced from the magneto-optical Kerr spectra,  for Nd$_2$Mo$_2$O$_7$ and electron-doped counterpart Nd$_2$(Mo$_{1-x}$Nb$_x$)$_2$O$_7$ in Figs.~\ref{2-3_Kerr}(d) and (e)~\cite{2009PRLIguchi_Kerr}.
The large dc ($\omega  =0$) Hall conductivity $\sigma _{xy}$ at 10 K in Nd$_2$Mo$_2$O$_7$, which represents the topological Hall effect of SSC origin, appears to be dominated by the resonance spectral features of $\sigma _{xy}(\omega )$ around $\hbar \omega =0.2$ eV. It was interpreted in terms of the 2D model of a gapped Weyl node (band-anticrossing point) with the energy parameters, such as the Fermi energy ($E_{\mathrm{F}}$), the distance of the anti-crossing point from $E_{\mathrm{F}}(\mu )$, and a half of the energy gap ($m$), as shown in Fig.~\ref{2-3_Kerr}(a).
The interband transition at $\hbar \omega =E_{\mathrm{MIR}}$ ($\sim 0.2$ eV) can pick up the Berry curvature in this scheme. The calculated $\sigma _{xy}(\omega )$ spectra based on such a simplified quasi-2D model, in which the three-dimensionality arising from the $k_z$ dependence of the mass $m$ is approximately taken into consideration as a broadening factor due to the distribution of $m$,  are also shown in broken lines to fit the observed spectra in Figs.~\ref{2-3_Kerr}(d) and (e). The gap magnitude $m$ and the energy distance $\mu $ gives the comparable energy ($\sim 80$ meV) in this simple model for Nd$_2$Mo$_2$O$_7$ ($x=0$), while the photon energy of the maximum $\sigma ''_{xy}(2\mu )$ gives the good measure of the observed peak $E_{\mathrm{MIR}}$ ($\sim 0.2$ eV). The higher dc Hall conductivity than the $\omega  \rightarrow 0$ extrapolation of the optical Hall conductivity may indicate the contribution from the lower energy Weyl-node structures below 0.1 eV.
In the case of the electron-doped compound ($x=0.06$), the value of $2\mu $ as estimated by the observed $\sigma ''_{xy}$ peak is about 0.4 eV, signaling the upshift of $E_{\mathrm{F}}$ relative to the Weyl node. The systematic change of the dc topological Hall conductivity $\sigma _{xy}(\mathrm{dc})$, the $\sigma ''_{xy}$ peak photon-energy $E_{\mathrm{MIR}}$ ($\sim 2\mu $) and the optical Hall conductivity $\sigma '_{xy}(\omega  =E_{\mathrm{MIR}_\hbar })$ is plotted as a function of the hole ($y$) and electron ($x$) doping $\delta =x-y$ in Fig.~\ref{2-3_Kerr}(b), which is in accord with the expected behavior. Here, note that the electron doping (Nb doping) appears to cause the localization due to the strong disorder and leads to the rapid reduction of the topological Hall response with $y$, in particular in the dc limit.

\subsubsection{Weak coupling case: field-induced scalar spin chirality.}

\begin{figure}[tb]
\centering
\includegraphics[width=0.95\columnwidth]{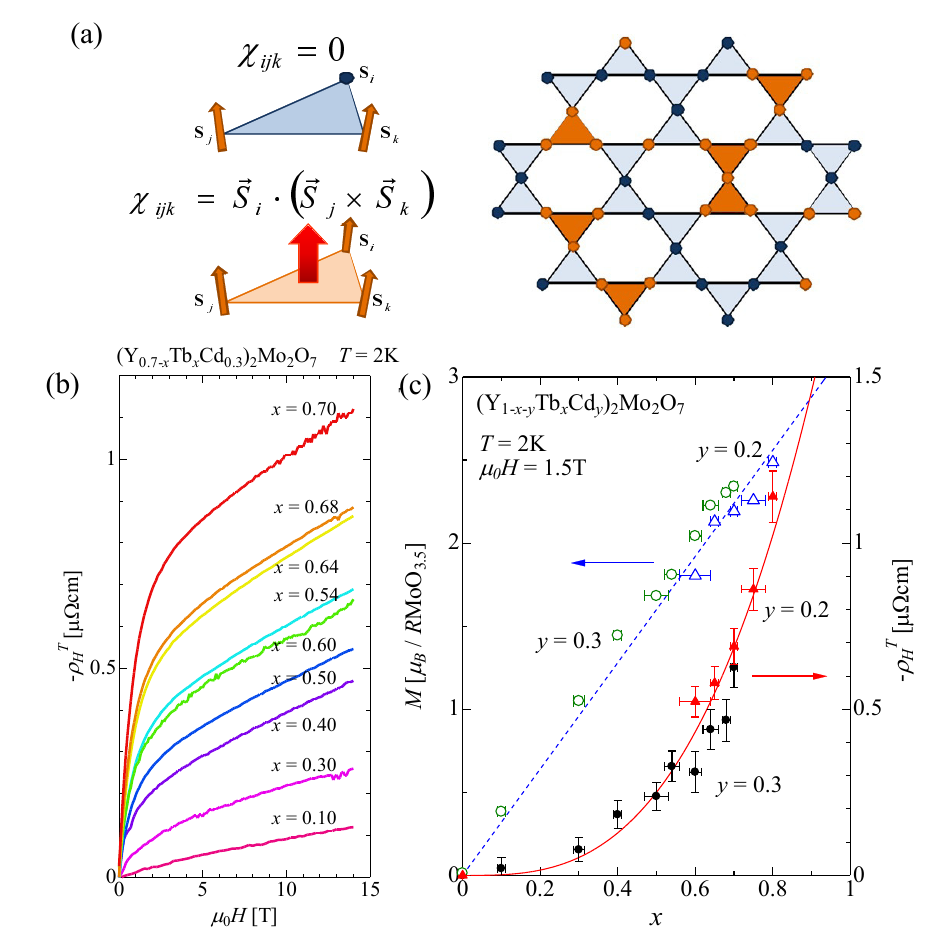}
\caption{\label{2-3_THE_Ueda}
(a) Scalar spin chirality (SSC) on two-dimensional kagome lattice as the 2D projection of the pyrochlore lattice. In the kagome/pyrochlore lattice with the diluted $R$ = Tb magnetic ions, SSC becomes nonzero only when the three neighboring sites are simultaneously occupied by Tb moments. (b) The magnetic-field ($H$) dependence of topological Hall resistivity $\rho _{mathrm
H}^{\mathrm{T}}$ for Tb content $x=0.1$-$0.7$ and hole doping $y=0.3$ in (Y$_{0.7-x}$Tb$_x$Cd$_{0.3}$)$_2$Mo$_2$O$_7$.
(c) The $x$ dependence of topological Hall resistivity (closed circles: $y=0.3$ and closed triangles: $y=0.2$) and magnetization (open circles: $y=0.3$ and open triangle: $y=0.2$) at $\mu _0 H=1.5$ T. Dashed and solid lines represent the references proportional to $x$ and $x^3$, respectively.
Reproduced with permission from \cite{2012PRLUedaMo}, Copyright (2012) by the American Physical Society.
}
\end{figure}

In the above examples, such as based on Nd$_2$Mo$_2$O$_7$, the ferromagnetic order of Mo moments is modified to gain the SSC due to the progressive order of the $R$ moments coupled antiferromagnetically to the Mo moments in decreasing temperature, which produces the topological Hall effect. Here, we show the contrastive case of the weak coupling~\cite{2002JPSJTatara}, where the originally paramagnetic Mo electrons can gain the Berry curvature via the coupling with the Ising-like $R$ ({\it e.g.}, Tb) moment SSC order induced by external magnetic fields.

We show the case of hole doped (Y$_{1-x-y}$Tb$_x$Cd$_y$)$_2$Mo$_2$O$_7$ ($y=0.3$, 0.2) in Fig.~\ref{2-3_THE_Ueda}. The parent compound (Y$_{1-y}$Cd$_y$)$_2$Mo$_2$O$_7$ is viewed as hole ($y$) doped in the Mott-insulating Y$_2$Mo$_2$O$_7$ via partial substitution of Y$^{3+}$ with Cd$^{2+}$ [see Fig.~\ref{2-2_hole_phasediagram}(c)] and shows the paramagnetic barely-metallic state down to the lowest temperature.
Then, as schematically shown in Fig.~\ref{2-3_THE_Ueda}(a), the nonmagnetic Y sites are further substituted with Tb ($x$) with the $4f$ moment of Ising character.
Considering that the mean free path of the conduction electron is comparable to the atomic spacing, the SSC is anticipated to be produced due to the noncoplanar neighboring Tb moments in proportion to the site occupancy of Tb on the neighboring triangular sites in the (Y, Tb, Cd) tetrahedron in pyrochlore lattice, {\it i.e.}, $\propto x^3$.
Figure~\ref{2-3_THE_Ueda}(b) shows the topological Hall component $\rho _{\mathrm{H}}^{\mathrm{T}}$ of the Hall resistivity which is deduced by subtracting the anomalous Hall component of spin-orbit-coupling origin observed for the $x=0$ (no Tb) compound from the Hall resistivity data for the respective-$x$ compounds. As shown in Fig.~\ref{2-3_THE_Ueda}(c), the magnetization value at relatively low fields ({\it e.g.}, at 1.5 T) is enhanced linearly with $x$. Considering the Ising-like nature of Tb moment, like Nd moment, the 2-in 2-out (3-in 1-out) configuration is dominant (less dominant) in a relatively low field region, which is responsible for the observed topological Hall resistivity  $\rho _{\mathrm{H}}^{\mathrm{T}}$ of the Mo conduction electrons via the coupling to the Tb moments with SSC.
This assignment is in accord with the observed scaling of $\rho _{\mathrm{H}}^{\mathrm{T}}$ at 1.5 T to the probability ($x^3$) of the neighboring three Tb-sites occupancy in the pyrochlore lattice, while the magnetization $M$ at 1.5 T is linear to $x$ [Fig.~\ref{2-3_THE_Ueda}(c)].

\begin{figure}[tb]
\centering
\includegraphics[width=0.95\columnwidth]{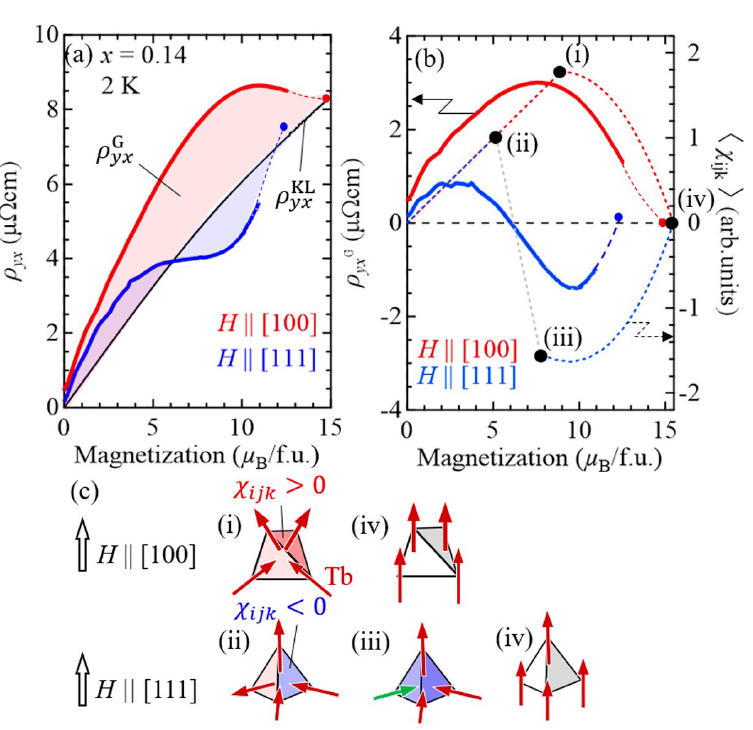}
\caption{\label{2-3_GHE_Fukuda}
(a) Magnetization dependence of the Hall resistivity composed of the spin-orbit-coupling-induced Karplus-Latttinge type anomalous Hall component $\rho _{yx}^{\mathrm{KL}}$ and geometrical (or topological) Hall component $\rho _{yx}^{\mathrm{G}}$. The solid lines denote the data obtained with the dc field, and the circles are the values at the maximum pulse field. The dotted lines are the smoothed data taken by the pulse field measurement. (b) Magnetization dependence of the geometrical Hall effect (GHE) or topological Hall effect and the simulated scalar spin chirality (SSC) $\chi _{ijk}=\textbf{S}_i\cdot \textbf{S}_j\times \textbf{S}_k$ where $\textbf{S}_i$, $\textbf{S}_j$, and $\textbf{S}_k$ are spins on the nearest-neighbor sites $i$, $j$, and $k$. $\langle \chi _{ijk}\rangle $ is SSC averaged over four triangle planes on a single tetrahedron.
Red and blue dashed lines represent the simulated SSC of the single Tb tetrahedron based on the magnetization curves. (c) Possible magnetic structure and SSC at each point in panel (b).
Reproduced with permission from \cite{2022PRBFukuda}, Copyright (2022) by the American Physical Society.
}
\end{figure}

The above investigated compounds are polycrystalline in nature as prepared under high pressure condition. The sample homogeneity was good and the transport characterization was well done, but the field-anisotropic behavior of the topological Hall effect, such as corresponding to the external field applied parallel to [100] vs. [111], cannot be resolved. Such a field-anisotropic topological Hall effect was successfully investigated in the weak coupling regime by using the single crystal of slightly holed doped pyrochlore (Tb$_{1-x}$Ca$_x$)$_2$Mo$_2$O$_7$ ($x=0.14$) which maintains the paramagnetic metallic state down to the lowest temperature~\cite{2022PRBFukuda}.
Figure~\ref{2-3_GHE_Fukuda}(a) shows the Hall resistivity $\rho _{yx}$ as a function of the magnetization $M$ for $H$//[100] and $H$//[111]. Note here that the Tb moment shows Ising-like character with the single-ion anisotropy along the local $\langle 111\rangle $ direction ({\it i.e.}, toward the center of the Tb tetrahedron) in the relatively low field region below 5 T, but increases above the Ising limit of 2-in 2-out ($H$//[100]) or 3-in 1-out ($H$//[111]) with further increasing $H$. For $H$//[100], for example, the application of $\mu _{0}H\sim 35$ T realizes the almost saturated magnetization expected for the fully aligned Tb moment along the field direction. Thus Tb$_2$Mo$_2$O$_7$ system can offer the opportunity to examine the magneto-transport properties for both collinear and noncollinear magnetic states on the single sample with use of high magnetic field up to 35 T. 
In general, the Hall resistivity is expressed as $\rho _{yx}=R_{0}\mu _{0}H+R_{\mathrm{S}}\rho _{xx}^{n} M_z+\rho _{yx}^{\mathrm{G}}$, where $R_0$ is the ordinary Hall coefficient, $R_{\mathrm{S}}\rho _{xx}^{n}$ is the anomalous Hall coefficient with the scaling factor $n$($\sim 1.4$) in the case of diffusive metal~\cite{2007PRLMiyasato}.
$\rho _{yx}^{\mathrm{G}}$ is geometrical (topological) contribution such as induced by SSC. In the present compound, the ordinary Hall effect is negligible. The intrinsic (Karplus-Luttinger type) anomalous Hall resistivity $\rho _{yx}^{\mathrm{KL}}=R_{\mathrm{S}}=\rho _{xx}^{n}M$ is shown in a black solid line in Fig.~\ref{2-3_GHE_Fukuda}(a), in which the correction due to the $M$ dependence of $\rho _{xx}^n$ is considered.
The distinct field-anisotropic behavior in $\rho _{yx}$ is observed for $H$//[100] (a red curve) and $H$//[111] (a blue curve). The typical topological contribution due to the exchange field from the SSC can be defined by the deviation of the observed $\rho _{yx}$ from $\rho _{yx}^{\mathrm{KL}}$; the extracted topological contribution, $\rho _{yx}^{\mathrm{G}}=\rho _{yx}-\rho _{yx}^{\mathrm{KL}}$, is shown in Fig.~\ref{2-3_GHE_Fukuda}(b) as a function of $M$. For  $H$//[100], the positive $\rho _{yx}^{\mathrm{G}}$ increases with $M$ up to the nearly full 2-in 2-out configuration attained around 9 $\mu _{\mathrm{B}}$/f.u. [at the point (i)], and then decreases toward zero at the fully aligned moment 15 $\mu _{\mathrm{B}}$/f.u. [at the point (iv)]; the corresponding variation of the Tb moment arrangement and the SSC is shown in Fig.~\ref{2-3_GHE_Fukuda}(c).
On the other hand, $\rho _{yx}^{\mathrm{G}}$ for $H$//[111] shows unique $M$ dependence. Below $M\sim 2$ $\mu _{\mathrm{B}}$/f.u., $\rho _{yx}^{\mathrm{G}}$ increases as in the case of $H$//[100], but decrease towards the negative value as the $M$ is further increased. It takes a negatively maximal value at $M\sim 9$ $\mu _{\mathrm{B}}$/f.u. at the point (iii) and then approaches zero at the larger magnetization value. The successive change of the Tb moment arrangement is assigned to (ii)$\rightarrow $(iii)$\rightarrow $(iv) in the course of increase of $M$ as shown in Fig.~\ref{2-3_GHE_Fukuda}(c). According to the neutron diffraction study~\cite{2010PRBEhlers} the Tb-Tb interaction in Tb$_2$Mo$_2$O$_7$ is ferromagnetic at weak magnetic fields, like the canonical spin-ice system. At the intermediate field applied along [111], the apical spins, whose easy axes are along the field direction, are fixed while the other three spins obey the ice rule, forming the so-called kagome ice state~\cite{Hiroi2003} with the 2-in 2-out configuration (see Sect.~\ref{sec:2_c-f interaction}) in state (ii).
Further increase of the magnetization, the Tb moment configuration is turned to the 3-in 1-out ordered state (iii) and then finally reaches the collinearly aligned state (iv) with vanishing SSC. The most important observation is the different sign of $\rho _{yx}^{\mathrm{G}}$ around the points of (i) and (iii), which corresponds to the positive SSC for the 2-in 2-out and the negative SSC for the 3-in 1-out. These Tb moment SSC is transmitted to the Berry curvature sign of the Mo conduction band via the $f$-$d$ coupling in this weak coupling regime.
Note the similar sign change of the topological Hall effect is also observed in Nd$_2$Mo$_2$O$_7$ for $H$//[111] in the strong coupling regime, as accompanied by the field-induced change of SSC, as described in Fig.~\ref{2-3_Mo_Hall2}.

\subsection{Summary}
\label{sec:3_summary}

In this section, we primarily focus on two emergent phenomena; quantum phase transitions in Sects.~\ref{sec:3_MIT} and ~\ref{sec:3_highpressure_holedoping} and topological Hall effect arising from SSC in Sect.~\ref{sec3:Spin chirality and geometrical Hall effect}.
In the former sections, we show that a variety of metal-insulator transitions are realized by the chemical/physical pressure and hole doping.
Among them, the ferromagnetic metal phase and the spin-glass insulating phase compete under the external pressure or hole doping, giving rise to the enigmatic metal with no long-range magnetic order.
This implies the possible realization of intriguing states such as non-Fermi liquid and chiral spin liquid, which would be worth exploring in future research.
In the latter section, we mainly discuss the topological Hall effect and its related optical responses.
Especially we provide a detailed and up-to-date discussion on the topological Hall effect, including the effect of hole doping and dilution of rare-earth Ising-type magnetic moments in the weak/strong coupling regimes.

\section{Ruthenates ($R$,$A$)$_2$Ru$_2$O$_7$}
\label{sec:4_Ruthenates}

Ruthenium compounds are known to exhibit a variety of interesting physical properties such as unconventional superconductivity~\cite{2003RMPMakenzie}, Mott transition~\cite{2000PRLNakatsuji}, and Kitaev spin liquid~\cite{2009PRLJackeli,2014PRBPlumb,2016NMBanerjee}, owing to several competing parameters including spin-orbit coupling, Hund's coupling, as well as electron correlation.
A representative material is perovskite ruthenium oxide, which has trivalent or tetravalent valence of Ru ions, or equivalently 4 or 5 electrons in $4d$ orbitals ($4d^{4}$ or $4d^{5}$).
Pyrochlore-type oxide can be viewed as the higher-valence counterparts ($4d^{3}$ or $4d^{4}$), and hence their physical properties have attracted much interest.
Thus far, pyrochlore ruthenates have been found to exhibit a wide range of interesting phenomena since there are a variety of choices for $A^{2+}$ or $R^{3+}$ cations.
In this section, we present the experimental findings in the following categories: trivalent rare earths $R^{3+}$ (Sect.~\ref{sec:4_rare-earth ruthenates}), divalent elements $A^{2+}$ = Ca, Cd, and Hg (Sect.~\ref{sec:4_divalent ruthenates}), mixed cations $R_{1-x}A_{x}$ (Sect.~\ref{sec:4_filling-controlled ruthenates}), and heavy elements $A$ = Tl, Pb, and Bi (Sect.~\ref{sec:4_heavy A ruthenates}).



\subsection{Electronic and magnetic properties in rare-earth ruthenates}
\label{sec:4_rare-earth ruthenates}

\subsubsection{Basic properties.}


\begin{figure}[h]
\begin{center}
\includegraphics[width=0.7\columnwidth]{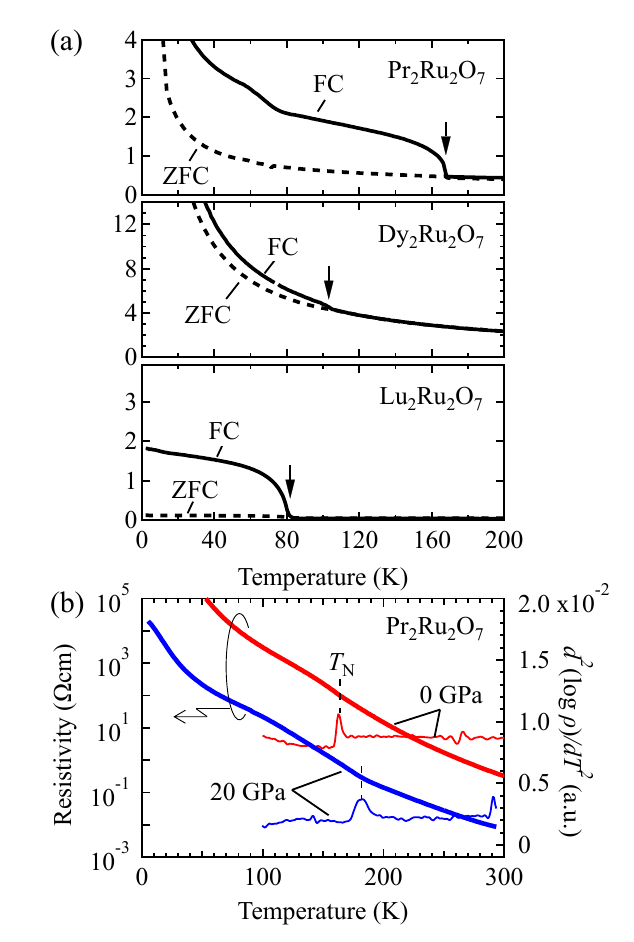}
\caption{
(a) Temperature dependence of magnetization for several $R_{2}$Ru$_{2}$O$_{7}$. The solid curves denote the magnetization on the field-cooling (FC) process and the dashed curves denote that on the zero-field-cooling (ZFC) process, respectively. The arrows indicate the magnetic transition temperature.
(b) Temperature dependence of resistivity (thick curves) for Pr$_{2}$Ru$_{2}$O$_{7}$ at ambient pressure (red) and 20 GPa (blue). The right axis shows the second-order temperature derivative of resistivity $d^{2}(\mathrm{log}\rho )/dT^{2}$ (thin curves).
Reproduced with permission from \cite{2020PRBKaneko}, Copyright (2020) by the American Physical Society.
}
\label{3-1_basic_final}
\end{center}
\end{figure}

The $R^{3+}$ site of pyrochlore ruthenates ($R_2$Ru$_2$O$_7$) contains all rare-earth and Y ions but La and Ce.
The nominal Ru$^{4+}$ ($4d^{4}$) indicates that two electrons are accommodated in $e_{g}'$ orbitals while one electron is occupied in $a_{1g}$ orbital (see Sect.~\ref{sec:2_4dcase}).
The transition temperature systematically decreases as the $R$ ionic radius decreases, from 162 K for $R$ = Pr compound with the largest $R$ ionic radius, to 80 K for $R$ = Lu with the smallest~\cite{2001JPCSIto}.
It is presumably due to the fact that the change of $R$ ionic radius modulates the Ru-O-Ru bond angle that affects the supertransfer interaction between neighboring Ru sites via O $2p$ states.
Figure~\ref{3-1_basic_final}(a) shows the temperature dependence of magnetization for respective $R$ compounds~\cite{2020PRBKaneko}.
The magnetization on the field-cooling (FC) process increases rapidly and deviates from that on zero field-cooling (ZFC) process below the transition temperature.
Such a temperature dependence can be attributed to the magnetic transition from the paramagnetic to the canted antiferromagnetic state, as suggested by the neutron experiment~\cite{2001JPCSIto}; see the next Sect.~\ref{sec:4_ruthenates magnetic structure} for details.
We note that the magnetization strongly depends on $R$ ions.
For example, the difference between ZFC and FC is large for nonmagnetic $R$ = Lu compounds, whereas the Curie-Weiss like temperature dependence seems dominant for $R$ = Dy.
For $R$ = Pr compound, an additional anomaly is observed at temperatures around 80 K which is much lower than Ru ordering temperature $\tn $.
These are attributable to the coexistent order of the $R$ magnetic moment; on the one hand, the magnetization includes the contribution of $R$ moments, and on the other hand, the Ru ordering configuration is changed by the interaction with the $R$ moments (see Sect.~\ref{sec:4_ruthenates magnetic structure}).

The resistivity shows insulator-like behavior for all $R_2$Ru$_2$O$_7$.
We show the representative data of Pr$_2$Ru$_2$O$_7$ in Fig.~\ref{3-1_basic_final}(b)~\cite{2020PRBKaneko}.
The resistivity increases monotonically with decreasing temperature.
A small anomaly appears at $\tn $, as can be seen more clearly in $d^{2}(\mathrm{log}\rho )/dT^{2}$ on the right ordinate in Fig.~\ref{3-1_basic_final}(b).
By applying a pressure of 20 GPa, the resistivity is reduced by two orders of magnitude while keeping the semiconducting like temperature dependence.
Further, $\tn $ increases by $\sim 20$ K under the pressure of 20 GPa, meaning that $\tn $ increases continuously with a large decrease of the effective electronic correlation tuned by both varying $R$ ionic radius and applying a pressure.


\begin{figure}[h]
\begin{center}
\includegraphics[width=0.95\columnwidth]{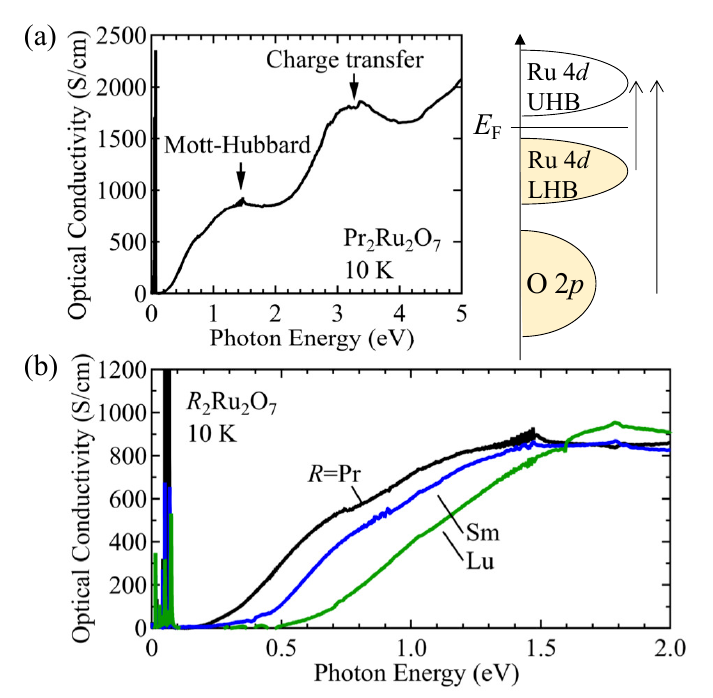}
\caption{
(a) Optical conductivity for Pr$_2$Ru$_2$O$_7$ at 10 K. The right panel shows the schematic diagram of the electronic structure for Pr$_2$Ru$_2$O$_7$. UHB and LHB stand for the upper and lower Hubbard bands, respectively.
(b) Optical conductivity spectra for $R$ = Pr, Sm, Lu compounds at 10 K. 
Reproduced with permission from \cite{2020PRBKaneko}, Copyright (2020) by the American Physical Society.
}
\label{3-1_optics}
\end{center}
\end{figure}

Spectroscopic study provides an important insight into the electronic state of $R_2$Ru$_2$O$_7$.
Figure~\ref{3-1_optics}(a) shows the optical conductivity spectra of Pr$_2$Ru$_2$O$_7$ below 5 eV.
There are two hump structures centered at 1.5 eV and 3.2 eV, which are attributed to the optical transitions to the upper Hubbard band from the lower Hubbard band and O $2p$ band, respectively~\cite{2020PRBKaneko}.
Similar spectra are also reported for Y$_2$Ru$_2$O$_7$~\cite{2001PRBLee}.
Figure~\ref{3-1_optics}(b) shows the magnified view of optical conductivity spectra below 2 eV for several $R$ compounds.
No charge excitation is observed in the low energy region, in accord with the insulator-like dc transport properties in Fig.~\ref{3-1_basic_final}(b).
Extrapolating the rising part of the optical conductivity with a straight line, the magnitude of the charge gap is estimated to be approximately 0.28 eV for the $R$ = Pr compound.
The charge gap increases systematically with decreasing $R$ ionic radius or with increasing the electron correlation, reaching $\sim 0.6$ eV for the $R$ = Lu compound.
The respective energy scales are consistent with the results of photoemission spectroscopy on Sm$_2$Ru$_2$O$_7$~\cite{2006PRBOkamoto}.

\begin{figure}[h]
\begin{center}
\includegraphics[width=0.95\columnwidth]{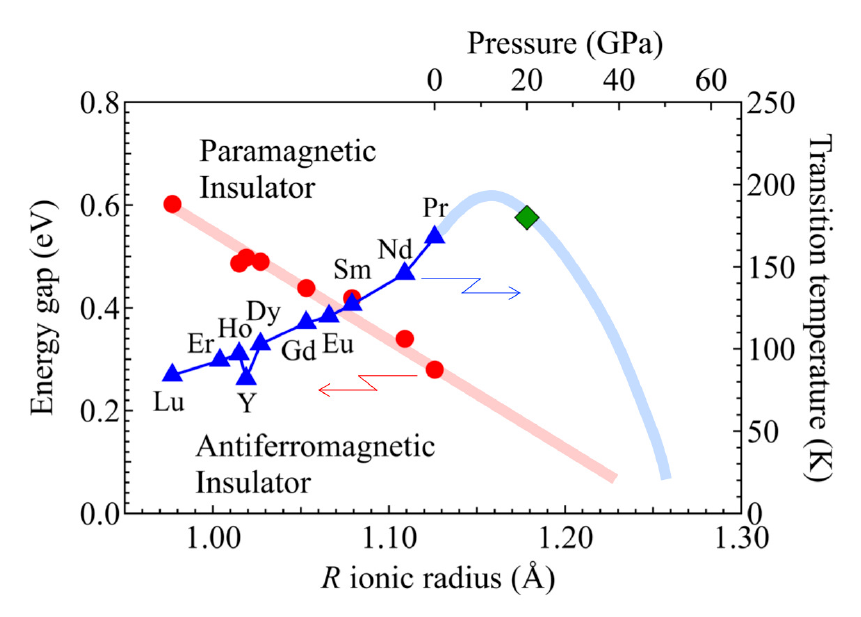}
\caption{
Phase diagram of $R_2$Ru$_2$O$_7$ plotted on the basis of Ref.~\cite{2020PRBKaneko}. The antiferromagnetic transition temperature (blue triangles) and the magnitude of the charge gap (red circles) are plotted as functions of $R$ ionic radius.
The green square represents the transition temperature for Pr$_2$Ru$_2$O$_7$ at 20 GPa.
The upper scale for the hydrostatic pressure on Pr$_2$Ru$_2$O$_7$
is calculated by an empirical relation observed in $R_2$Ir$_2$O$_7$~\cite{2015PRBUeda}.
}
\label{3-1_phasediagram}
\end{center}
\end{figure}

The above results are summarized in Fig.~\ref{3-1_phasediagram} where the abscissa is the $R$ ionic radius, and the charge gap (left) and $\tn $ (right) are plotted in the ordinates \cite{2020PRBKaneko}.
The magnitude of the external pressure is obtained from the scaling relation between the $R$ ionic radius and the external pressure, which is known for $R_{2}$Ir$_{2}$O$_{7}$ containing the same transition-metal ionic radius \cite{2015PRBUeda}.
The size of the charge gap decreases almost linearly with increasing $R$ ionic radius, implying that the change in $R$ ionic radius can be regarded as a good indicator of the effective electronic correlation $U/t$.
On the other hand, $\tn $ apparently increases as the $R$ ionic radius and external pressure increases.
This indicates that the super-exchange interaction ($J\propto t^2/U\propto W^2/U$) may play a dominant role in the antiferromagnetic order, as widely observed in strongly correlated systems~\cite{1968MRBGoodenough}.
If $U/t$ is further reduced by, for instance, applying pressure over 20 GPa on $R$ = Pr, $\tn $ would reach a maximum at a certain point and then sharply decreases, accompanied by a Slater-like metal-insulator transition (Fig.~\ref{3-1_phasediagram}).

\subsubsection{Magnetic structure.}
\label{sec:4_ruthenates magnetic structure}

Magnetic interactions between $R$ and Ru sublattices play an important role in the magnetic states of both $R$ and Ru spins, resulting in a variety of magnetic ordering patterns in different $R$ compounds \cite{2022IOPHuebsch}.


We firstly introduce the magnetic state of Y$_2$Ru$_2$O$_7$ that hosts nonmagnetic Y ions.
The absolute value of the Curie-Weiss temperature $|\theta _{\mathrm{CW}}|$ is 1250 K while the transition temperature $T_{\mathrm{N}}$ is 75 K, resulting in the huge magnetic frustration index $f=\theta _{\mathrm{CW}}/T_{\mathrm{N}}\sim 17$.
Neutron scattering offers an insight into the magnetic structure~\cite{2001JPCSIto}.
As the temperature is lowered below $T_{\mathrm{N}}$, the peak intensities of 111 and 220 reflections become larger with no superlattice reflections.
Since no sign of lattice distortion is observed in infrared~\cite{2004PRBLee} and Raman~\cite{2006PRBBae} spectroscopy, the increases of the peak intensities in neutron scattering manifest a long-range magnetic order with the magnetic propagation vector $\textbf{q}=0$.
The integrated intensities coincide well with the calculated structure factor assuming a tilted antiferromagnetic ordering pattern with the Ru magnetic moment of 1.36 $\mu _{\mathrm{B}}$.
Furthermore, the inelastic neutron scattering and M\"{o}ssbauer spectroscopy suggests the spin excitation gap of 11 meV, reflecting easy-axis or easy-plane anisotropy~\cite{2006PRBKmiec,2008PRBDuijn}.
Considering that the antiferromagnetic interaction is dominant, the all-in all-out magnetic configuration is plausible \cite{1998JPCMBramwell}.

\begin{figure}[h]
\begin{center}
\includegraphics[width=0.95\columnwidth]{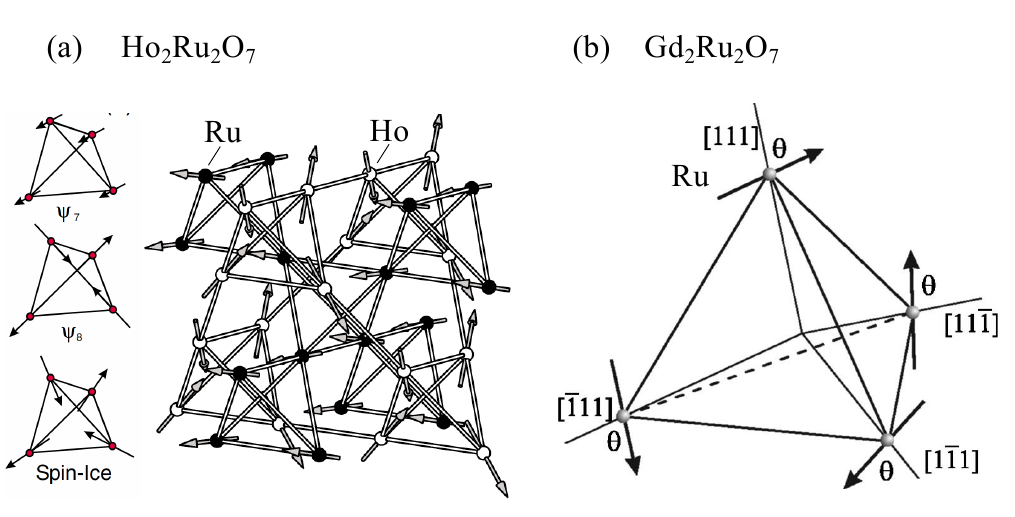}
\caption{
Schematic magnetic structure of (a) Ho$_2$Ru$_2$O$_7$ and (b) Gd$_2$Ru$_2$O$_7$.
(a) Reprinted figure with permission from \cite{2004PRLWiebe}, Copyright (2004) by the American Physical Society.
(b) Reprinted figure with permission from \cite{2007PRBGurgul}, Copyright (2007) by the American Physical Society.
}
\label{3-1_magstructure}
\end{center}
\end{figure}

On the other hand, magnetic $R$ compounds exhibit different spin configurations.
Neutron scattering measurements reveal that Ho$_2$Ru$_2$O$_7$ undergoes two magnetic transitions at 95 K and 1.4 K \cite{2004PRLWiebe}.
The magnetic Bragg peak with $\textbf{q}=0$ allows the ordering pattern belonging to the irreducible representation $\Gamma _{9}$ (T$_{1g}$) of the space group $Fd\overline{3}m$.
Thus, based on the calculation of the magnetic structure factor for the linear combination of two structures $\Psi _{7}$ and $\Psi _{8}$ [Fig. \ref{3-1_magstructure}(a)], Ru spin configuration is close to the 2-in 2-out state below 95 K.
Additional Bragg peaks show up below 1.4 K, indicative of the ordering of Ho spins which follows the representation $\Gamma _{9}$ as well.
Figure \ref{3-1_magstructure}(a) shows the possible magnetic ordering patterns of Ho and Ru at the base temperature, which explain the neutron data quite well.
In this configuration, the ordering of Ho spins relaxes the frustration of Ru moments and thereby makes it almost collinear ferromagnetic state.
On the other hand, the Ho moments are tilted from the ideal 2-in 2-out state because of the large modulation of the crystal fields due to the Ru spin ordering.
As a result, the Ho and Ru sublattices are antiferromagnetically aligned.
Gd pyrochlore oxides with nonmagnetic transition-metal ions frequently show noncollinear antiferromagnetic long-range ordering near 1 K \cite{2002JPSJMatsuhira}.
For example, in Gd$_2$Ti$_2$O$_7$, Gd magnetic moments show long-range ordering at 1.1 K and 0.75 K with the magnetic propagation vector $\textbf{q}$ = (0.5,0.5,0.5)~\cite{2001PRBChampion,2004JPCMStewart,2021npjQMPaddison}.
Because of the magnetic interaction between Gd and Ru spins, the magnetism of Gd$_2$Ru$_2$O$_7$ is more complicated.
Ru spins undergo a long-range order at 113 K in Gd$_2$Ru$_2$O$_7$~\cite{2002JMCTaira}.
Additionally, below 35 K, a broad peak is observed in specific heat while the magnetization deviates from the Curie-Weiss law, implying the freezing of Gd magnetic moments.
$^{155}$Gd M\"{o}ssbauer spectroscopy uncovers that the hyperfine field at Gd site ({\it i.e.}, Gd magnetic moment) gradually increases with decreasing temperature below 113 K and rapidly increases below 35 K~\cite{2007PRBGurgul}.
The direction of Gd moments is perpendicular to the local three-fold rotational symmetry axis $\langle 111\rangle$.
On the other hand, in the spectroscopy with $^{99}$Ru, the exact direction cannot be determined because the relationship between the hyperfine field and the Ru spin direction is not trivial.
Considering that the Ru spin is nearly perpendicular to $\langle 111\rangle$ in nonmagnetic $R$ compound Y$_2$Ru$_2$O$_7$~\cite{2006PRBKmiec},  Ru spins can be tilted by 72$^{\circ }$ from $\langle 111\rangle$ [Fig.~\ref{3-1_magstructure}(b)].

The above magnetic structures are well reproduced by the first-principles calculation devising the combination scheme of cluster-multipole and spin-density-functional theory \cite{2022IOPHuebsch}.

\subsection{Electronic and magnetic properties in divalent $A$ = Ca, Cd, Hg ruthenates}
\label{sec:4_divalent ruthenates}

$A^{2+}_{2}$Ru$^{5+}_{2}$O$_7$ hosts Hund's-rule coupled three electrons in the $4d$ $t_{2g}$ orbitals and hence no orbital degeneracy (Sect.~\ref{sec:2_4dcase}).
Although it is a half-filled state, the electron correlation in this system seems weaker than that in $R^{3+}_{2}$Ru$^{4+}_{2}$O$_7$; for instance, the resistivity of $A^{2+}_{2}$Ru$^{5+}_{2}$O$_7$ is much smaller than that of $R^{3+}_{2}$Ru$^{4+}_{2}$O$_7$.
This counter-intuitive behavior can be attributed to the strong hybridization of Ru $4d$ $t_{2g}$ and O $2p$ orbitals owing to the high-valence Ru$^{5+}$ ions, which induces a variety of physical properties.

\begin{figure}[h]
\begin{center}
\includegraphics[width=0.70\columnwidth]{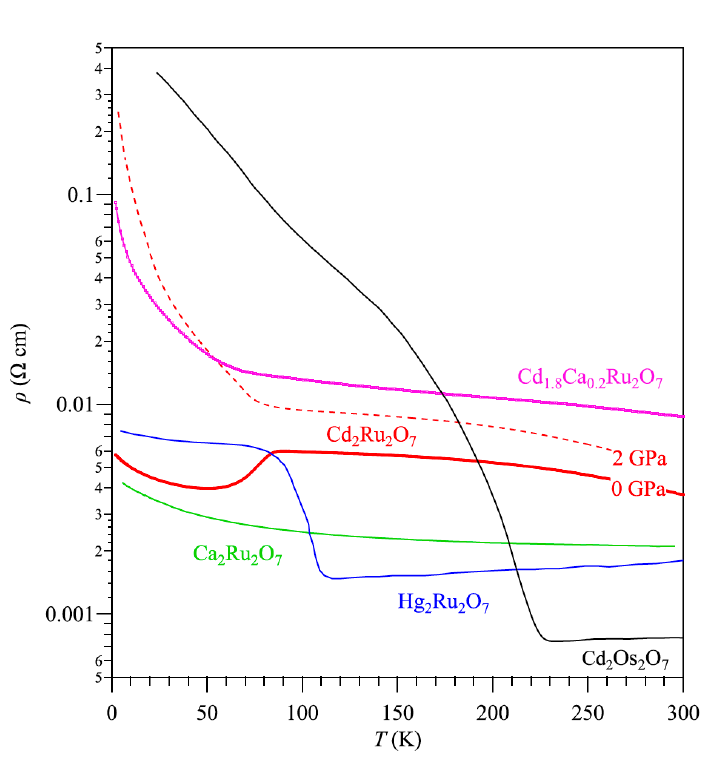}
\caption{
Temperature dependence of resistivity for several $A^{2+}$ ruthenates and the $5d$ counterpart Cd$_2$Os$_2$O$_7$.
The red curves show the resistivity for Cd$_2$Ru$_2$O$_7$ under ambient pressure (solid curve) and 2 GPa (dashed curve).
Reprinted figure with permission from \cite{2018PRBJiao}, Copyright (2018) by the American Physical Society.
}
\label{3-2_resistivity}
\end{center}
\end{figure}

Figure~\ref{3-2_resistivity} shows the temperature dependence of resistivity for polycrystalline $A^{2+}$ = Ca, Cd, and Hg ruthenates as well as $5d$ counterpart Cd$_2$Os$_2$O$_7$ (Os$^{5+}$: $5d^{3}$) for comparison \cite{2018PRBJiao}.
The resistivity monotonically increases as the temperature decreases for Ca$_2$Ru$_2$O$_7$, while those for Cd$_2$Ru$_2$O$_7$ and Hg$_2$Ru$_2$O$_7$ show some anomalies at low temperatures despite isovalent $A$ ions.
Similar transport properties are observed in single crystals grown by hydrothermal synthesis~\cite{1998MRBWang,2006JPSJMunenaka,2007JMCKlein}.
Such a difference is attributed to the cation covalency (Ca$<$Cd$<$Hg), rather than the Ru-O-Ru bond angle distortion~\cite{2010PRBMiyazaki}.

For Ca$_2$Ru$_2$O$_7$, the magnetization shows a typical Curie-Weiss temperature dependence at high temperatures and an anomaly at 25 K, despite no discernible anomaly in resistivity.
The estimated magnetic moments are 0.25-0.36 $\mu _{\rm B}$/Ru and the Curie-Weiss temperature is $\sim -5$ K.
The nonlinear susceptibility divergently increases as the temperature approaches to the transition temperature, indicative of the spin glass transition~\cite{2006JPSJMunenaka}.
Spin-glass-like behavior is also observed by $\mu $SR spectroscopy~\cite{2010PRBMiyazaki}.
However, there remain several issues about the origin of this spin state.
One is that the Curie-Weiss temperature is quite small compared to the transition temperature, which is quite different from the case of the isostructural spin-glass compound Y$_2$Mo$_2$O$_7$ \cite{2997PRLGingras}.
The other is that the $\mu $SR spectra observed in Ca$_2$Ru$_2$O$_7$ is similar to those for dilute alloy systems where the randomness of Ruderman-Kittel-Kasuya-Yosida (RKKY) interaction among localized magnetic moments play a key role.
These indicate the distinct origin and nature of spin-glass state in Ca$_2$Ru$_2$O$_7$.
As shown in Fig.~\ref{3-2_resistivity}, Hg$_2$Ru$_2$O$_7$ undergoes the phase transition at around 100 K.
At this temperature, the lattice constant and susceptibility also show discontinuous changes and hence it appears to be a structural phase transition, similar to the case of Cd$_2$Ru$_2$O$_7$ and Tl$_2$Ru$_2$O$_7$~\cite{2006NMLee}; see also Sect.~\ref{sec:4_heavy A ruthenates}.
However, in contrast to Tl$_2$Ru$_2$O$_7$, Ru$^{5+}$ ions have no orbital degeneracy and is hence Jahn-Teller inactive in Hg$_2$Ru$_2$O$_7$.
There are several suggestions for the mechanism of the metal-insulator transition.
One is the charge transfer between Ru-O and Hg-O sublattices which gives rise to the orbital-selective Mott transition \cite{2009PRBCraco}.
The other is the charge disproportion of Ru ions driven by the strong covalency originating from the Ru $4d$ - O $2p$ - Hg $5d$ hybridized state \cite{2015PRBBaidya}.
The latter suggests the noncoplanar magnetic structure as the ground state, which could be a subject for future study.

\subsection{Filling-control metal-insulator transitions}
\label{sec:4_filling-controlled ruthenates}

Band-filling effects are known to play an important role in electronic properties in Mott systems, as exemplified in high-$T_{\rm c}$ cuprate superconductors \cite{2006RMPLee}.
In multi-orbital systems, Hund's-rule coupling, which places two electrons on different orbitals with ferromagnetically coupled spins, markedly modulates the effective electronic correlation and thereby induces nontrivial magnetic and electronic properties \cite{2013ARCMPGeorges}.
In particular, the Hund's-rule coupling causes a significant effect near the half-filled state like Ru$^{5+}$ compounds.
Therefore, pyrochlore ruthenates offer an ideal platform to study the Hund's-rule coupling physics.




\begin{figure}[h]
\begin{center}
\includegraphics[width=0.8\columnwidth]{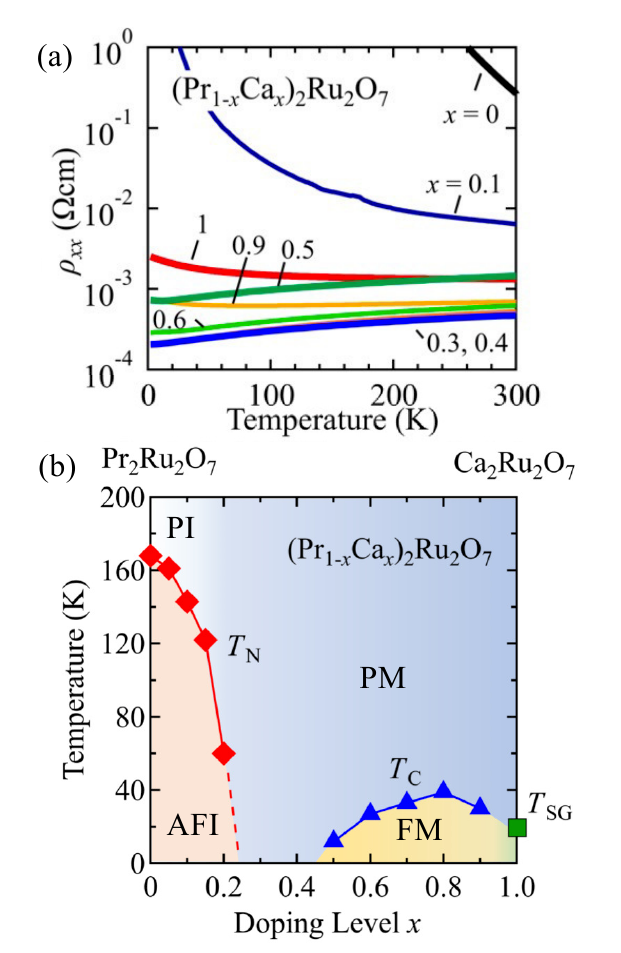}
\caption{
(a) Temperature dependence of resistivity for several (Pr$_{1-x}$Ca$_{x}$)$_{2}$Ru$_{2}$O$_{7}$. 
(b) Phase diagram of (Pr$_{1-x}$Ca$_{x}$)$_{2}$Ru$_{2}$O$_{7}$ in the plane of doping level $x$ and temperature. PI stands for paramagnetic insulator, PM stands for paramagnetic metal, AFI stands for antiferromagnetic insulator, and FM stands for ferromagnetic metal.
$\tn $ denotes the antiferromagnetic transition temperature, $T_{\rm c}$ denotes the ferromagnetic transition temperature, and $T_{\rm SG}$ denote the spin-glass critical temperature.
Reproduced figure with permission from \cite{2021PRBKaneko}, Copyright (2021) by the American Physical Society.
}
\label{3-3_basic}
\end{center}
\end{figure}

One route towards the filling control in pyrochlore ruthenates is to substitute $R^{3+}$ cations with $A^{2+}$ ones, which is the solid-solution in different sites from the transition metal cations and hence effectively changes the number of electrons between 4 and 3 in Ru $4d$ $t_{2g}$ orbitals.
Figure~\ref{3-3_basic}(a) shows the temperature dependence of resistivity for (Pr$_{1-x}$Ca$_{x}$)$_{2}$Ru$_{2}$O$_{7}$~\cite{2020PRBKaneko}.
The resistivity markedly decreases while the antiferromagnetic transition is gradually suppressed as the hole-doping level $x$ increases.
Eventually the metal phase is realized in the whole temperature range at $x=0.3$ - $0.9$.
The resistivity shows the slight increase with lowering temperature for $x=1$, consistent with the transport properties of Ca$_2$Ru$_2$O$_7$ single crystals~\cite{2006JPSJMunenaka}.
Figure~\ref{3-3_basic}(b) shows the phase diagram in the plane of $x$ and temperature.
It is noteworthy that a ferromagnetic metal phase shows up at low temperatures for $0.5<x<0.9$~\cite{2021PRBKaneko}, which is discussed below.

\begin{figure}[h]
\begin{center}
\includegraphics[width=0.95\columnwidth]{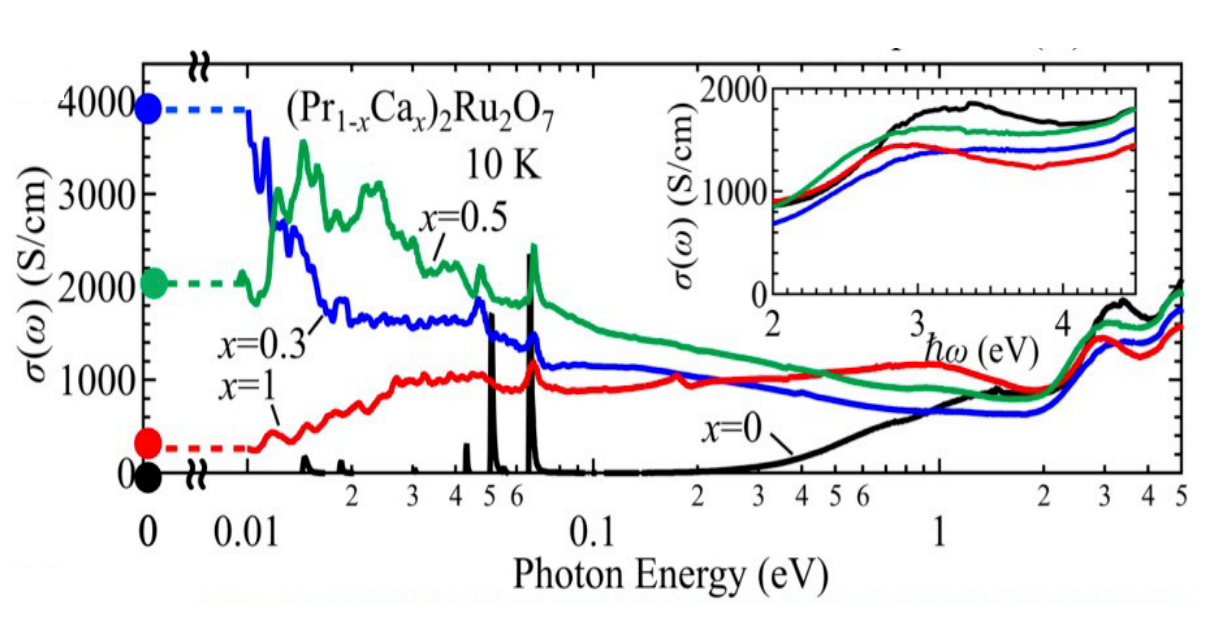}
\caption{
Optical conductivity spectra at 10 K for several $x$ compounds of (Pr$_{1-x}$Ca$_{x}$)$_{2}$Ru$_{2}$O$_{7}$.
The inset shows the magnified view of optical conductivity around 3 eV.
Reproduced figure with permission from \cite{2020PRBKaneko}, Copyright (2020) by the American Physical Society.
}
\label{3-3_optics}
\end{center}
\end{figure}

Figure~\ref{3-3_optics} shows the optical conductivity spectra for several $x$ compounds~\cite{2020PRBKaneko}.
Pr$_2$Ru$_2$O$_7$ ($x=0$) shows a charge gap of about 0.3 eV as discussed above.
As $x$ increases, the spectral weight shifts from higher to lower energy across $\sim 1$ eV, resulting in the Drude like peak below 0.02 eV for $x=0.3$.
Such a dramatic change of spectral weight is quantitatively similar to the filling-control Mott transition observed in $3d$ electron systems~\cite{1995PRLKatsufuji}.
On the other hand, at $x=1$, the Drude response disappears in the low energy region, in agreement with the incoherent dc transport.
Additionally, as shown in the inset of Fig.~\ref{3-3_optics}, the spectral peak around 3 eV, which corresponds to the charge transfer excitation, shows a red-shift with increasing $x$, reflecting high valence Ru ions.
The observed spectral weight transfer over such a large energy scale ($\sim 4$ eV) indicates the significant change of electronic structure relevant to $4d$-$2p$ states governed by $4d$ electron correlation and $p$-$d$ transfer integral, distinct from a simple charge gap opening/closing across semiconductor-to-semimetal transitions.

\begin{figure}[h]
\begin{center}
\includegraphics[width=0.95\columnwidth]{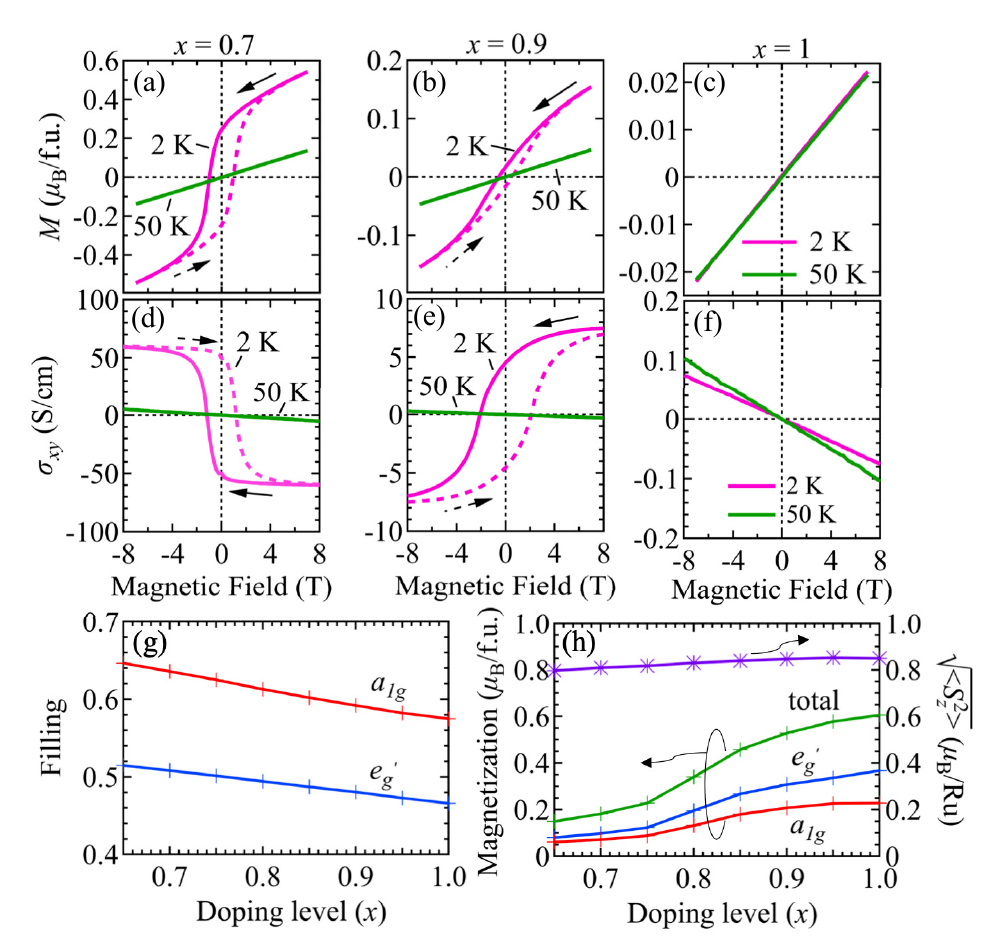}
\caption{
Magnetic field dependence of (a)-(c) magnetization and (d)-(f) Hall conductivity for (Pr$_{1-x}$Ca$_{x}$)$_{2}$Ru$_{2}$O$_{7}$ ($x$ = 0.7, 0.9, 1) at 2 and 50 K, respectively.
Theoretical results for (g) the orbital occupancy and (h) magnetization plotted against $x$.
Red and blue crosses represent the results for $a_{\rm 1g}$ and $e_{\rm g}'$ orbitals, respectively. Purple asterisks in (h) represent the expectation value of the local moment.
Reprinted figures with permission from \cite{2021PRBKaneko}, Copyright (2021) by the American Physical Society.
}
\label{3-3_hall}
\end{center}
\end{figure}

The top panels of Fig.~\ref{3-3_hall} show magnetization and Hall conductivity for representative $x$ compositions~\cite{2021PRBKaneko}.
At 50 K, the magnetization increases in proportion to the magnetic field, whereas the spontaneous component shows up at 2 K for $x$ = 0.7 and 0.9.
In spite of the ferromagnetic-like state, the magnetization continuously increases, indicating that the Ru and Pr spins are antiferromagnetically coupled.
The Hall conductivity also produces an anomalous Hall effect with a large spontaneous component compared to the normal Hall effect at 50 K.
On the other hand, the spontaneous component vanishes and the negative normal Hall component remains for $x=1$, consistent with the spin-glass-like state reported in single crystals~\cite{2006JPSJMunenaka}.

The possible origin of the ferromagnetic metallic phase is discussed on the basis of dynamical mean-field theory~\cite{2021PRBKaneko}.
Figures~\ref{3-3_hall}(g) and (h) show the $x$ dependence of the calculated values of band filling in $a_{1g}$ and $e_{g}'$ orbitals, magnetization, and local moment $\sqrt{S_{z}^{2}}$, respectively.
As $x$ increases, the numbers of electrons in both $a_{1g}$ and $e'_{g}$ orbitals decrease monotonically in a similar manner.
This means that both orbitals have itinerant features, which is different from the double-exchange description between the localized $a_{1g}$ and itinerant $e'_{g}$ proposed for Mo oxides.
Nonetheless, the ferromagnetic on-site coupling between different orbitals (Hund's-rule coupling) still plays an important role in the ferromagnetic state.
In fact, if the Hund's coupling went off, the ferromagnetic state would not appear as the ground state.
It is consistent with the fact that the magnetization decreases as the number of electrons increases from Ru$^{5+}$ ($S=3/2$) to Ru$^{4+}$ ($S=1$) ({\it i.e.}, the Hund's coupling strength decreases).
The calculated magnetization values are about 0.2-0.6 $\mu _{\mathrm{B}}$, which is reasonably close to the experiment.
In contrast, $\sqrt{S_{z}^{2}}$ is considerably larger ($\sim 0.8$ $\mu _{\mathrm{B}}$), reflecting the dynamical spin fluctuations of itinerant electrons.
However, this calculation does not describe the spin glass-like behavior at $x=1$, which may be due to competing ferromagnetic and antiferromagnetic interactions or geometrical frustration.

\subsection{$A_{2}$Ru$_{2}$O$_{7}$ with heavy $A$ ions ($A$ = Tl, Pb, Bi)}
\label{sec:4_heavy A ruthenates}

Here we briefly introduce the physical properties of pyrochlore ruthenates with $A$ ions in the sixth row in the period table.
In contrast to other $A$ elements, $6s$ or $6p$ orbitals are close to the Ru $4d$ and O $2p$ bands near the Fermi energy and thereby have a strong influence on the electronic properties~\cite{2000JPSJIshii}, similar to the high-$T_{c}$ cuprates containing Tl and Bi~\cite{1988PRLKrakauer,1988PRBMattheiss,1988PCYu} and pyrochlore Tl$_2$Mn$_2$O$_7$ showing colossal magnetoresistance~\cite{1996NatureShimakawa,1996ScienceSubramanian} (Cu/Mn $3d$ and O $2p$).

\subsubsection{Tl$_{2}$Ru$_{2}$O$_{7}$.}
The physical properties of Tl$_{2}$Ru$_{2}$O$_{7}$ strongly depend on the growth condition and stoichiometry~\cite{1999JMCTakeda}.
Stoichiometric samples, which are obtained under high oxygen pressure, shows a metal-to-insulator transition accompanied by the structural change to the orthorhombic $Pnma$ space group at 120 K.
Elastic and inelastic neutron scattering experiments suggest the possible emergence of Haldane chains~\cite{2006NMLee}.
Below 120 K, the Ru-O bond distance splits into short and long bonds, leading to the splitting of the $t_{2g}$ orbitals.
As a result, the exchange interaction is active only along one dimensional zigzag Ru bond chains.
Furthermore, the inelastic neutron scattering reveals the spin excitation gap of approximately 11 meV, which is consistent with the NMR study~\cite{2002JPSJSakai}.
This may indicate that one-dimensional Haldane chain is spontaneously realized in this $S=1$ system.

\subsubsection{Bi$_{2}$Ru$_{2}$O$_{7}$ and Pb$_{2}$Ru$_{2}$O$_{6.5}$.}


Single crystals of Bi$_2$Ru$_2$O$_7$ and Pb$_2$Ru$_2$O$_{6.5}$ are grown by Bi$_2$O$_3$-V$_2$O$_5$ and PbO flux, respectively~\cite{2006PRBTachibana}.
The former crystalizes the typical pyrochlore type structure ($Fd\overline{3}m$) while the latter is noncentrosymmetric structure ($F\overline{4}3m$) where the ordered O' vacancy is accompanied with the displacement of Pb ions.
While Bi$_2$Ru$_2$O$_7$ shows the incoherent transport properties due to the structural instability originating from Bi-$6s^2$ lone-pair electrons, Pb$_2$Ru$_2$O$_{6.5}$ is a good metal with a small residual resistivity of $\sim 5$ $\mu \Omega $cm likely due to the extended Pb $6p$ orbitals which strongly hybridize with Ru $4d$ orbitals.

\subsection{Summary}
\label{sec:4_summary}

Ruthenium oxides exhibit a wide range of physical properties owing to a variety of elements in the $A$ site.
For instance, the compounds containing trivalent rare-earth ions are insulators, whereas many compounds with divalent cations are metallic in nature.
Moreover, the admixture of $A$-site cations with different valence states, {\it i.e.,} effectively controlling the band filling, gives rise to novel phases such as the ferromagnetic metal state.
The heavy $A$-site compounds also exhibit various physical properties including high electrical conductivity and structural phase transitions, due to the hybridization of $6s$ or $6p$ orbitals.
The family of ruthenium compounds has continued to expand, with recent examples including newly-synthesized In$_2$Ru$_2$O$_7$~\cite{2024SAKrajewska} and thin-films~\cite{2024CMZhang,2024PRMRabinovich}.
This diverse class of materials provides a new platform for exploring the complex physical phenomena emerging from the interplay of strong correlations, Hund's coupling, and spin-orbit coupling.

\section{$5d$ pyrochlores Cd$_2$Re$_2$O$_7$ and Cd$_2$Os$_2$O$_7$}
\label{sec:5_5dsystems}

In this section, we briefly discuss two different $5d$ pyrochlores, Cd$_2$Re$_2$O$_7$ and Cd$_2$Os$_2$O$_7$. In the case of Cd$_2$Re$_2$O$_7$, each Re$^{5+}$ ion hosts two $5d$ electrons in the $j_{\mathrm{eff}}=3/2$ bands split from the $t_{2g}$ orbitals by the spin-orbit coupling. Meanwhile, in Cd$_2$Os$_2$O$_7$, each Os$^{5+}$ ion carries three $5d$ electrons in the $j_{\mathrm{eff}}=3/2$ manifold. In both cases, as discussed in Sect.~\ref{sec:2_5dcase}, the system undergoes keen competition among the spin-orbit coupling, the trigonal distortion, and electron correlation, which leads to a variety of phases and transitions between them as shown below.

\subsection{Cd$_2$Re$_2$O$_7$.}
\label{sec:5_Cd2Re2O7}

Cd$_2$Re$_2$O$_7$ shows superconductivity below $T_{\rm c}\simeq 1$~K~\cite{Hanawa2001,Sakai2001,Jin2001}. 
This is the only superconductor found in the $\alpha$-type pyrochlore oxides to date. 
The nature of the superconducting state is of weakly-correlated $s$-wave type, while a mixing of $p$-type is anticipated through the spin-orbit coupling since the lattice structure breaks the spatial inversion symmetry at low temperature as discussed below. 

\begin{figure}[tb]
\centering
\includegraphics[width=0.95\columnwidth]{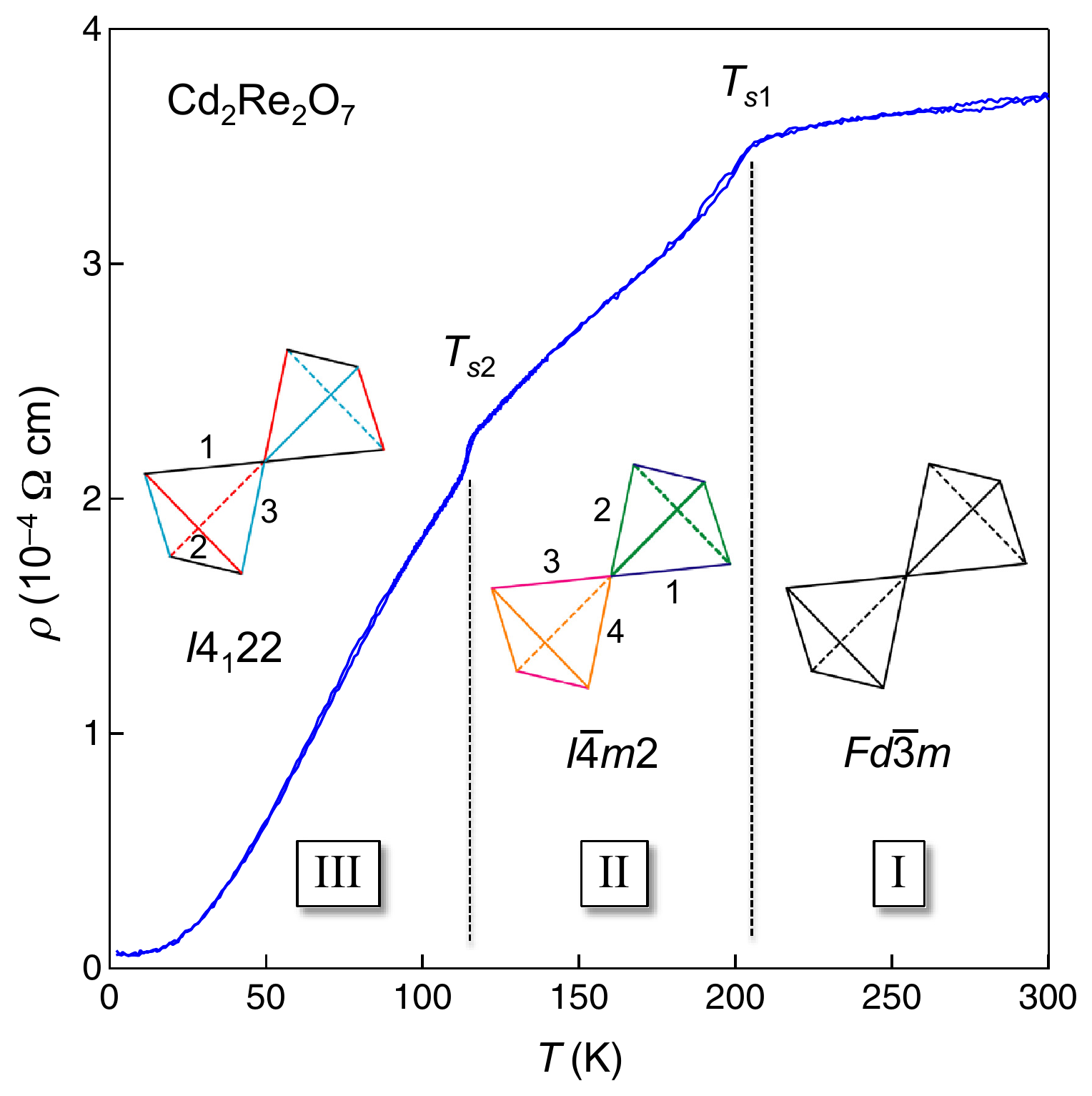}
\caption{\label{fig:Cd2Re2O7_resistivity}
Temperature dependence of the resistivity for Cd$_2$Re$_2$O$_7$. 
There are three phases I, II, and III, separated by the phase transitions at $T_{{\rm s}1}\simeq 200$~K and $T_{{\rm s}2}\simeq 120$ K. 
The insets show the schematics of two neighboring tetrahedra with corresponding space group. 
Reproduced with permission from Ref.~\cite{Hiroi2018}. $\copyright $ (2018) The Physical Society of Japan.
}
\end{figure}

At ambient pressure, besides the superconducting transition, Cd$_2$Re$_2$O$_7$ exhibits two structural phase transitions at $T_{{\rm s}1}\simeq 200$~K and $T_{{\rm s}2}\simeq 120$~K (Fig.~\ref{fig:Cd2Re2O7_resistivity})~\cite{Yamaura2002}, though the symmetry of the low-temperature phases as well as the nature of the phase transitions have been debated for a long time~\cite{Hiroi2018}. 
In the high-temperature phase above $T_{{\rm s}1}$, called phase I, the symmetry is cubic of space group $Fd\overline{3}m$. 
At $T=T_{{\rm s}1}$, the spatial inversion symmetry is broken with a weak tetragonal lattice distortion~\cite{Yamaura2002,Castellan2002}.
The space group of the phase II below $T_{{\rm s}1}$ is most likely $I\overline{4}m2$, while the lower symmetry was also argued~\cite{DiMatteo2017,Harter2017,Harter2018}. 
With further lowering temperature down to $T_{{\rm s}2}$, a weak first-order transition is observed to the phase III whose space group is assigned as $I4_122$, though some experiments indicated the absence of the phase transition~\cite{Jin2002,Lu2004,Petersen2006}. 
From the symmetry point of view, the phase transition from phase I to II at $T=T_{{\rm s}1}$ is of second order as $I\overline{4}m2$ is a subgroup of $Fd\overline{3}m$, whereas the transition from phase II to III is of first order as $I\overline{4}m2$ and $I4_122$ belong to different branches of the symmetry reduction from $Fd\overline{3}m$. 
These structural changes were discussed based on the Landau theory~\cite{Sergienko2003,Ishibashi2010}. 

Cd$_2$Re$_2$O$_7$ shows metallic conduction throughout the phases I, II, and III, as shown in Fig.~\ref{fig:Cd2Re2O7_resistivity}. 
While the resistivity depends on temperature very weakly in the high-temperature phase I, it starts to decrease in the phase II below $T_{{\rm s}1}$ and shows a small change at the transition to phase III at $T=T_{{\rm s}2}$~\cite{Hanawa2001,Hiroi2002}. 
The magnetic susceptibility also shows a decrease below $T_{{\rm s}1}$, whereas it shows almost no change at $T=T_{{\rm s}2}$~\cite{Hanawa2001,Hiroi2002}. 
These results indicate a reduction of the density of states and an increase of carrier density below $T_{{\rm s}1}$. 
From the first-principles calculations, it was suggested that the electronic state in the phase I is a compensated metal, and the small structural change in the phase II may modulate the electronic bands and Fermi surfaces considerably~\cite{Singh2002,Harima2002,Huang2009}, inferred as the band Jahn-Teller mechanism~\cite{Hiroi2003}. 
The modulations of the Fermi surfaces were also studied experimentally, in comparison with first-principles calculations~\cite{Matsubayashi2018,Hirose2022}. 

\begin{figure}[tb]
\centering
\includegraphics[width=0.95\columnwidth]{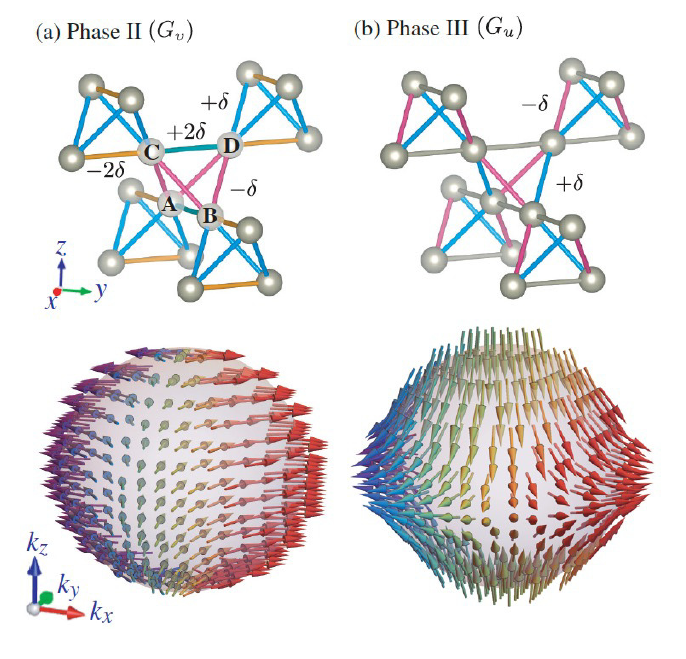}
\caption{\label{fig:ETQ}
Schematic pictures of the bond modulations and the spin polarizations on a spherical Fermi surface by the electric toroidal quadrupoles postulated for the phases II (a) and III (b) in Cd$_2$Re$_2$O$_7$.
Reproduced with permission from \cite{Hayami2019}, Copyright (2019) by the American Physical Society.
}
\end{figure}

However, since the structural changes are very small at $T_{{\rm s}1}$ and $T_{{\rm s}2}$, these phase transitions have been considered to be of electronic origins. 
Fu pointed out that, in the presence of strong spin-orbit coupling in the $5d$ electrons of Re ions, electron interactions may give rise to Fermi surface instabilities toward various types of parity-breaking phases~\cite{Fu2015}. 
Unconventional responses resulting from the induced order parameters were also predicted~\cite{Fu2015,Norman2015}. 
Several different scenarios for the actual order parameters in the phases II and III have been discussed, {\it e.g.}, tensor order~\cite{Petersen2006}, nematic order~\cite{Harter2017}, and higher-order multipoles~\cite{DiMatteo2017,Norman2020}. 
Amongst others, from the symmetry argument relying on the x-ray diffraction measurement, Hayami {\it et al.} proposed the possibility of even-rank electric multipoles~\cite{Hayami2019}. 
In this scenario, the cluster-type electric toroidal quadrupoles of $x^2-y^2$ ($G_v$) and $3z^2-r^2$ ($G_u$) forms, which can be induced by spontaneous bond or spin-current orders, are suggested to be the order parameters in the phase II and III, respectively; see Fig.~\ref{fig:ETQ}. 
The electric toroidal quadrupole orders are supported by the detailed measurement of the anisotropies in the crystal structure and resistivity~\cite{Tajima2020}, magnetic torque~\cite{Matsubayashi2020,Uji2020}, quantum oscillations and electronic structure calculations~\cite{Hirose2022}, and high-resolution x-ray diffraction~\cite{2023JPSMHirai}.

\begin{figure}[tb]
\centering
\includegraphics[width=0.95\columnwidth]{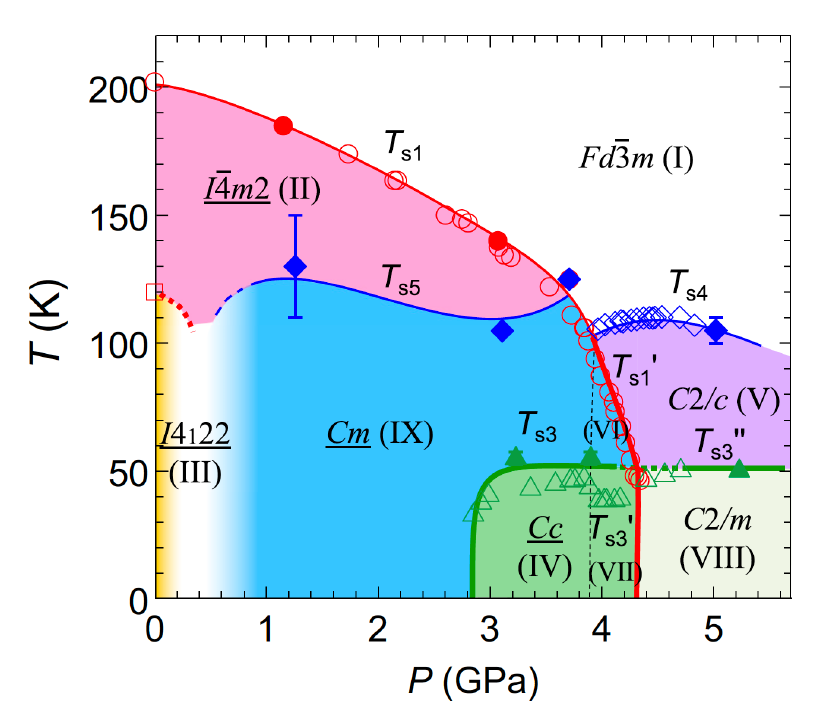}
\caption{\label{fig:Cd2Re2O7_pressure}
Temperature-pressure phase diagram for Cd$_2$Re$_2$O$_7$ obtained by synchrotron x-ray diffraction. 
The red line denotes the phase boundary by inversion symmetry breaking. 
The space group for each phase is shown. 
Reproduced with permission from \cite{Yamaura2017}, Copyright (2017) by the American Physical Society.
}
\end{figure}

In an external hydrostatic pressure, Cd$_2$Re$_2$O$_7$ exhibits a surprising variety of phases~\cite{Kobayashi2011,Yamaura2017}. 
The phase diagram obtained by synchrotron x-ray diffraction is shown in Fig.~\ref{fig:Cd2Re2O7_pressure}. 
The critical temperature $T_{{\rm s}1}$ between phase I and II decreases with increasing pressure, and vanishes at $P=P_{\rm c} \simeq 4$~GPa, apparently forming a quantum critical point. 
The phase boundary corresponding to the inversion symmetry breaking (red curve in Fig.~\ref{fig:Cd2Re2O7_pressure}) surrounds the parity-breaking phases, suggesting the enhancement of parity fluctuations around the quantum critical point that may induce unconventional superconductivity~\cite{Kozii2015,Wang2016,Hiroi2018}. 
Experimentally, the superconducting temperature is increased to $\sim 2.5$~K and the upper critical field becomes about $27$ times larger with increasing the pressure toward $P_{\rm c}$~\cite{Hiroi2002b,Kobayashi2011}. 
These behaviors could be a hallmark of parity mixing and its fluctuations that are enhanced under pressure toward the quantum critical point. 
In both sides of this phase boundary, many different phases appear under pressure~\cite{Kobayashi2011,Yamaura2017,Hiroi2018}. 
By further increasing the pressure, the system exhibits metal-to-nonmetal transition around $20$~GPa; the resistivity obeys $T^2$ behavior in a wide range of temperature around the critical pressure, indicating the importance of electron correlation effects~\cite{Malavi2016}.

\subsection{Cd$_2$Os$_2$O$_7$.}
\label{sec:5_Cd2Os2O7}

\begin{figure}[tb]
\centering
\includegraphics[width=0.95\columnwidth]{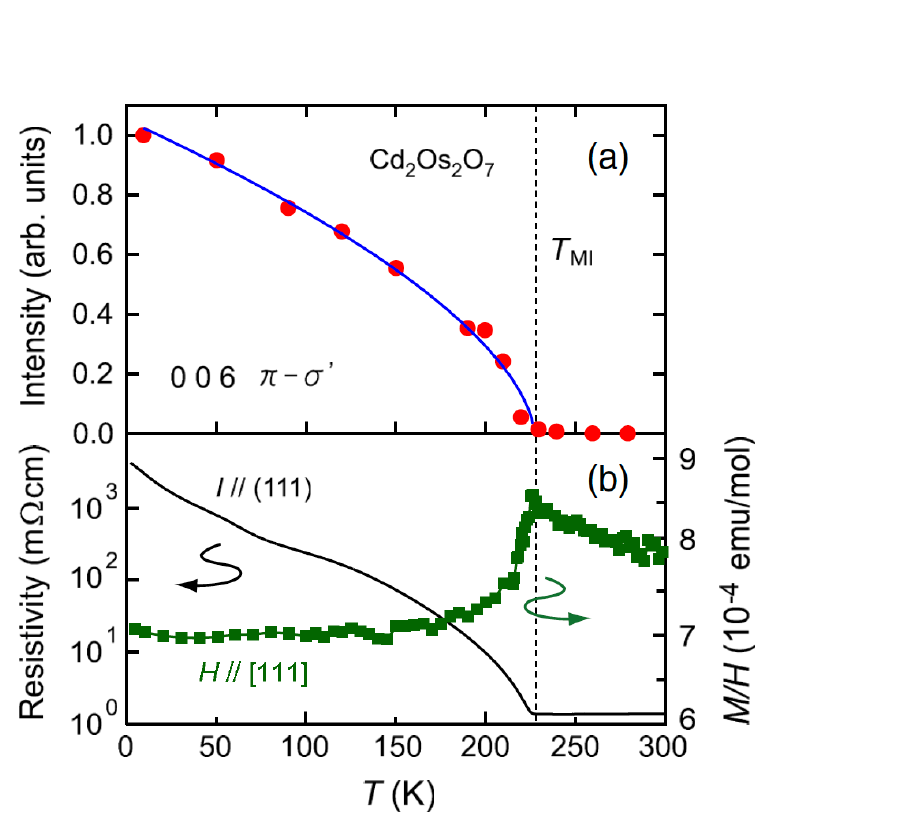}
\caption{\label{fig:Cd2Os2O7}
Temperature dependences of integrated intensity corresponding to the all-in all-out order (a) and the resistivity and susceptibility (b) for Cd$_2$Os$_2$O$_7$. 
Reproduced with permission from \cite{Yamaura2012}, Copyright (2012) by the American Physical Society.
}
\end{figure}

Cd$_2$Os$_2$O$_7$ exhibits a phase transition from metal to insulator around $225$~K with decreasing temperature~\cite{Sleight1974}. 
The origin of the metal-insulator transition has been controversial for a long time~\cite{Mandrus2001,Padilla2002,Harima2002,Singh2002,Hiroi2015}. 
Among them, the magnetic origin has been discussed intensively~\cite{Mandrus2001,Padilla2002,Koda2007}, but it was difficult to identify the magnetic order by neutron scattering experiments since Cd ions are neutron absorbers~\cite{Reading2001}. 
The problem was resolved by using resonant x-ray scattering, which showed that a magnetic long-range order of all-in all-out type develops concomitantly with the metal-insulator transition~\cite{Yamaura2012}; see Fig.~\ref{fig:Cd2Os2O7}. 
The stabilization of the all-in all-out order was also confirmed theoretically by using first-principles calculations, which pointed out the importance of the spin-orbit coupling and the trigonal distortion as well as electron correlations~\cite{Shinaoka2012}. 
It was also shown that, in the high-temperature paramagnetic phase, spin fluctuations of all-in all-out type develop dominantly through the intricate coupling among charge, spin, and orbital degrees of freedom~\cite{Uehara2015}. 
Recent optical measurement showed that the system crossovers from metal to insulator well below the magnetic ordering temperature, suggesting the existence of an intermediate all-in all-out metallic state~\cite{Sohn2015}. 
Such a state was also found in the first-principles calculations between the paramagnetic metal and the all-in all-out insulator~\cite{Shinaoka2012}.

\begin{figure}[tb]
\centering
\includegraphics[width=0.95\columnwidth]{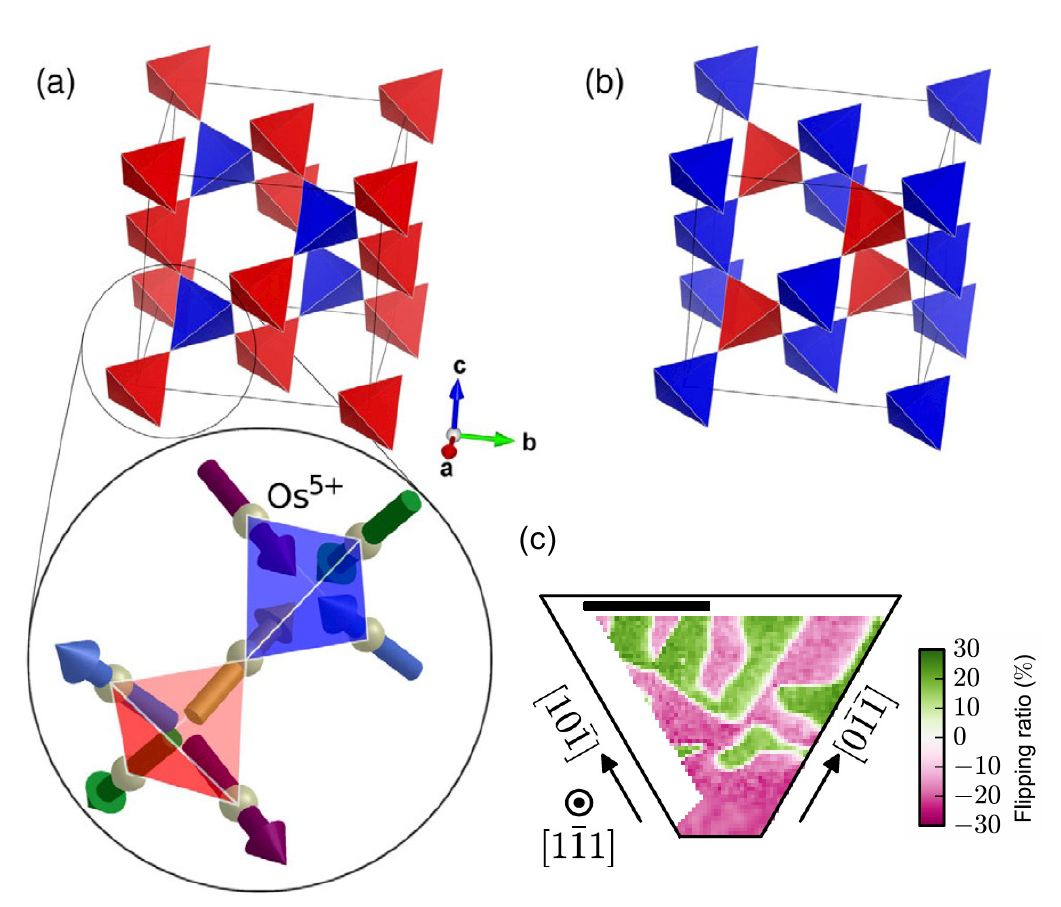}
\caption{\label{fig:Cd2Os2O7_aiao}
Resonant magnetic x-ray diffraction imaging on Cd$_2$Os$_2$O$_7$.
(a) All-in all-out and (b) all-out all-in magnetic order. Both configurations are time-reversal symmetric of each other.
All Os spins are located at the vertices of the tetrahedra and point either towards the center of the blue tetrahedra or away from the center of the red tetrahedra, as shown in the magnified region.
(c) Flipping ratio map at the 0010 reflection for an incident x-ray beam with left-handed and right-handed circular polarizations. The scale bar is 100 $\mu $m across.
Reproduced with permission from \cite{Tardif2015}, Copyright (2015) by the American Physical Society.
}
\end{figure}

Interestingly, the all-in all-out ordered state has unique domain walls. 
In this state, there are two domains, all-in all-out and all-out all-in, which are related with each other by time-reversal symmetry [Figs.~\ref{fig:Cd2Os2O7_aiao}(a) and (b)].
The coexistence of the two domains were indeed visualized by the circular polarized resonant x-ray diffraction imaging technique [Fig.~\ref{fig:Cd2Os2O7_aiao}(c)]~\cite{Tardif2015}. 
The domain walls can be magnetic and carry uncompensated magnetic moments perpendicular to the walls~\cite{Arima2013,Hiroi2015}. 
It was shown that the magnetic domain walls are conductive and contribute significantly to the electronic and magnetic properties in the low-temperature insulating phase, such as weak ferromagnetism and unusual temperature dependence of the electrical resistivity~\cite{Hirose2017}. 
Similar magnetic domain walls appear in the all-in all-out states in iridium pyrochlores, but they bring about more intriguing properties due to the peculiar electronic band structure, as discussed in the next section.

\subsection{Summary}
\label{sec:5_summary}

In this section, we introduce two $5d$ systems; Cd$_2$Re$_2$O$_7$ and Cd$_2$Os$_2$O$_7$.
The former becomes a superconductor at low temperatures, while the latter exhibits antiferromagnetic insulating behavior, despite having only a single electron difference in the $5d$ orbitals.
This contrast highlights the complexity of strongly correlated electron systems.
Notably, Cd$_2$Re$_2$O$_7$ is the only $\alpha $-type pyrochlore oxide known to exhibit superconductivity, and it also undergoes a parity-breaking phase transition due to the instability of the Fermi surface in the presence of strong spin-orbit coupling, sparking considerable interest.
Recently, related materials such as Pb$_2$Re$_2$O$_{7-\delta }$~\cite{2024PRMNakayama} and Hg$_2$Os$_2$O$_7$~\cite{2022JPCMKataoka} have been reported, which will further advance our understanding of strongly correlated electron phenomena affected by strong spin-orbit coupling.

\section{Iridates ($R$,$A$)$_2$Ir$_2$O$_7$}
\label{sec:6_Iridates}

In this section, we describe experimental and theoretical studies of pyrochlore iridates.
Because of the high vapor pressure and slow chemical reaction rate of iridium oxides, it is difficult to synthesize high-quality crystals with few impurities or defects and hence their physical properties were not well reproduced.
In addition, polycrystals tend to be porous which limits the measurement methods.
Therefore, the physical properties of pyrochlore iridates were not deeply understood.

Recently, the importance of the relativistic spin-orbit coupling is recognized in layered perovskite iridates \cite{2009ScienceKim}.
In the atomic limit, all the five $5d$ electrons in Ir$^{4+}$ occupy the $t_{2g}$ orbital due to the dominant crystal field splitting.
The on-site spin-orbit coupling gives rise to a further energy splitting into an effective pseudospin $j_{\mathrm{eff}}=1/2$ doublet and a $j_{\mathrm{eff}}=3/2$ quadruplet.
For the sufficiently strong spin-orbit coupling in iridates, the half-filled $j_{\mathrm{eff}}=1/2$ single band state is realized (see Sect.~\ref{sec:2_5dcase}), offering an ingredient for novel correlated-electron physics such as spin-orbit-assisted Mott insulator~\cite{Kim2008} and Kitaev spin liquid~\cite{Jackeli2009}.
Among them, $R_2$Ir$_2$O$_7$ is the first material in which the Weyl semimetal state was suggested to be realized~\cite{2011PRBWan}.
It is characterized by pairs of linear dispersions in three-dimensional bulk and gapless surface states.
Unlike Dirac semimetal having fourfold degenerate touching points, Weyl points are doubly degenerate between nondegenerate conduction and valence bands in the absence of time-reversal or inversion symmetry.
Remarkably, each of paired points behaves like a source or sink of Berry curvature, giving rise to a number of interesting physical phenomena \cite{2018RMPArmitage}.
Such a fascinating theoretical proposal in strongly-correlated systems has stimulated experimental studies on pyrochlore iridates.

Here, we introduce recent experimental and theoretical findings as follows; bandwidth control metal-insulator transitions (Sect.~\ref{sec:6_Iridates MIT}), band calculations (Sect.~\ref{sec:6_Iridates theory}), magneto-transport properties (Sect.~\ref{sec:6_Iridates magnetotransport}), and filling-control effect (Sect.~\ref{sec:6_Iridate filling-control}).


\subsection{Bandwidth control metal-insulator transitions}
\label{sec:6_Iridates MIT}

\subsubsection{Crystal growth.}

There are several growth techniques to circumvent the off-stoichiometry.
One is that starting materials are well-grounded and sintered several times with extra amounts of IrO$_2$~\cite{2011JPSJMatsuhira}.
The other is the high pressure synthesis which suppresses atomic defects or vacancies because of a closed environment~\cite{2012PRLUedaIr}.
The latter produces dense samples which enable us to measure a wide range of physical properties.

\begin{figure*}[h]
\begin{center}
\includegraphics[width=1.8\columnwidth]{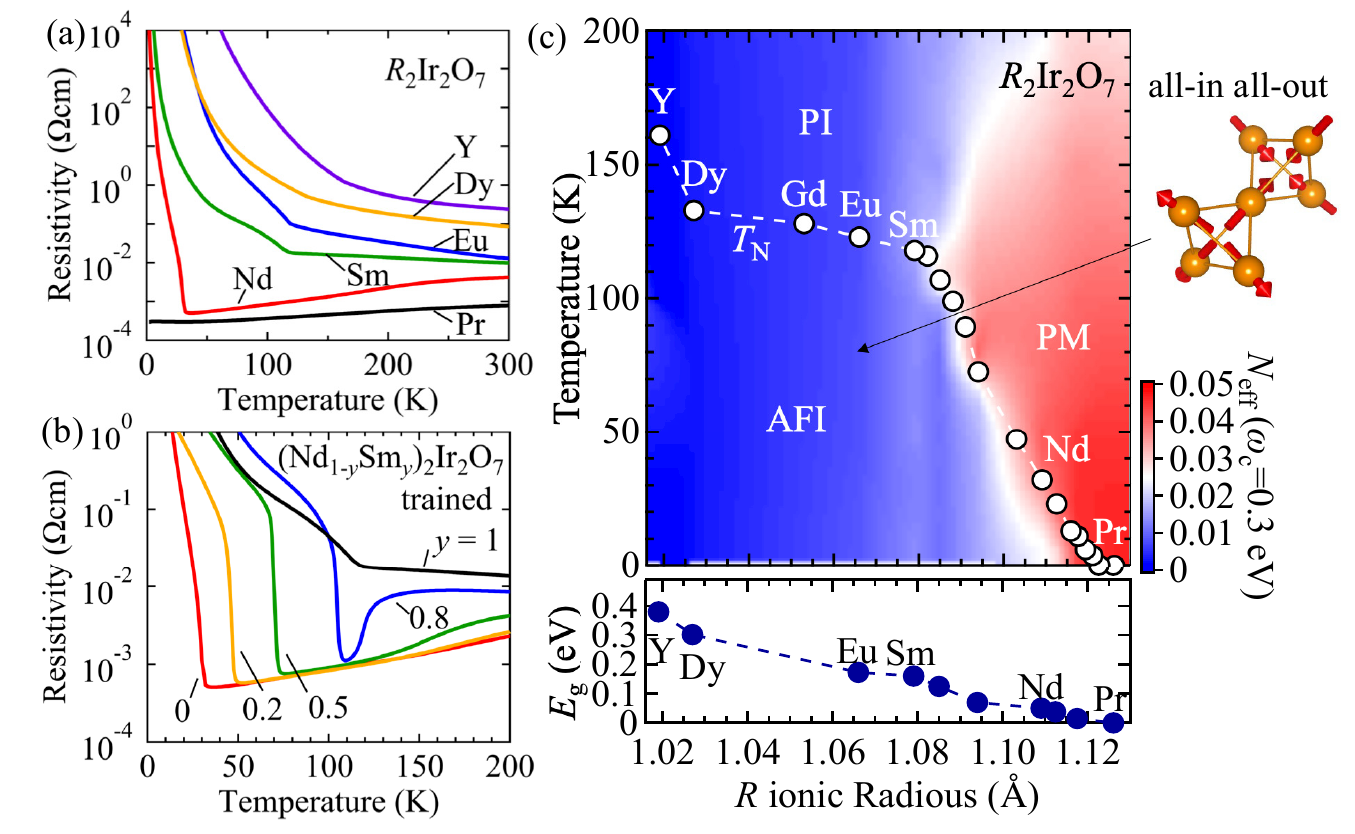}
\caption{
(a) Temperature dependence of resistivity for $R_2$Ir$_2$O$_7$.
(b) Temperature dependence of resistivity for (Sm$_{y}$Nd$_{1-y}$)$_{2}$Ir$_{2}$O$_{7}$.
(c) Phase diagram of $R_2$Ir$_2$O$_7$ in the plane of $R$ ionic radius and temperature. The bottom panel shows the magnitude of charge gap estimated by optical spectroscopy.
Reproduced with permission from \cite{2016PRBUeda}, Copyright (2016) by the American Physical Society.
}
\label{5-1_Ir_phasediagram}
\end{center}
\end{figure*}

Single crystals are synthesized using flux method with KF \cite{2007MRBMillican} or CsCl \cite{2021MCPVLASKOVA,2023JPCSStasko}.
In both cases, Ir loss is inevitable and thus the obtained crystals are subject to slightly off-stoichiometric.
Actually, the anti-mixing of Ir site with Pr ion (Pr$_{2+\delta }$Ir$_{2-\delta }$O$_{7}$) occurs in single crystals~\cite{2022PRBUeda}, leading to the suppression of the long-range order of Pr magnetic moments in Pr$_2$Ir$_2$O$_7$ or the decrease of magnetic transition temperature of $R_2$Ir$_2$O$_7$, although the overall features are similar to those of stoichiometric polycrystals.
Hydrothermal synthesis has also been reported~\cite{2018SSCSleight}, although the transport properties are totally different from those of both single and polycrystals grown by other methods, and hence further verification is necessary in the future.

\subsubsection{Bandwidth control metal-insulator transitions.}

As in molybdenum and ruthenium oxides, in which the change of $R$ ionic radius affects the one-electron bandwidth via the modulation of the $B$-O-$B$ bond angle, a bandwidth control metal-insulator transition is also observed in iridates.
Figures~\ref{5-1_Ir_phasediagram}(a) and (b) show the temperature dependence of resistivity for several $R$ compounds~\cite{2016PRBUeda}.
The resistivity of Pr$_2$Ir$_2$O$_7$ shows metallic like temperature dependence in the whole temperature region. On the other hand, the resistivity of Nd$_2$Ir$_2$O$_7$ abruptly increases below the antiferromagnetic transition temperature $\tn $.
For compounds with smaller $R$ ionic radius such as Sm$_2$Ir$_2$O$_7$, the resistivity increases with lowering temperature even above $\tn $.
Notably, a crossover from paramagnetic insulator to paramagnetic metal occurs in between; for $y=0.8$ in the solid-solution mixed compound (Sm$_{y}$Nd$_{1-y}$)$_{2}$Ir$_{2}$O$_{7}$, the resistivity gradually increases with lowering temperature, sharply drops at around 120 K by one order of magnitude, and then increases rapidly below $\tn $ [Fig.~\ref{5-1_Ir_phasediagram}(b)].
This behavior reminds us of the reentrant transitions among paramagnetic insulator, paramagnetic metal, and antiferromagnetic insulator as observed in $3d$-electron and organic electron-correlated systems~\cite{1992PRBTorrance,1993PRBMcWhan,2005NatureKagawa}.
Figure~\ref{5-1_Ir_phasediagram}(c) exhibits the phase diagram based on the transport, magnetization, and optical measurements.
The magnetic ordering configuration is the antiferromagnetic-like all-in all-out state where all spins on the vertices of a tetrahedron point inwards or outward the center as shown in the right side of Fig.~\ref{5-1_Ir_phasediagram}(c).
The bottom panel of Fig.~\ref{5-1_Ir_phasediagram}(c) shows the charge gap estimated by optical measurements on polycrystals grown under high pressure~\cite{2016PRBUeda}. The amplitude of the charge gap changes almost linearly as the $R$ ionic radius increases,
indicating that the $R$ ionic radius is an appropriate parameter to control one-electron bandwidth ($W$) or the effective electronic correlations ($U/W$).
The phase diagram is overall similar to the prototypical Mott transition systems V$_2$O$_3$ (Mott-Hubbard type) \cite{1993PRBMcWhan} and $R$NiO$_3$ (charge-transfer type) \cite{1992PRBTorrance}.
In the weak correlated regime ({\it i.e.}, large $R$ ionic radius), the paramagnetic metal phase is stabilized.
As $U/W$ increases, the antiferromagnetic order and associated metal-to-insulator transition take place.
Eventually the paramagnetic insulator phase shows up in the strongly correlated regime.
One difference is that there is no first-order phase transition with the critical point near the paramagnetic metal-insulator boundary \cite{1996RMPGeorges}.
This is presumably due to the weak frustration inherent to the all-in all-out magnetic structure and hence high magnetic transition temperature, which may bury the critical point.
These results indicate that the pyrochlore iridates are rare $5d$ electron systems that exhibit a typical Mott transition.

Another route for the bandwidth control is the hydrostatic pressure, which effectively increases the one-electron bandwidth by decreasing the lattice constant without introducing chemical disorders.
Pressure-induced insulator-to-metal transitions are observed in several compounds including Eu$_2$Ir$_2$O$_7$ \cite{2012PRBTafti}, Sm$_2$Ir$_2$O$_7$ \cite{2020PRBWang,2024npjCoak}, and Nd$_2$Ir$_2$O$_7$ \cite{2011PRBSakata,2015PRBUeda}.
Among them, the physical pressure dependence of resistivity in Nd$_2$Ir$_2$O$_7$ is similar to the chemical substitution of Nd with Pr, although the lattice deformation is different~\cite{2020PRBWang}.
In fact, $\tn $ is lowered with increasing pressure ($\Delta P$) and increasing the Pr composition ($\Delta x$) in the form of (Nd$_{1-x}$P$_{x}$)$_{2}$Ir$_{2}$O$_{7}$, scaled to each other ($\Delta P\sim 0.65$ GPa corresponds to $\Delta x\sim 0.1$)~\cite{2015PRBUeda}.
Another important point is that the reentrant-transition like behavior is also observed in Eu$_2$Ir$_2$O$_7$ under pressure~\cite{2012PRBTafti}, indicating that Mott transition in this system is intrinsic.

\subsubsection{Spectroscopic study on electronic states.}

\begin{figure*}[h]
\begin{center}
\includegraphics[width=1.8\columnwidth]{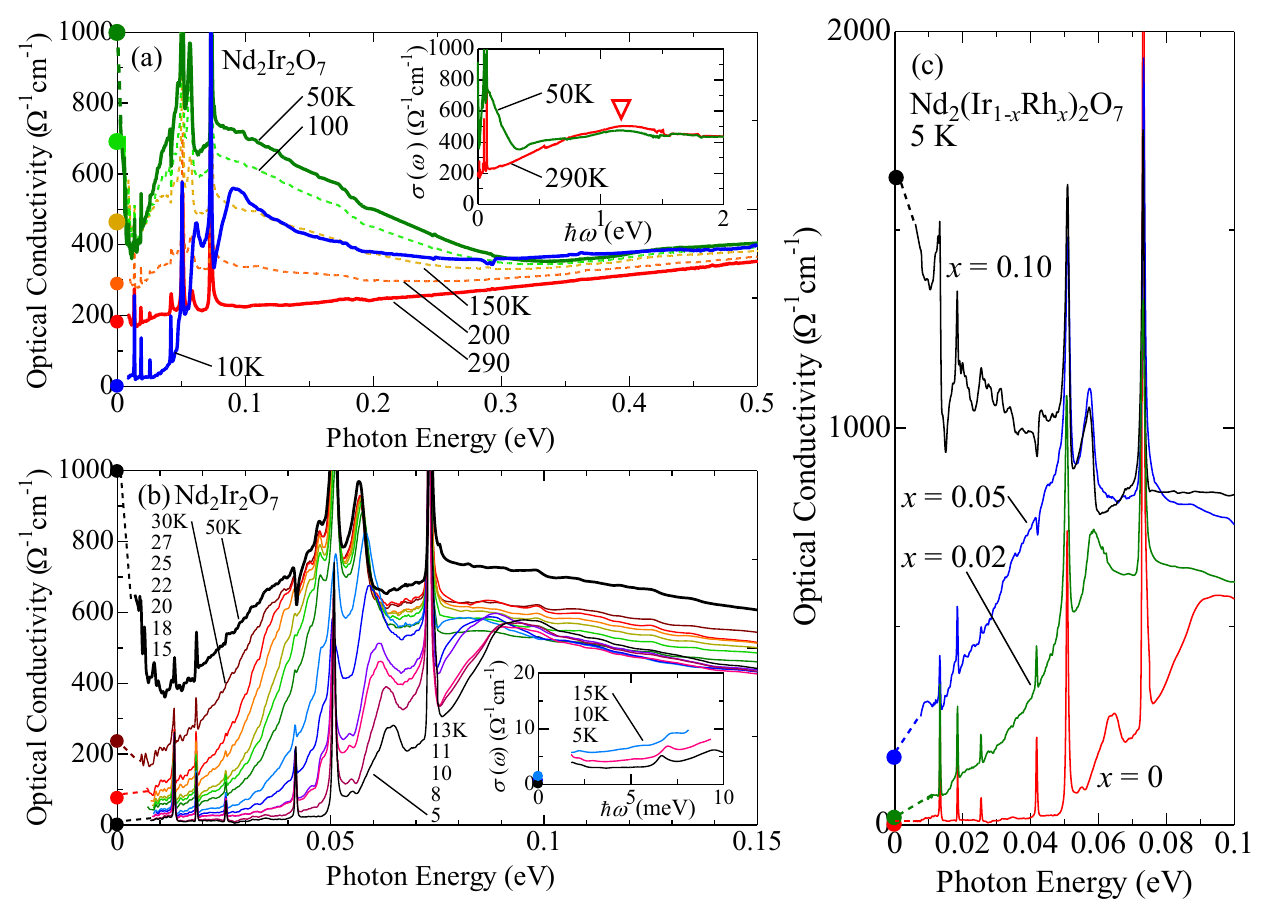}
\caption{
(a) Optical conductivity spectra at various temperatures for Nd$_2$Ir$_2$O$_7$.
The filled circles denote dc conductivities. The inset shows the spectra at 290 K and 50 K up to 2 eV. The triangle indicates the absorption band around 1.2 eV.
(b) Optical conductivity spectra below 50 K. The inset shows the magnified view of the spectra in the far-infrared region as deduced by time-domain terahertz spectroscopy.
(c) Optical conductivity spectra for Nd$_{2}$(Ir$_{1-x}$Rh$_{x}$)$_{2}$O$_{7}$ at 5 K. The red, green, blue, and black line indicates the spectra for $x=0$, $0.02$, $0.05$, and $0.10$, respectively.
The filled circles denote the dc conductivity and the dashed lines are the guide to the eyes.
Reproduced with permission from \cite{2012PRLUedaIr}, Copyright (2012) by the American Physical Society.
}
\label{5-1_nd2ir2o7_optics}
\end{center}
\end{figure*}

Figure~\ref{5-1_nd2ir2o7_optics} shows the optical conductivity spectra of Nd$_2$Ir$_2$O$_7$ at several temperatures~\cite{2012PRLUedaIr}.
The optical conductivity shows the broad hump like structure centered at around 1.2 eV as shown in the inset of Fig.~\ref{5-1_nd2ir2o7_optics}(a), which is assigned to the optical transitions between upper and lower Hubbard bands.
The spectral weight below 0.5 eV significantly increases with decreasing temperature.
At 50 K, the peak like structure develops at 0.1 eV and very sharp Drude-like response shows up below 0.02 eV.
With further decreasing temperature below $\tn $ [Fig.~\ref{5-1_nd2ir2o7_optics}(b)], the Drude response rapidly decays and the absorption peak shifts to higher energy, leading to the formation of the charge gap.
At the lowest temperature (5 K), the magnitude of the gap is estimated to be $\sim 50$ meV.
It is quite large compared to the energy scale of $\tn $, indicating that the electron correlation plays an important role in the metal-insulator transition, rather than the Slater mechanism.
Similar temperature-dependent spectra are also obtained in Nd$_2$Ir$_2$O$_7$ single crystals~\cite{2020NPWang}.
Figure~\ref{5-1_nd2ir2o7_optics}(c) shows the optical conductivity spectra at 5 K for Rh doping samples where the spin-orbit coupling is effectively reduced by doping $4d$ ions~\cite{2012PRLUedaIr}.
The excitation peak between the Hubbard bands shifts to the lower energy as $x$ increases, and eventually the gap closes at $x=0.05$, reminiscent of the bandwidth control insulator-to-metal transition.
It is attributable to the fact that the effective dilution of the spin-orbit coupling reduces the splitting width of $j_{\rm eff}=1/2$~\cite{2012PRBLee,2020NPZwartsenberg}.
It is reconciled with the theoretical result suggesting that the on-site Coulomb interaction and spin-orbit coupling work cooperatively for the insulating state in iridates~\cite{2010PRLWatanabe}.
It is noteworthy that the optical conductivity below 0.1 eV is nearly proportional to the photon energy for the gap-closed $x=0.02$ or $0.05$ compounds.
This behavior is observed in several compounds whose electronic states are characterized by linear dispersion in three-dimensional space~\cite{2016PRBTabert,2014NPOrlita}.
Actually, linear optical conductivity with respect to energy is also observed in nonstoichiometric Eu$_2$Ir$_2$O$_7$~\cite{2015PRBSushkov}.
This suggests that the filling control through off-stoichiometry may be another route to realize Weyl semimetals.

\begin{figure}[h]
\begin{center}
\includegraphics[width=0.95\columnwidth]{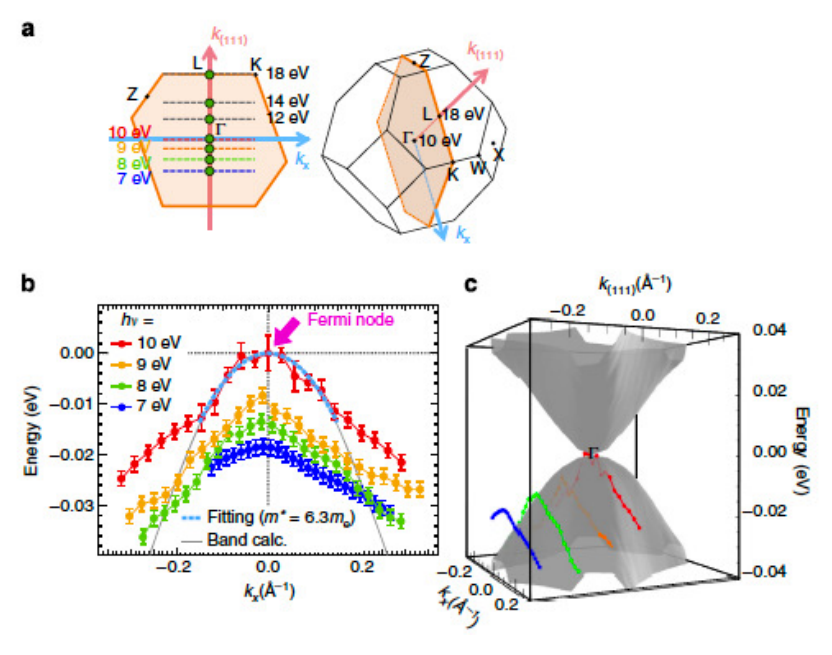}
\caption{
Angle-resolved photoemission spectroscopy (ARPES) on Pr$_2$Ir$_2$O$_7$.
(a) Brillouin zone, showing a momentum sheet along which ARPES data were measured.
The momentum cuts at $h\nu =10$ and $18$ eV crosses the $\Gamma $ and $L$ points in the first Brillouin zone.
(b) Energy dispersions along the $k_x$ direction measured at $h\nu =7$, $8$, $9$ and $10$ eV.
The corresponding momentum cuts are indicated in (a) by dashed coloured lines.
The band dispersion obtained by the first-principles band calculation is superimposed (grey curve).
The data close to $E_F$ is fitted by a parabolic function, $e(k)\propto k^2$ (light-blue
dotted curve).
The estimated effective mass at $\Gamma $, $m_{\mathrm{eff}}=6.3m_{0}$ ($m_{0}$: free electron mass), is in agreement with the band calculation.
(c) The calculated band dispersion in the $k_x-k_{(111)}$ sheet, painted with orange in (a).
On it, the ARPES data in (b) are plotted.
Reprinted by permission from Springer Nature Customer Service Centre GmbH: Nature Communications.~\cite{2015NCommKondo}. $\copyright $ 2015.
}
\label{5-1_arpes}
\end{center}
\end{figure}

It is theoretically suggested that the band structure of Pr$_2$Ir$_2$O$_7$ hosts the quadratic band touching (QBT) which is characterized by the $\Gamma _{8}$ representation of the double group of $O_{h}$ and located just at the Fermi energy~\cite{2013PRLMoon}.
This state is observed in HgTe and $R$PtBi and is demonstrated as a key ingredient for the versatile topological electronic states~\cite{2006ScienceBernevig,2010NMChadov,2010NMLin}.
Figure~\ref{5-1_arpes} shows the result of the angle-resolved photoemission spectroscopy (ARPES) on Pr$_2$Ir$_2$O$_7$~\cite{2015NCommKondo}.
The momentum is shifted along [111] direction across the zone center by changing the exciting photon energy [Fig.~\ref{5-1_arpes}(a)].
Figure~\ref{5-1_arpes}(b) shows the $k_{x}$ points and photon energy of the quasiparticle peaks.
A quadratic peak is observed at $k_{x}=0$ for all photon energies.
As the photon energy ($h\nu $) increases, the energy of the peak systematically increases and reaches nearly zero at $h\nu =10$ eV where the momentum is close to the $\Gamma $ point.
Figure~\ref{5-1_arpes}(c) shows the calculated band dispersion and ARPES data in the sheet of $k_{x}$ and $k_{(111)}$ with energy.
The experimental data is in good agreement with the calculation for the effective mass $m^{*}=6.3m_{0}$, indicative of the strong mass renormalization due to the electron correlation in this system.
Similar band structure is also observed in the paramagnetic metal phase of Nd$_2$Ir$_2$O$_7$~\cite{2016PRLNakayama}.
With cooling below $\tn $, the charge gap opens and its size reaches $\sim 40$ meV, consistent with the optical spectroscopy~\cite{2012PRLUedaIr}.
Theoretical study suggests that the all-in all-out magnetic order breaks time-reversal symmetry and lifts the band degeneracy of the quadratic touching node, resulting in the emergence of Weyl points.
As the magnetic order parameter increases (or experimentally the temperature decreases), the Weyl point moves along the threefold rotational axis and causes a pair-annihilation at the Brillouin zone boundary, eventually leading to the gap opening.
However, in the ARPES experiment, the signature of Weyl semimetal state is missed in the intermediate temperature, probably because the temperature window of Weyl semimetal phase is strictly narrow, as revealed by precise transport measurements as discussed later.

\subsubsection{Neutron and X-ray scattering experiments.}

The magnetic properties are strongly correlated to the electronic states in iridates.
Therefore, neutron and x-ray magnetic scattering experiments play an important role in elucidating the mechanism of the metal-insulator transitions and magnetotransport properties.

Neutron scattering experiments involve difficulties because iridium atoms strongly absorb neutrons.
The powder neutron diffraction experiments using a large volume of Nd$_2$Ir$_2$O$_7$ polycrystals uncovers the magnetic scattering with the propagation vector $\textbf{q}$ = (0,0,0) below 15 K~\cite{2012JPSJTomiyasu}.
This temperature, which is lower than $\tn $, may reflect the neutron scattering from Nd $4f$ moments.
The experimental data are in good agreement with the calculated magnetic structure factor of the all-in all-out configuration with Nd magnetic moments of $2.3\pm 0.4$ $\mu _{\rm B}$.
The inelastic scattering reveals the splitting of Nd$^{3+}$ Kramers ground doublet possibly originating from the exchange interaction with surrounding Ir moments which form all-in all-out configuration below $\tn $.
It is also confirmed in the later $\mu $SR study~\cite{2013PRBGuo} and the neutron scattering study by another group~\cite{2016PRBGuo}.
X-ray scattering study is performed on single crystal of Eu$_2$Ir$_2$O$_7$ by using linearly polarized x-rays at the Ir $L_3$ absorption edge~\cite{2013PRBSagayama}.
In spite of pyrochlore structure where ($4n+2$ 0 0) is the forbidden nuclear reflections, the $\sigma $-$\pi '$ signal at (10 0 0) begins to increase below $\tn $, indicative of the long-range order of Ir moments with $\textbf{q}$ = (0,0,0).
Since no magnetostriction occurs below $\tn $, the all-in all-out magnetic structure is the plausible magnetic ground state.

Resonant inelastic x-ray scattering (RIXS) equipments have developed remarkably in recent years~\cite{2011RMPAment,2024NRMPGroot}.
The high-resolution RIXS spectra reveal the clear magnon dispersion in Sm$_2$Ir$_2$O$_7$~\cite{2016PRLDonnerer}, Eu$_2$Ir$_2$O$_7$~\cite{2018PRLChun}, and nonstoichiometric Tb$_2$Ir$_2$O$_7$~\cite{2024PRBFaure}.
The obtained magnon dispersion in Sm$_2$Ir$_2$O$_7$, which is similar to that in Eu$_2$Ir$_2$O$_7$, is well fitted with a simple model assuming the all-in all-out state with the Heisenberg exchange interaction $J=27.3$ meV and the Dzyaloshinskii-Moriya (DM) interaction $D=4.9$ meV~\cite{2016PRLDonnerer}.
Quantum chemistry calculation suggests that, in Nd$_2$Ir$_2$O$_7$ and Tb$_2$Ir$_2$O$_7$, the DM interaction becomes larger than the Heisenberg interaction due to the small Ir-O-Ir bond angle~\cite{2018PRMYadav}. This leads to a unique flat magnon dispersion, which may be worth investigating in future studies.
On the other hand, the high-energy RIXS spectra of Eu$_2$Ir$_2$O$_7$ allow us to see the energy level scheme of Ir $5d$ states~\cite{2014PRBHozoi,2015PRBUematsu}.
As mentioned in Sect.~\ref{sec:2_LatticeStructure}, the IrO$_6$ octahedron in pyrochlore oxides is trigonally distorted and thereby the $t_{2g}$ manifold is splitted in a different manner from the case in Sr$_2$IrO$_4$ (see Sect.~\ref{sec:2_5dcase}).
Several key parameters for the electronic structure are estimated by analyzing the spectra; the trigonal crystal field splitting is $0.45$ eV, the spin-orbit coupling $0.5$ eV, and the cubic crystal field splitting $3.5$ eV.


\subsection{Band calculation and possible topological electronic states}
\label{sec:6_Iridates theory}

\begin{figure}[tb]
\centering
\includegraphics[width=0.95\columnwidth]{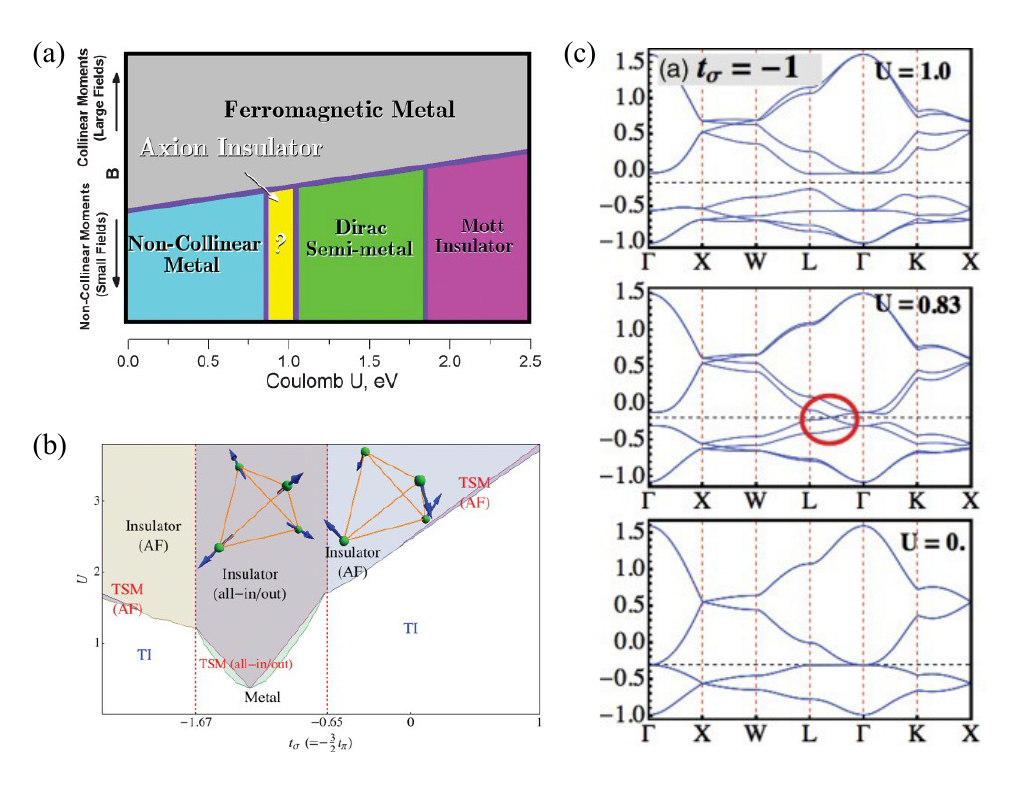}
\caption{\label{fig:5-2_TSM}
(a) Schematic phase diagram for Ir pyrochlore oxides predicted in Ref.~\cite{Wan2011}.
The horizontal and vertical axes represent the Coulomb interaction $U$ and an external magnetic field $B$, respectively. 
Reproduced with permission from \cite{Wan2011}, Copyright (2011) by the American Physical Society.
(b) Mean-field phase diagram of $R_2$Ir$_2$O$_7$ as a function of Hubbard repulsion $U$ and direct hopping between Ir atoms for the $\sigma $ ($\pi $) overlap $t_{\sigma }$ ($t_{\pi }$)~\cite{2012PRBKrempa}.
AF stands for antiferromagnetic order, TI stands for topological insulator, and TSM stands for topological semimetal with Weyl-type fermions.
(c) Evolution of the band structure for different $U$ at $t_{\sigma }=-1$.
The horizontal dashed line is the Fermi level.
The red circle indicates the presence of Weyl points.
Reproduced with permission from \cite{2012PRBKrempa}, Copyright (2012) by the American Physical Society.
}
\end{figure}

Synergy between electron correlations and spin-orbit coupling may give rise to unconventional topological states of matter beyond topological insulators and semimetals in noninteracting or weakly-correlated systems. 
Such a possibility in the iridium pyrochlore oxides was first pointed out in Ref.~\cite{Pesin2010}, where the possible topological Mott insulating state was discussed. 
Subsequently, it was shown that the topological Mott insulator can be destabilized by trigonal distortions of IrO$_6$ octahedra, resulting in an insulator-to-metal transition~\cite{Yang2010}. 
In 2011, Wan {\it et al.} showed that, based on first-principles calculations for Y$_2$Ir$_2$O$_7$, the system may exhibit a Weyl semimetal as well as an axion insulating state in the vicinity of the metal-insulator transition [Fig.~\ref{fig:5-2_TSM}(a)]~\cite{Wan2011}. 
This intriguing possibility has been studied theoretically from several perspectives, such as the $f$-$d$ coupling between magnetic moments of $R$ ions and conduction electrons of Ir ions~\cite{Chen2012,Lee2013}, effects of external pressure, temperature, and magnetic field~\cite{2012PRBKrempa,Witczak-Krempa2013}, quantum many-body effects~\cite{Go2012,Shinaoka2015}, and quantum criticality~\cite{Savary2014}. 

Here we introduce the theoretical result deduced from an extended Hubbard model which includes several hopping terms allowed in a pyrochlore lattice~\cite{2012PRBKrempa}.
In addition to the nearest-neighbor hopping mediated by oxygens, the direct hopping terms between Ir atoms ($t_{\sigma }$ for the $\sigma $ overlap between $t_{\mathrm{2g}}$ orbitals, and $t_{\pi }$ for the $\pi $ overlap) are included.
Figure~\ref{fig:5-2_TSM}(b) shows the phase diagram for $t_{\sigma }=-2t_{\pi }/3$ which intersects all magnetic and electronic phases.
Although many tuning parameters allow a variety of phases in Fig.~\ref{fig:5-2_TSM}(b), there are some similarities between Fig.~\ref{fig:5-2_TSM}(a) and Fig.~\ref{fig:5-2_TSM}(b). For instance, as $U$ increases, the metal phase turns into the Weyl semimetal phase, and eventually reaches Mott insulator phase in both phase diagrams, although the axion insulator phase is not observed in Fig.~\ref{fig:5-2_TSM}(b).
It is noteworthy that the Weyl semimetal phase is limited in a small parameter region for the latter phase diagram, which is in fact close to the experimental result.
Figure~\ref{fig:5-2_TSM}(c) shows the electronic band structure for the three phases along the evolution of $U$ at $t_{\sigma }=-1$ in Fig.~\ref{fig:5-2_TSM}(b).
As seen in the bottom panel of Fig.~\ref{fig:5-2_TSM}(c), the semimetal phase shows a unique electronic structure; distinct from the compensated state with hole and electron pockets, the quadratic conduction and valence bands touch each other at a single point of $\Gamma $, leading to the zero-gap semiconducting state.
Such a band-inverted state breaks into various topological states including topological insulators and Weyl semimetals in the presence of cubic or time-reversal symmetry breaking perturbations~\cite{2013PRLMoon}.
In the case of $R_2$Ir$_2$O$_7$, the all-in all-out magnetic order, which breaks the time-reversal symmetry while preserving cubic symmetry, lifts the band degeneracy and thereby brings about linear crossing points along the $\Gamma $ - $L$ lines, as shown in the middle panel of Fig.~\ref{fig:5-2_TSM}(c).
With further increasing $U$, the Weyl nodes migrate towards the Brillouin zone boundary and eventually cause pair annihilations [the top panel of Fig.~\ref{fig:5-2_TSM}(c)] to be the antifferomagnetic insulator.

The topological properties of Weyl semimetals manifest themselves in electronic and magnetic properties, such as the Fermi arc on the surface of the system through the bulk-edge correspondence, and the anomalous Hall and Nernst effects~\cite{Murakami2007,Armitage2018}. 
In addition, for the all-in all-out state in pyrochlore iridates, there are two magnetic domains with magnetic domain walls between them, as in the case of Cd$_2$Os$_2$O$_7$ in Sect.~\ref{sec:5_Cd2Os2O7}. 
Interestingly, it was pointed out that the magnetic domain walls may also host a peculiar metallic state with helical two-dimensional Fermi surfaces~\cite{Yamaji2014,Yamaji2016}.

\subsection{Magneto-transport properties near the quantum phase transition}
\label{sec:6_Iridates magnetotransport}

\subsubsection{Anomalous Hall effect in Pr$_2$Ir$_2$O$_7$.}
\label{sec:6_Iridates AHE}

\begin{figure}[h]
\begin{center}
\includegraphics[width=1\columnwidth]{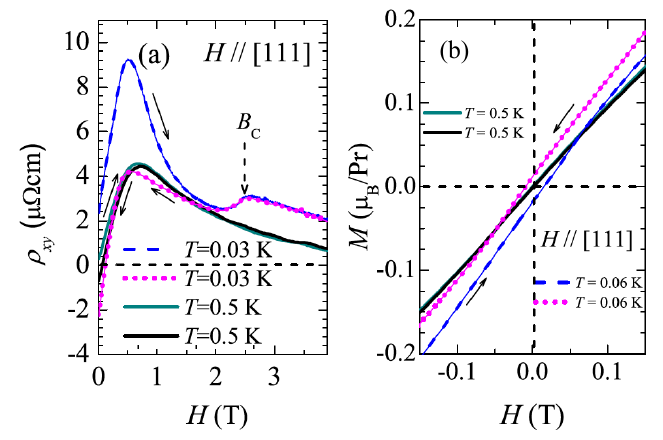}
\caption{
Magnetic field dependence of (a) Hall conductivity at 0.03 K and 0.5 K, and (b) magnetization at 0.06 K and 0.5 K for Pr$_2$Ir$_2$O$_7$.
The orientation of the field is the [111] crystallographic direction.
The dashed arrow represents the metamagnetic transition field of 2.3 T.
The arrows indicate up and down field sweep sequences.
Reproduced with permission from \cite{2011PRLBalicas}, Copyright (2011) by the American Physical Society.
}
\label{5-3_pr2ir2o7-nakatsuji}
\end{center}
\end{figure}

Figure~\ref{5-3_pr2ir2o7-nakatsuji} shows the magnetic field dependence of Hall conductivity (0.03 K and 0.5 K) and magnetization (0.06 K and 0.5 K) for Pr$_2$Ir$_2$O$_7$~\cite{2010NatureMachida,2011PRLBalicas}.
Both Hall conductivity and magnetization for $H$ // [111] show a kink at 2.3 T which is attributed to the change of Pr magnetic configuration from 2-in 2-out to 3-in 1-out, as observed in dipolar spin ice systems such as Dy$_2$Ti$_2$O$_7$~\cite{2003PRLSakakibara}.
Below 2.3 T, large spontaneous Hall conductivity is observed despite a vanishingly small spontaneous magnetization.
It was argued that this feature reflects a chiral spin liquid state where the time-reversal symmetry is spontaneously broken on a macroscopic scale in the absence of magnetic dipole long-range order~\cite{2010NatureMachida,1989PRBWen,2020PRXSzasz}.
As for this observation, we may take account of another possibility to observe a topological Hall effect even in a minimal magnetism.
Recently, the electronic bands in iridates are proven to host topological nature as introduced in the previous section~\cite{2015NCommKondo,2016PRLNakayama}.
Indeed, since Weyl points can be regarded as a source or sink of Berry curvature, a minimal magnetization produces a giant anomalous Hall effect~\cite{2015NatureNakatsuji,2016SANayake}.
In this regard, the observed Hall effect of Pr$_2$Ir$_2$O$_7$ may reflect the nontrivial Berry curvature in momentum space.
In fact, the zero-gap semiconducting state is realized in the paramagnetic metal phase in Pr$_2$Ir$_2$O$_7$, which can turn into various topological electronic states by the time-reversal symmetry breaking~\cite{2015NCommKondo}.

\begin{figure*}[h]
\begin{center}
\includegraphics[width=1.8\columnwidth]{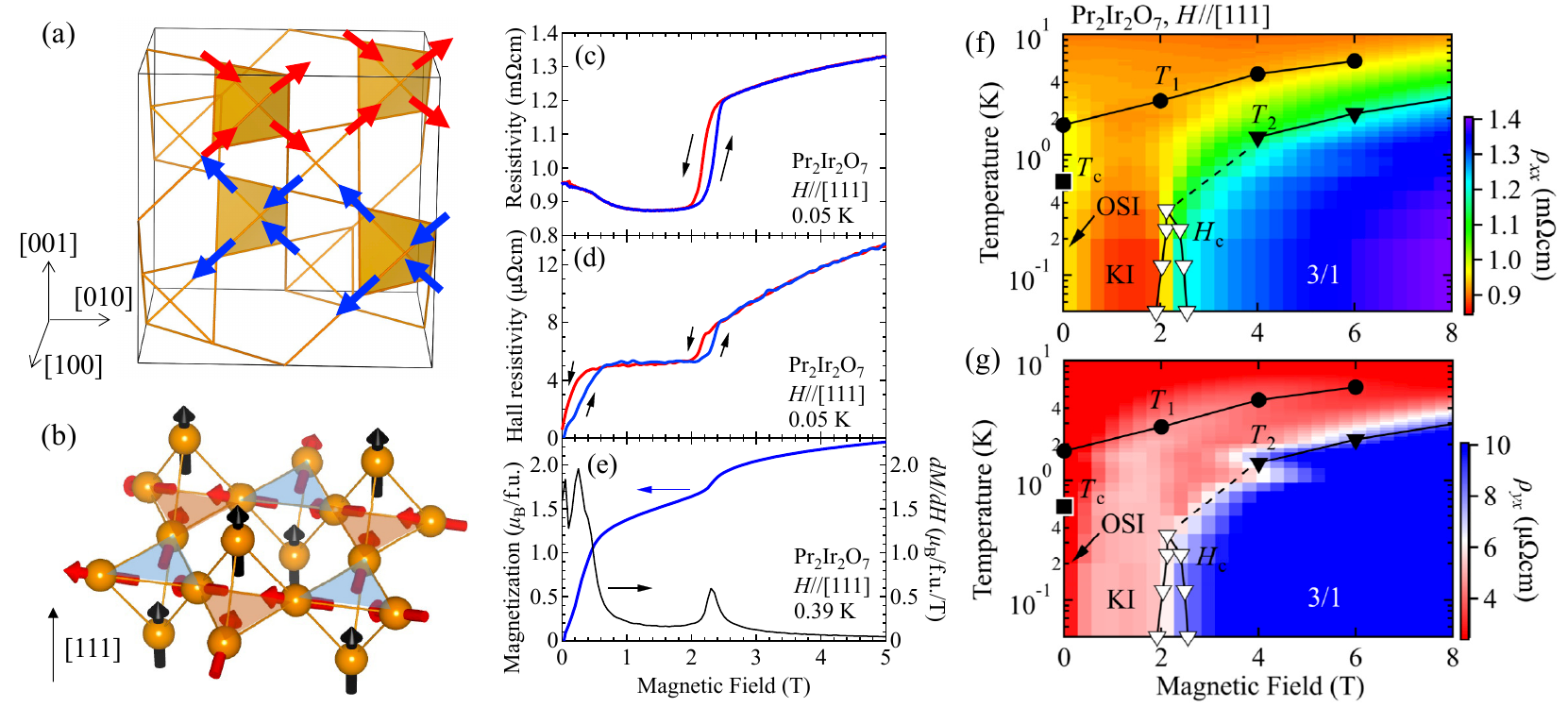}
\caption{
(a) Schematic picture of the magnetic configuration for the ordered spin ice state. The [010] components of the red spins are positive while those of the blue spins negative.
(b) Schematic picture of the magnetic configuration for the kagome ice state.
Magnetic-field dependence of (c) resistivity at 0.05 K, (d) Hall resistivity at 0.05 K, and (e) magnetization and its field derivative at 0.39 K along the [111] crystalline direction for Pr$_2$Ir$_2$O$_7$.
Contour plot of (f) resistivity and (g) Hall resistivity, respectively.
The solid square indicates the transition temperature $T_{\rm c}$ at zero field.
The solid circles (triangles) denote the temperature $T_1$ ($T_2$) determined from specific heat, and the open triangles denote the critical magnetic field $H_{\rm c}$ determined from (c) and (d).
OSI stands for ordered spin ice, KI stands for kagome ice, and 3/1 stands for the 3-in/1-out state.
Reproduced with permission from \cite{2022PRBUeda}, Copyright (2022) by the American Physical Society.
}
\label{5-3_pr2ir2o7-ueda}
\end{center}
\end{figure*}

As mentioned above, iridium oxides are prone to crystal defects which significantly affect their physical properties.
The long-range ordering of Pr magnetic moments is observed in polycrystalline Pr$_2$Ir$_2$O$_7$, which was missed in single crystals~\cite{2015PRBMacLaughlin}.
The specific heat of Pr$_2$Ir$_2$O$_7$ polycrystals shows a sharp peak at 0.9 K.
Moreover, the neutron diffraction uncovers the spin propagation vector $\textbf{q}$ = (0,0,1) below 0.9 K.
This result is consistent with the ``ordered spin ice state" which is theoretically proposed in Ref.~\cite{2001PRLMelko}.
The magnetic configuration is depicted in Fig. \ref{5-3_pr2ir2o7-ueda}(a); (i) the magnetic moments are aligned following the spin-ice rule (namely 2-in 2-out configuration); (ii) the net magnetization of a tetrahedron is aligned ferromagnetically in the (001) plane perpendicular to $\textbf{q}$; (iii) the planes are stacked antiferromagnetically with the wavelength and direction defined by $\textbf{q}$.
Note that this configuration is expected in canonical spin-ice systems such as Dy$_2$Ti$_2$O$_7$ and Ho$_2$Ti$_2$O$_7$, but the key interactions are fundamentally different.
In titanates, the nearest-neighbor exchange and long-range dipolar interactions play a major role, whereas RKKY interaction is dominant in the itinerant Pr$_2$Ir$_2$O$_7$~\cite{2012JPSJIshizuka}.

The relation between crystal defects and physical properties was closely examined in Pr$_2$Ir$_2$O$_7$ single crystals~\cite{2022PRBUeda}.
Precise structure analysis using synchrotron x-rays reveals that Ir sites are partially replaced by Pr ions.
A sample with small antisite-mixing shows the long-range magnetic ordering and the anomalies in magnetization and specific heat at 0.6 K.
Furthermore, it exhibits the field-induced magnetic transitions to the kagome-ice [see Fig.~\ref{5-3_pr2ir2o7-ueda}(b)] and 3-in 1-out state, similar to the typical spin-ice systems.
On the other hand, these behaviors are strongly suppressed in crystals with heavily antimixing sites.
Figures~\ref{5-3_pr2ir2o7-ueda}(c-e) show the magnetic field dependence of resistivity, Hall resistivity, and magnetization at the lowest temperature for Pr$_2$Ir$_2$O$_7$ with minimized antimixing.
(Note the large difference between the Hall resistivity shown in Fig.~\ref{5-3_pr2ir2o7-nakatsuji}(a)~\cite{2010NatureMachida} and Fig.~\ref{5-3_pr2ir2o7-ueda}(d)~\cite{2022PRBUeda} albeit the single-crystal samples for both cases.)
The resistivity and Hall resistivity at 0.05 K in the present better-stoichiometric crystal clearly exhibit a plateau-like magnetic field dependence in the intermediate field region, and a remarkable jump and hysteresis at 2.3 T.
The magnetization also shows a plateau and jump in the same field region, although these signatures are somewhat blurred due to the higher temperature (0.39 K).
These behaviors reflect the liquid-gas like metamagnetic transitions among the spin-ice state, kagome-ice state, and 3-in 1-out state as observed in insulating counterparts.
Figures~\ref{5-3_pr2ir2o7-ueda}(f) and (g) show the contour plot of resistivity and Hall resistivity in the plane of magnetic field and temperature.
The critical temperatures and fields are determined by transport and specific heat measurements.
Overall phase diagram is similar to that of Dy$_2$Ti$_2$O$_7$~\cite{Sakakibara2003}.
These behaviors suggest that the ordered spin ice state is likely realized in the metallic Pr$_2$Ir$_2$O$_7$ at zero field.
Another interesting point is the hydrostatic pressure dependence of Hall effect~\cite{2022PRBUeda}.
The resistivity at zero magnetic field hardly changes as the pressure is applied.
In contrast, the Hall conductivity for 3-in 1-out state dramatically decreases by half at 1.5 GPa, which cannot be explained by a subtle change of the longitudinal resistivity.
The theoretical calculation suggests that the pressure enlarges the strength of $f$-$d$ coupling between localized Pr magnetic moments and Ir conduction electron, promoting the pair annihilation of the Weyl points~\cite{2022PRBUeda}.

\subsubsection{Magnetic field-induced metal-insulator transitions.}
\label{sec:6_Iridates QCP}

\begin{figure}[h]
\begin{center}
\includegraphics[width=0.7\columnwidth]{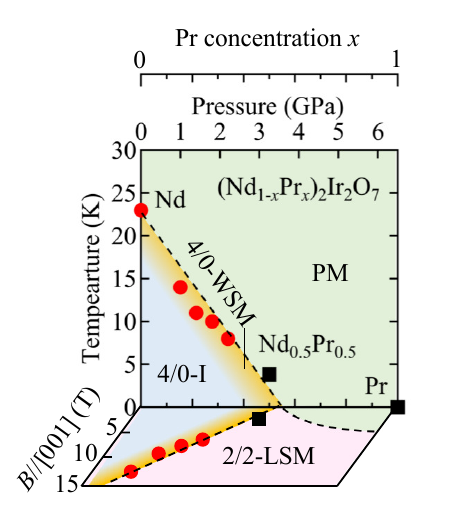}
\caption{
Phase diagram of (Nd$_{1-x}$Pr$_{x}$)$_2$Ir$_2$O$_7$ near the quantum critical point.
The horizontal axis is Pr concentration $x$ and hydrostatic pressure on Nd$_2$Ir$_2$O$_7$, on the basis of the scaling relation ($\Delta P\sim 0.65$ GPa corresponds to $x\sim 0.1$)~\cite{2015PRBUeda}.
The vertical axis is temperature, and the third axis is the magnetic field $B$ along the [001] crystalline direction.
The critical temperatures and fields are taken from Ref.~\cite{2017NCommUeda}.
4/0 stands for all-in all-out magnetic configuration, 2/2 stands for 2-in 2-out magnetic configuration, I stands for insulator, WSM stands for Weyl semimetal, LSM stands for line-node semimetal, and PM stands for paramagnetic metal.
}
\label{5-3_QCP_phasediagram}
\end{center}
\end{figure}

A quantum criticality and associated phenomena are a central issue in condensed matter physics.
Pyrochlore iridate exhibits an antiferromagnetic quantum critical point controlled by the application of hydrostatic pressure or the substitution of $R$ ions as shown in Fig.~\ref{5-3_QCP_phasediagram}.
For instance, in Nd$_{2}$Ir$_{2}$O$_{7}$, QBT breaks into eight Weyl points right below $\tn $, and causes the pair annihilation with further lowering temperature (see Sect.~\ref{sec:6_Iridates theory}).
As a pressure is applied to Nd$_2$Ir$_2$O$_7$, the all-in all-out type order is systematically suppressed and eliminated at around 4 GPa, which is close to the chemical composition $x=0.5$ in (Nd$_{1-x}$Pr$_{x}$)$_{2}$Ir$_{2}$O$_{7}$.
Notably, the quantum critical point separates the semimetal phase characterized by QBT and the linear-dispersing Weyl semimetal phase, giving rise to a nontrivial critical behavior~\cite{Savary2014}.
Moreover, other topological semimetals are realized by modulating the magnetic structure, apparently assembling towards the quantum critical point as well.
For instance, as shown in Fig.~\ref{5-3_QCP_phasediagram}, a line-node semimetal is realized by applying a magnetic field along the [001] direction which stabilizes the 2-in 2-out magnetic structure.
In the following, we present the experimental results of magnetotransports near the quantum phase transitions.

\begin{figure}[h]
\begin{center}
\includegraphics[width=0.95\columnwidth]{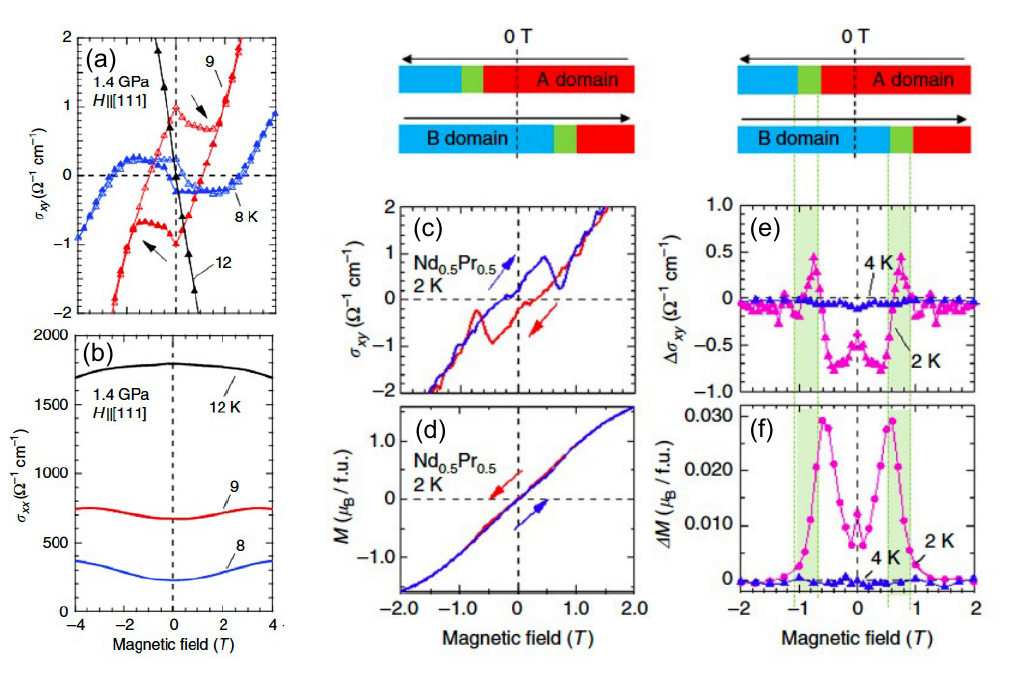}
\caption{
Magnetic field dependence of (a) Hall conductivity and (b) longitudinal conductivity for a magnetic field along the [111] crystallographic direction for Nd$_2$Ir$_2$O$_7$ at 1.4 GPa.
The open (closed) marks are Hall conductivities on increasing (decreasing) field process which is indicated by black arrows.
Magnetic field dependence of (c) Hall conductivity and (d) magnetization for a magnetic field along the [111] direction at 2 K for (Nd$_{0.5}$Pr$_{0.5}$)$_{2}$Ir$_{2}$O$_{7}$.
The blue (red) lines are Hall conductivity and magnetization on field-increasing (field-decreasing) process.
The difference of (e) Hall conductivity and (f) magnetization between field-increasing and field-decreasing process.
The magenta and blue denote 2 K ($<\tn $) and 4 K ($>\tn $), respectively.
The top pictures show the domain states in each process.
The red bars indicate A domain, blue ones are B domain, and green ones denote the domain flipping regions.
Reproduced by permission from Springer Nature Customer Service Centre GmbH: Nature Communications. \cite{2018NCommUeda} $\copyright $ 2018.
}
\label{5-3_nd2ir2o7_spontaneousHall}
\end{center}
\end{figure}

Figure~\ref{5-3_nd2ir2o7_spontaneousHall}(a) shows the magnetic field dependence of the Hall conductivity under pressure of 1.4 GPa in Nd$_2$Ir$_2$O$_7$ single crystals, which undergo the antiferromagntic transition at $\tn $ = 11 K (see Fig.~\ref{5-3_QCP_phasediagram}).
The spontaneous Hall effect is observed conspicuously at 9 K which is just below $\tn $, and is almost smeared out at 8 K, whereas the conductivity shows a minimal field dependence at each temperature [Fig.~\ref{5-3_nd2ir2o7_spontaneousHall}(b)].
It implies that the electronic structure dramatically changes below $\tn $, as observed in thermoelectric effect~\cite{2011JPSJMatsuhira,2022APLUeda}.
Such a unique temperature dependence of Hall effect is also discerned in (Nd$_{0.5}$Pr$_{0.5}$)$_{2}$Ir$_{2}$O$_{7}$ where the bandwidth is controlled by the substitution of $R$ ions and hence $\tn $ is markedly decreased down to $\sim 3$ K [Fig.~\ref{5-3_nd2ir2o7_spontaneousHall}(c)].
Moreover, it is noteworthy that the Hall conductivity is quite large ($\sim 0.5$ $\Omega ^{-1}$cm$^{-1}$) even though the spontaneous magnetization is tiny [$\sim 0.01$ $\mu _{\rm B}$ as shown in Fig.~\ref{5-3_nd2ir2o7_spontaneousHall}(d)].
These observations are attributable to the emergence of Weyl points.
As described in Sect.~\ref{sec:6_Iridates theory}, Weyl points begin to appear at the Brillouin zone center by time-reversal symmetry breaking and quickly immigrate towards the Brillouin zone boundary with increasing the order parameter, followed by the pair annihilation at the zone boundary~\cite{2012PRBKrempa}.
We note that the net magnetization should be strictly zero under zero magnetic field for the ideal all-in all-out state.
One possibility for the observed spontaneous component is that the magnetic Nd ions are mixed into the Ir sites as observed in Pr$_2$Ir$_2$O$_7$~\cite{2022PRBUeda}.
The other is an intrinsic origin that the cubic symmetry is broken by other degrees of freedom besides the magnetic dipole moment.
Currently, the magnetic torque experiment detects the breaking of the four-fold rotational symmetry below $\tn $ in Eu$_2$Ir$_2$O$_7$~\cite{2017NPLiang}.
Raman spectroscopy also uncovers the emergence of new peaks that cannot be explained by a simple reduction of the lattice symmetry~\cite{2019PRBUeda}.
Revealing the origin is a subject for future work.

\begin{figure*}[h]
\begin{center}
\includegraphics[width=1.8\columnwidth]{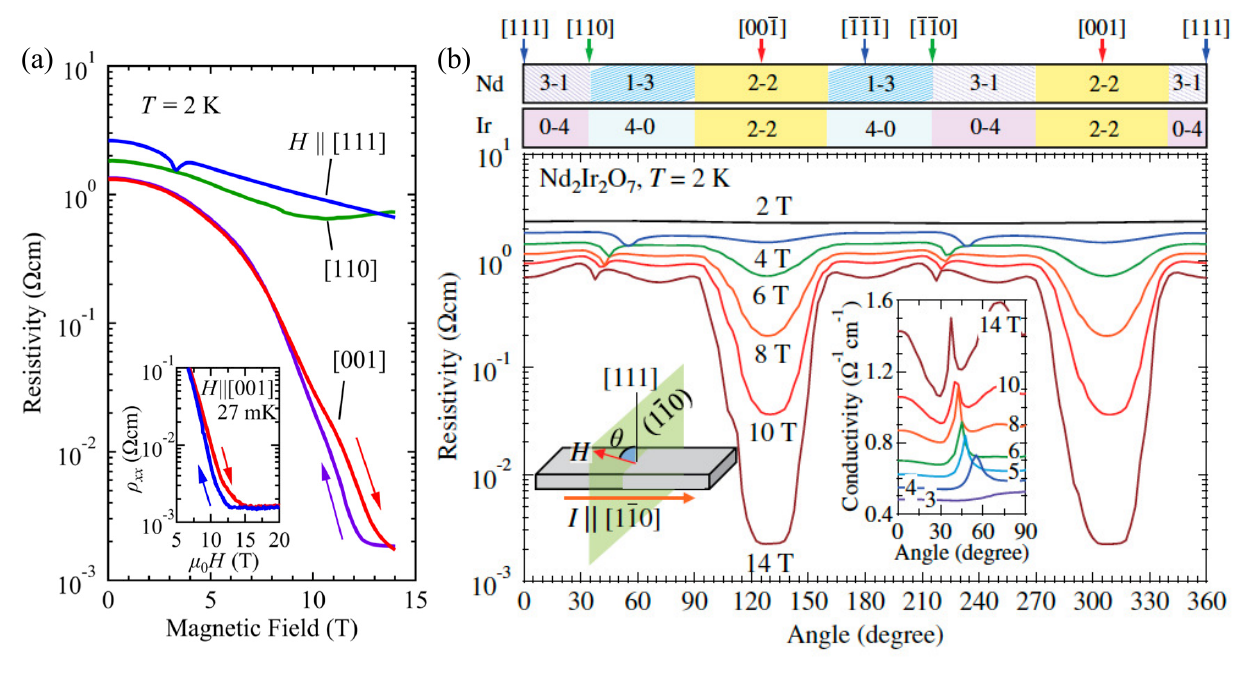}
\caption{
(a) Magnetic field dependence of resistivity at 2 K for several field
directions in Nd$_2$Ir$_2$O$_7$.
The blue curve denotes $B//[111]$, the green one $B//[111]$, and the red (purple) one $B//[001]$ on the increasing (decreasing) field processes, respectively.
The inset to (a) shows the field dependence of the resistivity for $B//[001]$ up to 20 T at 27 mK.
(b) The angle dependence of resistivity (on a logarithmic scale) for several fields at 2 K.
The two top tables denote the possible magnetic structures of the Nd $4f$ and Ir $5d$ moments in the strong field limit.
The numbers denote the magnetic moment configurations; 3-1 stands for 3-in 1-out state, 2-2 stands for 2-in 2-out state, and 0-4 stands for 0-in 4-out state. 0-4 is one domain state of the all-in all-out structure and 4-0 is the other domain state.
The left inset exhibits the experimental configuration.
The right inset shows the low angle dependence of conductivity (on a linear scale) for several fields.
Reproduced with permission from \cite{2015PRLUeda}, Copyright (2015) by the American Physical Society.
}
\label{5-3_nd2ir2o7-angle}
\end{center}
\end{figure*}

Figure~\ref{5-3_nd2ir2o7-angle}(a) shows the magnetic field dependence of the resistivity for three high-symmetry axes in Nd$_2$Ir$_2$O$_7$~\cite{2015PRLUeda}.
The resistivity decreases as the field increases for all field directions.
Remarkably, for $B//$[001], the resistivity decreases by three orders of magnitude with hysteresis between the increasing and decreasing field processes, indicating the observed magnetoresistance due to the first-order phase transition.
Figure~\ref{5-3_nd2ir2o7-angle}(b) displays the angular dependence of resistivity at 2 K.
The electric current direction is kept along $I//[1\overline{1}0]$ and the magnetic field is continuously rotated in the (1$\overline{1}$0) plane which involves several high-symmetry axes.
Whereas little angular dependence is observed at 2 T, two distinct features are discerned in the period of 180$^{\circ }$ at higher magnetic fields.
One is a sharp dip structure at $30^{\circ }$ - $60^{\circ }$.
The inset shows the conductivity (the inverse of resistivity) in the corresponding angle region.
The peak conductivity is nearly invariant to the magnetic field ($\ge $ 4 T) while the peak angle decreases as the magnetic field increases.
This is attributable to the switching of the all-in all-out magnetic domain and hence to the appearance of metallic domain walls in the insulating bulk (the details of domain wall conduction will be discussed in the next section). 
The other remarkable feature is the large decrease of resistivity around $135^{\circ }$ or $B$//[001].
The resistivity decreases systematically with increasing fields, in contrast to the contribution from domain walls.
The magnetic structure of Nd magnetic moments in the large field limit is shown in the upper part of Fig.~\ref{5-3_nd2ir2o7-angle}(b).
The angular region where the resistivity decreases coincides with the region where the Nd magnetic moments form the 2-in 2-out state, suggesting that the 2-in 2-out magnetic structure is responsible for the origin of the large negative magnetoresistance observed in Fig.~\ref{5-3_nd2ir2o7-angle}(a).
Similar behavior is later confirmed in Ref.~\cite{2016NPTian} as well.

\begin{figure}[h]
\begin{center}
\includegraphics[width=0.95\columnwidth]{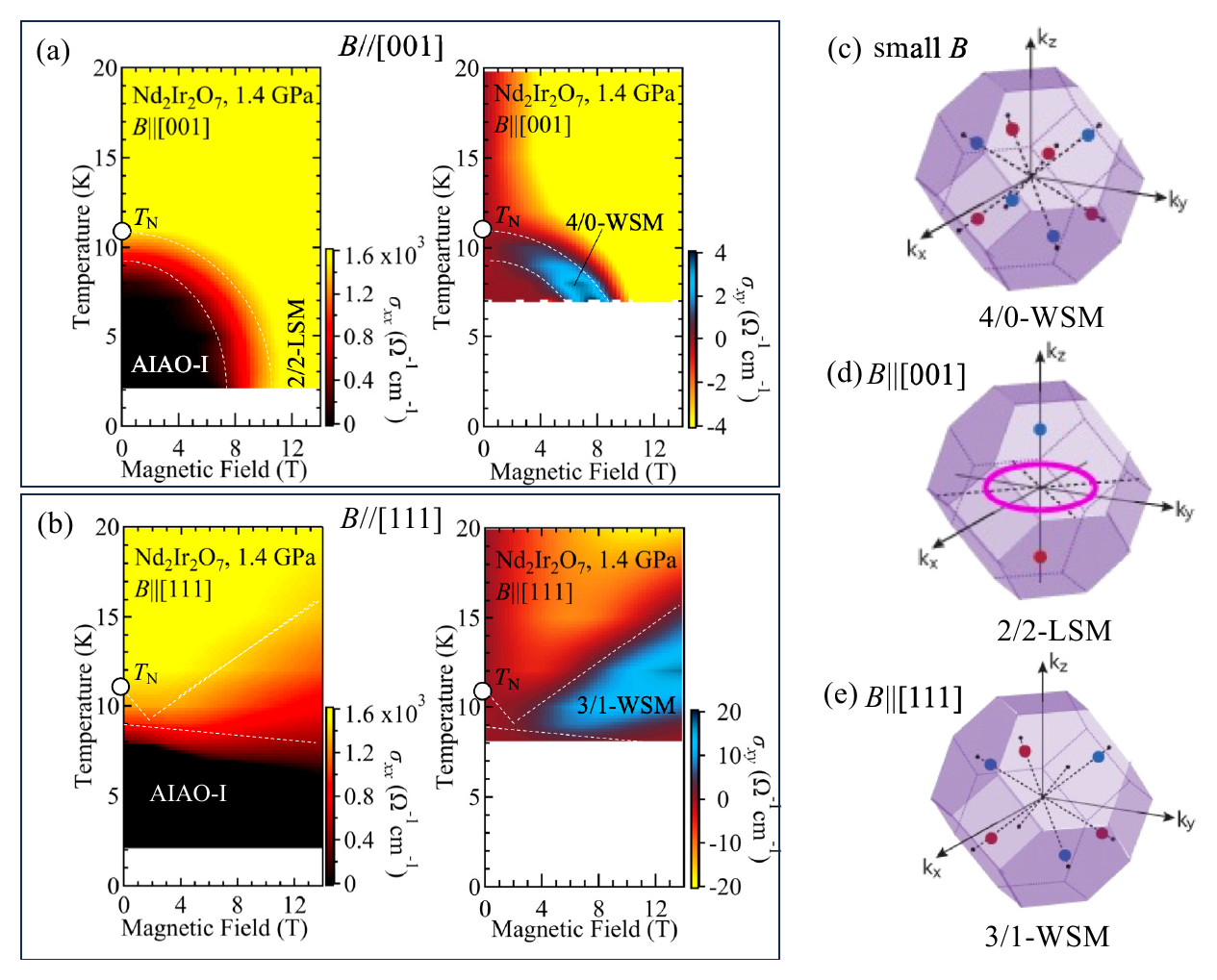}
\caption{
Counter plot of conductivity in Nd$_2$Ir$_2$O$_7$ at 1.0 GPa for (a) $B//$[001] and (b) $B//$[111].
AIAO-I stands for all-in all-out insulator, 2/2 LSM stands for 2-in 2-out line-node semimetal, and 3/1 WSM stands for 3-in 1-out Weyl semimetal.
Schematic picture of the distribution of Weyl points and line nodes in the three-dimensional momentum space for (c) WSM with AIAO in a small magnetic field, (d) 2/2-LSM in a magnetic field along the [001] direction ($B//[001]$), and (e) 3/1-WSM in a field along the [111] direction ($B//[111]$).
Red (blue) points denote the Weyl points with positive (negative) sign of the charge chirality and a purple ring denotes the line node.
Reproduced by permission from Springer Nature Customer Service Centre GmbH: Nature Communications. \cite{2017NCommUeda} $\copyright $ 2017.
}
\label{5-3_nd2ir2o7_pressure}
\end{center}
\end{figure}

As the hydrostatic pressure is applied on Nd$_2$Ir$_2$O$_7$ single crystals, the antiferromagnetic insulating phase is gradually suppressed (see Fig.~\ref{5-3_QCP_phasediagram})~\cite{2015PRLUeda}, like stoichiometric polycrystals~\cite{2011PRBSakata,2015PRBUeda}.
Moreover, the resistivity shows the strong magnetic field dependence as the quantum critical point approaches~\cite{2017NCommUeda}.
Figures~\ref{5-3_nd2ir2o7_pressure}(a) and (b) display the contour plot of conductivity at 1.0 GPa for $B//$[001] and $B//$[111].
In addition to the 2-in 2-out semimetallic state for $B//$[001], another semimetallic state shows up for $B//$[111] where the 3-in 1-out state is favored. 
Figure~\ref{5-3_nd2ir2o7_pressure}(c-e) show the schematic pictures of the distribution of Weyl points for respective magnetic configurations.
In the all-in all-out state, which maintains the cubic symmetry, four pairs of Weyl points show up on the three-fold rotational symmetry axes and cause pair annihilation at the zone boundary [Fig.~\ref{5-3_nd2ir2o7_pressure}(c)].
On the other hand, in the 2-in 2-out state, a line node in the (001) mirror plane and one pair on the [001] axis can be realized [Fig.~\ref{5-3_nd2ir2o7_pressure}(d)].
In the 3-in 1-out state, one three-fold rotational symmetry axis survives and hence the pair annihilation occurs only on that axis, resulting in the Weyl semimetal with three pairs of Weyl points [Fig.~\ref{5-3_nd2ir2o7_pressure}(e)].
Interestingly, by increasing the pressure, or equivalently tuning the effective bandwidth, the antiferromagnetic insulator phase shrinks and hence all the three topological states appear to merge into the quantum critical point~\cite{2017NCommUeda}.

The thermoelectric effect, which is sensitive to band structures near the Fermi level, also shows a large response near this magnetic field-induced metal-insulator transitions \cite{2022APLUeda}.
Especially, Nernst effect exhibits distinctively large values among oxides even comparable to high-$T_{c}$ cuprates, supporting the emergence of topologically nontrivial states.

\subsubsection{Anomalous metallic states on magnetic domain walls.}
\label{sec:6_Iridates DW}

The physics and applications of magnetic interfaces has been studied for many years.
As mentioned above, the bulk ground state of Nd$_2$Ir$_2$O$_7$ is an antiferromagnetic insulator with a charge gap of 50 meV.
However, anomalous metallic states at the boundary of magnetic domains were discovered by transport and spectroscopic measurements~\cite{2014PRBUeda,2015PRLUeda}, as well as by the real-space observations with microwave impedance microscopy~\cite{2015ScienceMa}.

\begin{figure}[h]
\begin{center}
\includegraphics[width=0.95\columnwidth]{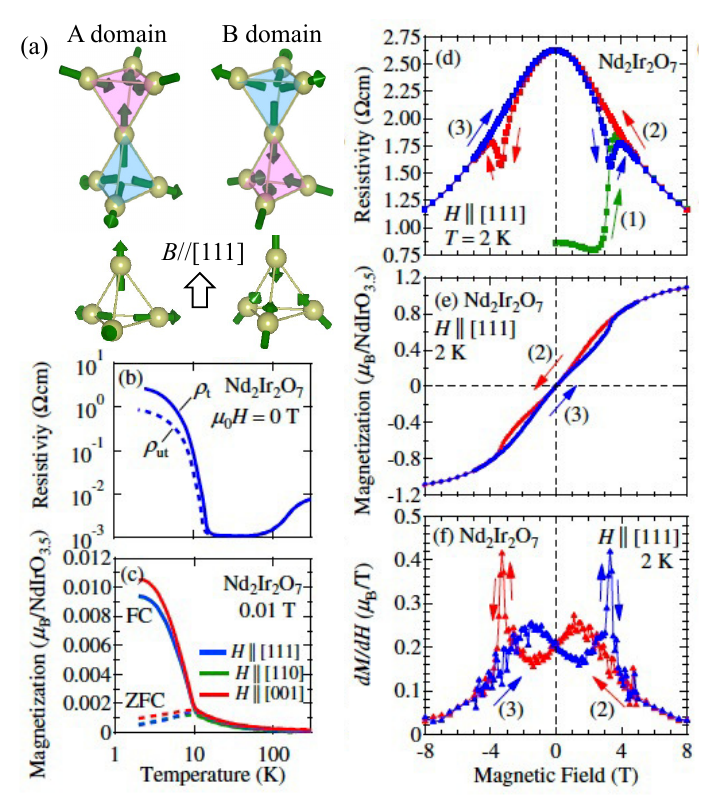}
\caption{
(a) Schematic picture of all-in all-out configuration for two domain states.
The lower panels show the spin canting under the magnetic field along [111] direction.
Temperature dependence of (b) the resistivity of the untrained state $\rho _{\rm ut}$ and that of the trained state $\rho _{\rm t}$ and (c) magnetization in zero field cooling and field cooling for Nd$_2$Ir$_2$O$_7$.
Magnetic field dependence of (d) resistivity, (e) magnetization, and
(f) field derivative of magnetization for $B//[111]$ in Nd$_2$Ir$_2$O$_7$.
The green curve denotes the virgin curve, the red one denotes the sweeping process (2), and the blue one denotes the process (3), respectively.
Reproduced with permission from \cite{2015PRLUeda}, Copyright (2015) by the American Physical Society.
}
\label{5-3_nd2ir2o7_DWtransport}
\end{center}
\end{figure}

The all-in all-out magnetic structure has two magnetic domains, as shown in Fig.~\ref{5-3_nd2ir2o7_DWtransport}(a).
Although the net magnetization is zero like antiferromagnets, the domain state can be aligned by the magnetic field along [111] direction~\cite{2012JPSJArima}.
This can be understood in terms of the spin canting in $B//[111]$, as shown in the lower panels of Fig.~\ref{5-3_nd2ir2o7_DWtransport}(a).
Because of the Zeeman energy gain, the three spins at the corner sites of the triangle are expected to be canted; they tilt toward the triangular plane (left panel) whereas they tilt away from it (right panel).
This leads to the difference of the total spin energy, and hence the magnetic field selects a unique domain.
Figure~\ref{5-3_nd2ir2o7_DWtransport}(b) shows the temperature dependence of the resistivity for Nd$_2$Ir$_2$O$_7$ measured in two ways.
On the one hand, the sample is initially cooled down under zero magnetic field and the resistivity is measured as temperature is elevated.
In this state, the sample is ``untrained" and the resistivity $\rho _{\rm ut}$ reflects the multidomain state.
On the other hand, the sample is cooled down to the lowest temperature with a magnetic field applied to the $[111]$ direction, followed by the resistivity measurement at zero field on the warming process.
The resistivity measured in this process, $\rho _{\rm t}$, corresponds to that of the single-domain state.
One can see that $\rho _{\rm t}$ is larger than $\rho _{\rm ut}$ below $\tn $ in Fig.~\ref{5-3_nd2ir2o7_DWtransport}(b).
The difference between $\rho _{\rm t}$ and $\rho _{\rm ut}$ is more pronounced for polycrystals, which are closer to the ideal stoichiometric ratio~\cite{2014PRBUeda}.
In most cases of ferromagnetic metals, the resistivity becomes lower by a few percent when the magnetic domains are aligned, because conduction electrons are scattered at magnetic domain walls.
In contrast, the domain walls in Nd$_2$Ir$_2$O$_7$ appear to have positive and large contribution to the electrical conductance.
Figures~\ref{5-3_nd2ir2o7_DWtransport}(d) and (e) show the magnetic field dependence of resistivity and magnetization at 2 K, respectively.
In the process~(1), the resistivity is measured as the magnetic field is applied after zero field cooling.
The resistivity clearly jumps at around 3.4 T.
As the magnetic field is decreased [process~(2)], the resistivity increases rapidly and reaches the maximum value at 0 T which is more than three times larger than that of the initial state.
As the field is further decreased, the resistivity shows a sharp dip structure at around -3.4 T.
The similar behavior is observed in the increasing field process [process~(3)].
The anomaly at $\pm $3.4 T is also observed in the field dependence of magnetization shown in Fig.~\ref{5-3_nd2ir2o7_DWtransport}(e).
Besides, the hysteresis is observed in the increasing and decreasing processes in the small field region between $\pm $3.4 T.
The magnetic field derivative of the magnetization shows clear peaks at $\pm $3.4 T, indicating that the all-in all-out magnetic domain is switched at this critical field [Fig.~\ref{5-3_nd2ir2o7_DWtransport}(f)].
These findings indicate that the domain walls are highly conductive in this material.

\begin{figure}[h]
\begin{center}
\includegraphics[width=0.95\columnwidth]{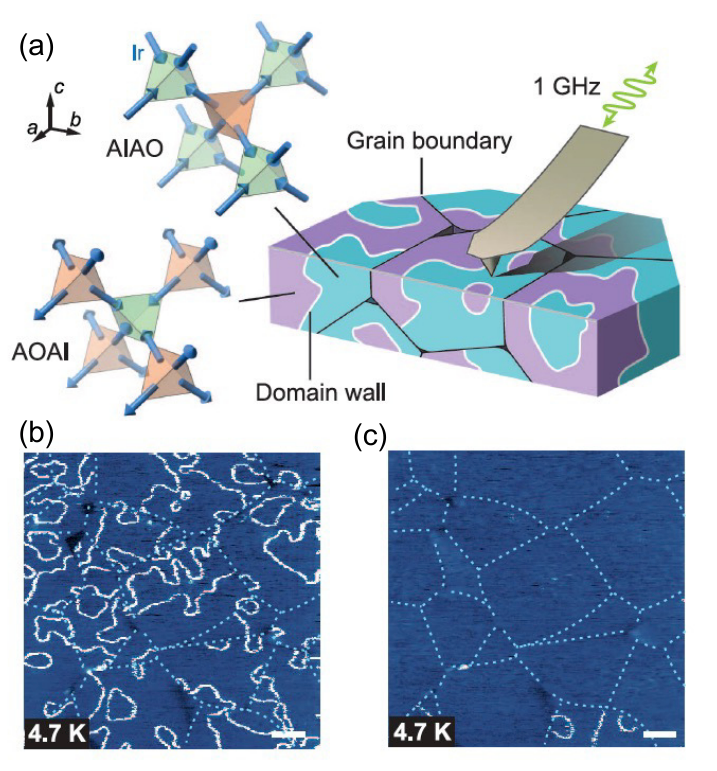}
\caption{
(a) Illustration of microwave impedance microscopy (MIM) scanning setup on polished polycrystal Nd$_2$Ir$_2$O$_7$, showing the spin configuration of all-in all-out (AIAO) order and its AIAO/AOAI variations.
(b) MIM image at 4.7 K after zero field cooling, showing randomized domains.
Domain walls can exist between the two variations.
A higher MIM signal corresponds to a higher local conductivity.
The dotted lines are grain boundaries and the dark spots are voids between grains, which can be identified in higher-temperature scans.
Curvilinear features much more conductive than the bulk are observed in all grains and are identified as AIAO magnetic domain walls.
They either form closed loops or terminate at the grain boundaries, sometimes in close proximity.
(c) Same region after a thermal cycle to 40 K and cooling back to 4.7 K in 9 T. Most grains become single-domain, agreeing with transport. Scale bars: 2 mm.
From \cite{2015ScienceMa}. Reprinted with permission from American Association for the Advancement of Science.
}
\label{5-3_nd2ir2o7_DWMIM}
\end{center}
\end{figure}

The metallic domain walls are directly observed by microwave impedance microscopy (MIM)~\cite{2015ScienceMa}.
MIM is a device that can measure the local impedance of a sample surface by microwaves and visualize the relative electrical conductivity [Fig.~\ref{5-3_nd2ir2o7_DWMIM}(a)]~\cite{2022NRPBarber}.
Figures~\ref{5-3_nd2ir2o7_DWMIM}(b) and (c) exhibit the MIM image of a polished surface of a Nd$_2$Ir$_2$O$_7$ polycrystal at 4.7 K.
The dotted white lines indicate the grain boundaries of the polycrystalline sample.
The metallic magnetic-domain walls (white area) are embedded in the insulating bulk (blue area) in the zero-field cooling state, whereas the domain walls are almost diminished after applying field.
Note that the domain walls remain in some grains where the out-of-plane crystallographic axis is close to the [001] axis.
Since the energy gain of two all-in all-out magnetic domains are identical under the external field along [001], the magnetic domain is not aligned by this specific field direction.
These observations are consistent with the transport properties~\cite{2014PRBUeda,2015PRLUeda}.

\begin{figure}[h]
\begin{center}
\includegraphics[width=0.95\columnwidth]{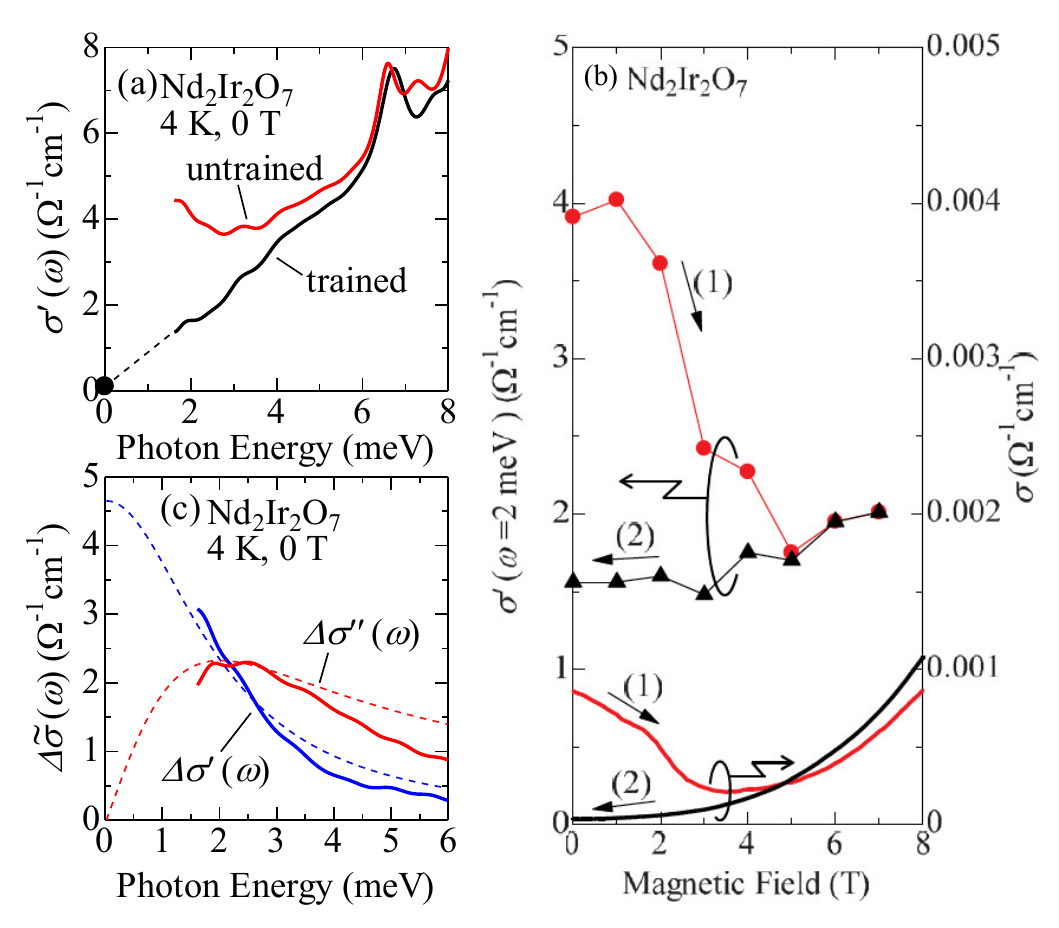}
\caption{
(a) The real part of optical conductivity spectra at 4 K for Nd$_2$Ir$_2$O$_7$.
The red (black) line indicates the spectra for the untrained (trained) state.
The filled circle denotes the dc conductivity of the trained state.
The broken line is the guide to the eye for the trained state.
(b) Magnetic field dependence of the optical conductivity at 2 meV (circles and triangles) and the dc conductivity at 4 K (solid lines).
(c) Complex optical conductivity $\Delta \tilde{\sigma }(\omega )$ at 4 K for the antiferromagnetic domain walls, as defined by the differential spectra between the untrained (multidomain) and trained (single-domain) states.
The blue (red) line indicates the real (imaginary) part of $\Delta \tilde{\sigma }(\omega )$.
The dashed lines denote the spectra fitted with the Drude formula.
Reproduced with permission from \cite{2014PRBUeda}, Copyright (2014) by the American Physical Society.
}
\label{5-3_nd2ir2o7_DWterahertz}
\end{center}
\end{figure}

The optical conductivity spectra were measured using terahertz time-resolved spectroscopy~\cite{2014PRBUeda}.
The advantage of this technique is that the real and imaginary parts of the optical conductivity can be measured directly without the Kramers-Kronig transformation.
Figure \ref{5-3_nd2ir2o7_DWterahertz}(a) shows the real part of the optical conductivity measured at 4 K and 0 T.
Similar to the transport properties in Fig. \ref{5-3_nd2ir2o7_DWtransport}, the conductivity spectra of untrained state and trained state are measured after zero-field cooling and field cooling, respectively.
The optical conductivity of the trained state monotonically decreases as the photon energy decreases and apparently well coincides with the dc conductivity.
On the other hand, the optical conductivity of the untrained state shows an upturn below 3 meV.
Figure \ref{5-3_nd2ir2o7_DWterahertz}(b) shows the magnetic field dependence of the optical conductivity at 2 meV (left axis) and dc conductivity (right axis).
The optical conductivity sharply decreases above 2 T as the field increases in process (1), whereas the conductivity slightly changes with decreasing field and reaches the smaller value than that of the initial state in process (2).
This behavior is similar to the transport properties plotted in the same figure.
These indicate that the difference of optical conductivity between trained and untrained state can be attributed to the presence of the metallic magnetic-domain walls.
To discuss the charge dynamics on the domain wall states, the difference of the optical conductivity $\Delta \sigma (\omega )=\sigma ^{\rm ut}(\omega )-\sigma ^{\rm t}(\omega )$ is plotted in Fig. \ref{5-3_nd2ir2o7_DWterahertz}(c).
Both real and imaginary parts can be well fitted by the Drude response as shown in the dashed lines.
In particular, the imaginary part of the optical conductivity takes a maximum value at about 2 meV, which corresponds to the inverse of the relaxation time $\hbar / \tau $.
It indicates that the conducting state on magnetic domain walls has a low scattering rate typical of good metals.

Finally, we discuss the origin of the metallic domain walls.
One possibility is that the antiferromagnetic order is disturbed at magnetic interfaces and hence the metallic state remains as in the paramagnetic state.
Or, if the magnetic state of the interface is ferromagnetic like such as 2-in 2-out pattern, the resistivity may be significantly small as observed in the magnetotransport previously.
Another and more plausible scenario is that the topological edge states are realized at the magnetic interface \cite{Yamaji2014}.
As mentioned earlier, Weyl nodes show up by the all-in all-out magnetic order but quickly disappear at low temperatures.
Interestingly, even after the pair annihilation, the surface state can remain protected in zero-mode or in-gap states at magnetic interfaces, resulting in the electrical conduction.
It is predicted that the helical spin textures are realized in the two-dimensional Fermi surface, which may be detected by low-energy magnetooptics \cite{Yamaji2016}.

\subsubsection{Magnetotransport properties of thin films.}

A major advantage of thin film fabrication is to expand physical properties and functions by introducing epitaxial strains and interface proximity effects~\cite{2012NMHwang}.
However, as mentioned above, growing pyrochlore iridates is challenging even for bulk crystals.
Single-phase Eu$_2$Ir$_2$O$_7$ thin films are obtained by the ``solid-phase epitaxy (SPE)" method, in which the amorphous thin film is firstly prepared at low temperature and then annealed ex situ to turn into the single-crystalline thin film~\cite{2015SRFujita}.
It is fabricated, for example, using Y-stabilized ZrO$_2$ (YSZ) (111) single crystal substrates, stretched by 0.7 \% along the out-of-plane [111] direction as compared with the Ir-oxide pyrochlore lattice.
This film reproduces the metal-insulator transition accompanied with the antiferromagnetic order at about 120 K.
Furthermore, below $\tn $, it shows a unique magnetoresistance which is odd with respect to the magnetic field, reflecting the all-in all-out magnetic structure~\cite{2012JPSJArima}.
An artificial all-in all-out magnetic domain boundary is created at the interface between Eu$_2$Ir$_2$O$_7$ and Tb$_2$Ir$_2$O$_7$~\cite{2016PRBFujita}.
Owing to the small Zeeman energy, the alignment of the magnetic domains in Eu$_2$Ir$_2$O$_7$ is achieved only by the field cooling.
On the other hand, in Tb$_2$Ir$_2$O$_7$, an external magnetic field can control the large Tb $4f$ magnetic moments and thereby the Ir magnetic domains via $f$-$d$ interactions, similar to Nd$_2$Ir$_2$O$_7$.
For instance, when the magnetic field is applied along the $[111]$ direction and the temperature is lowered, one magnetic domain is energetically stable while the other domain is diminished, resulting in the single domain state in both Eu$_2$Ir$_2$O$_7$ and Tb$_2$Ir$_2$O$_7$.
Then, as the magnetic field is applied to the opposite direction while keeping the low temperature, only the magnetic domain of Tb$_2$Ir$_2$O$_7$ is inverted, leading to a magnetic domain boundary at the interface between Eu$_2$Ir$_2$O$_7$ and Tb$_2$Ir$_2$O$_7$.
This artificial all-in all-out interface indeed produces the metallic contribution to the conductance.
Another type of heterostructure combining paramagnetic metal Bi$_2$Ir$_2$O$_7$ and magnetic insulator Dy$_2$Ti$_2$O$_7$ is reported in Ref.~\cite{2023NCZhang}.
In contrast to Bi$_2$Ir$_2$O$_7$ films, it shows the large negative magnetoresistance by $\sim 15$ \%.
Furthermore, a positive cusp-like anomaly shows up in $B$//[111] and $B$//[110], reflecting the ice-rule-breaking transition among 2-in 2-out and 3-in 1-out magnetic configurations.
It implies that the magnetic proximity effect plays a crucial role in the magnetotransport properties in such a heterostructure~\cite{2023NCZhang}.

\begin{figure*}[h]
\begin{center}
\includegraphics[width=1.8\columnwidth]{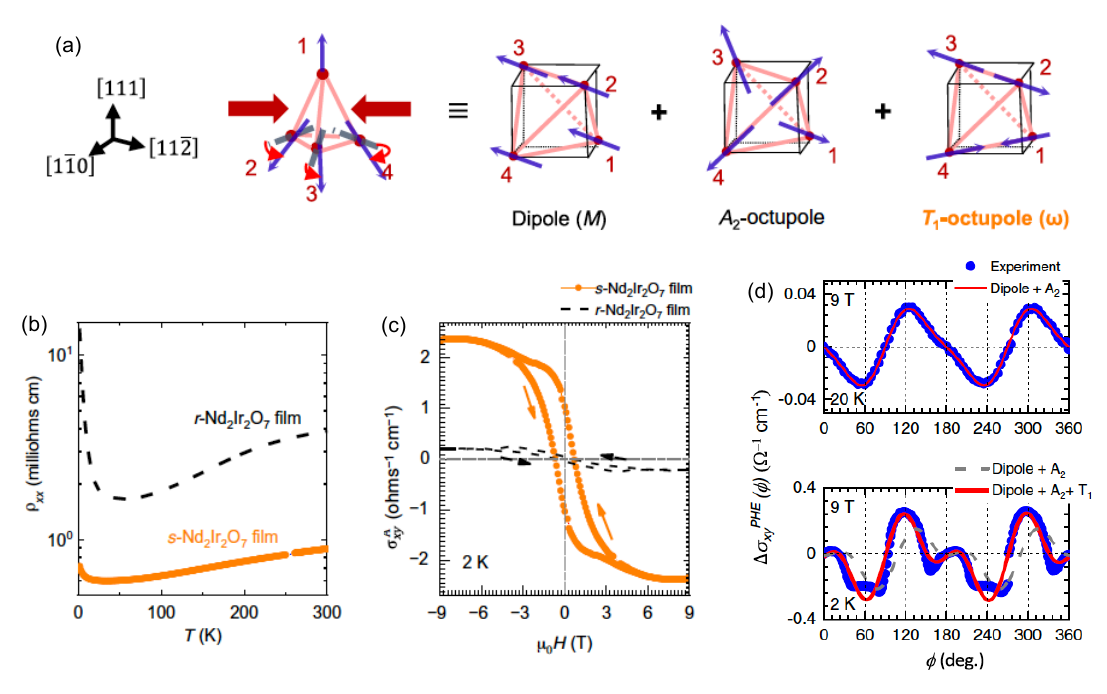}
\caption{
(a) Schematic diagram of the strained magnetic ground state in Nd$_2$Ir$_2$O$_7$ thin films.
Biaxial strain along the [111] direction will distort the all-in all-out configuration, which can be represented as the superposition of a cluster dipole ($M$), an $A_2$-octupole, and a $T_1$-octupole ($\omega $).
$M$ represents a ferromagnetic ordering, while $\omega $ represents an antiferromagnetic ordering other than all-in all-out.
(b) Temperature dependence of resistivity $\rho _{xx}$ for the strained-(s-) and relaxed-(r-)Nd$_2$Ir$_2$O$_7$ (NIO) films.
(c) Magnetic field dependence of anomalous Hall conductivity $\sigma _{xy}^{\rm A}$ for s-NIO (orange circles) and r-NIO (dashed lines).
From Ref.~\cite{2020SAWJKim}. Reprinted with permission from American Association for the Advancement of Science.
(d) Angle ($\phi $) dependence of planar Hall conductivity $\sigma _{xy}^{\rm PHE}$ at 20 K (upper panel) and 2 K (lower panel) for s-NIO.
Reproduced by permission from Springer Nature Customer Service Centre GmbH: Nature Communications. \cite{2022NCSong} $\copyright $ 2022.
}
\label{5-3_niofilm_octupole}
\end{center}
\end{figure*}

Application of biaxial strain via epitaxy is a powerful method for manipulating the charge, spin, and orbital degrees of freedom.
The epitaxial strain in Nd$_2$Ir$_2$O$_7$ thin films controls the magnetic multipole moments in the magnetically ordered phase~\cite{2020SAWJKim}.
While the all-in all-out state without strain is described as A$_2$-octupole in terms of a cluster multipole picture~\cite{2019PRBSuzuki}, it involves a superposition of dipole, $A_2$-octupole, and $T_1$-octupole in the presence of the biaxial strain [Fig.~\ref{5-3_niofilm_octupole}(a)].
In particular, the $T_1$-octupole breaks the cubic symmetry like dipole, and hence produces finite magnetization and Hall effects.
The lower panels of Fig. \ref{5-3_niofilm_octupole} show the transport properties of the relaxed (r-) and strained (s-) films of Nd$_2$Ir$_2$O$_7$.
For the unstrained thick film, the resistivity markedly increases below 30 K as in the bulk.
On the other hand, the increase of the resistivity is strongly suppressed in the strained thin film [Fig.~\ref{5-3_niofilm_octupole}(b)]. 
Furthermore, a large spontaneous Hall effect is observed while the net magnetization is not observed within the measurement accuracy ($\pm 0.01$ $\mu _{\mathrm B}$/NdIrO$_{3.5}$) in the strained films [Fig.~\ref{5-3_niofilm_octupole}(c)].
The characteristic angle dependence of planar Hall effect is also observed [Fig.~\ref{5-3_niofilm_octupole}(d)], which are consistently explained by the high-rank $T_{1}$-octupole~\cite{2022NCSong}.

Another advantage of thin films is that they can be engineered to approach the two-dimensional limit. This enhances quantum fluctuations, potentially leading to exotic magnetic ground states such as spin liquid states, particularly in pyrochlore oxides with frustrated lattices.
In a 30 nm thin film sample of Y$_2$Ir$_2$O$_7$~\cite{2024NCLiu}, the anomalous Hall effect is observed even in the absence of magnetic Bragg peaks, {\it i.e.,} the disappearance of long-range order.
Additionally, RIXS reveals a unique collective excitation with minimal dispersion that is energetically comparable to magnon excitations in other $R_2$Ir$_2$O$_7$ compounds.
These results suggest the feasibility of a chiral spin liquid characterized by finite SSC.

\subsection{Filling-control metal-insulator transitions and magnetotransport properties}
\label{sec:6_Iridate filling-control}



Different from the rare-earth pyrochlore iridates, Cd$_2$Ir$_2$O$_7$~\cite{2018PRBDai} and Ca$_2$Ir$_2$O$_7$~\cite{2022JPSJNakayama} with pentavalent Ir atoms show bad-metal like behaviors with no clear feature of phase transition.
Therefore, the partial substitution of $R^{3+}$ ions with $A^{2+}$ (Cd$^{2+}$ and Ca$^{2+}$) ions, namely the band filling, may give rise to a variety of electronic and magnetic phases.


\begin{figure}[h]
\begin{center}
\includegraphics[width=0.95\columnwidth]{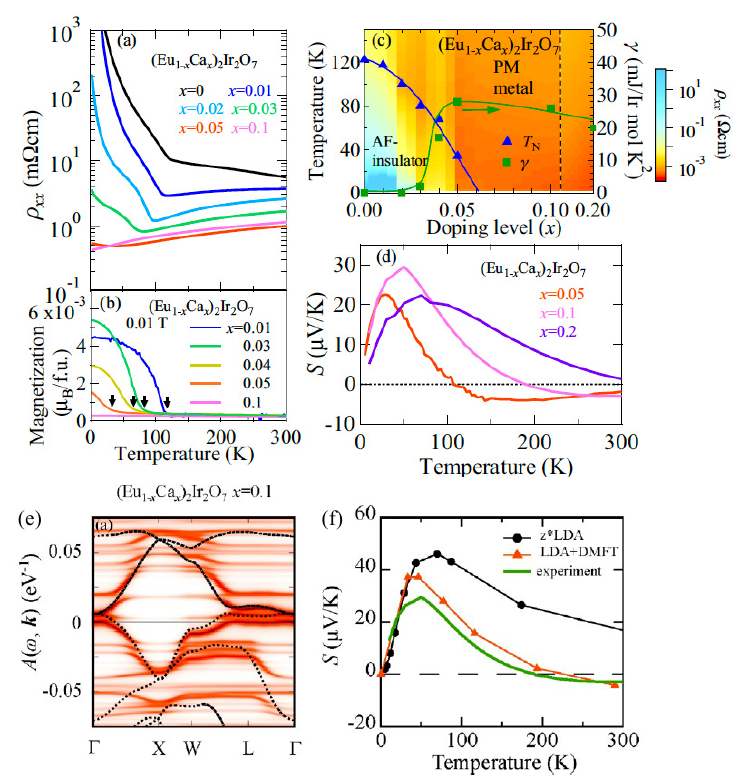}
\caption{
Temperature dependence of (a) resistivity and (b) magnetization for (Eu$_{1-x}$Ca$_{x}$)$_{2}$Ir$_{2}$O$_{7}$.
Arrows denote the transition temperature of antiferromagnetic ordering $\tn $.
(c) The contour map of resistivity in the $x$-$T$ plane as well as the $x$ dependence of $\tn $ and $\gamma $ for (Eu$_{1-x}$Ca$_{x}$)$_{2}$Ir$_{2}$O$_{7}$.
$\tn $ and $\gamma $ are denoted by triangle and square, respectively.
The solid line is shown as a guide to the eye.
(d) Temperature dependence of the Seebeck coefficient for (Eu$_{1-x}$Ca$_{x}$)$_{2}$Ir$_{2}$O$_{7}$.
(e) The band structure calculated by the LDA+DMFT ($U=1.3$ eV) for the paramagnetic metallic phase of (Eu$_{1-x}$Ca$_{x}$)$_{2}$Ir$_{2}$O$_{7}$ with $x=0.1$.
The temperature is set at 33 K. The black line denotes the LDA result with the renormalization factor $z=0.15$.
(b) Calculated Seebeck coefficient on the basis of the band structure by LDA (circle) and LDA+DMFT (triangle).
Reprinted with permission from \cite{2019PRBKaneko}, Copyright (2019) by the American Physical Society.
}
\label{5-4_ecio_kaneko}
\end{center}
\end{figure}

Figures~\ref{5-4_ecio_kaneko}(a) and (b) show the temperature dependence of resistivity and magnetization for (Eu$_{1-x}$Ca$_{x}$)$_{2}$Ir$_{2}$O$_{7}$~\cite{2019PRBKaneko}.
The resistivity of non-doped Eu$_{2}$Ir$_{2}$O$_{7}$ gradually increases as the temperature is lowered and shows a kink below $\tn $, at which the magnetization increases steeply.
As $x$ increases, the resistivity and $\tn $ systematically decreases.
Eventually, the paramagnetic metal state is realized for $x=0.10$.
Figure~\ref{5-4_ecio_kaneko}(c) exhibits the contour plot of resistivity in the plane of temperature and $x$.
The antiferromagnetic insulating phase is gradually suppressed as $x$ increases and apparently diminished at $x\sim 0.06$.
Near the filling control insulator-to-metal transition, the electronic specific heat coefficient $\gamma $ abruptly increases and exceeds 20 mJ/molK$^2$ at $x=0.05$, indicative of the enhanced electron mass characteristic of the filling-control Mott transition.
Figure~\ref{5-4_ecio_kaneko}(d) shows the temperature dependence of Seebeck coefficient for the paramagnetic metal $x=0.05$, 0.10, 0.20.
For $x=0.05$, the Seebeck coefficient is negative at high temperatures, turns its sign to positive below 100 K, and shows a peak at around 25 K.
As $x$ increases, the peak like structure shifts to higher temperature.
Similar temperature and hole-doping dependence of Seebeck coefficient is also observed in Ca-doped Pr$_2$Ir$_2$O$_7$ and Nd$_2$Ir$_2$O$_7$~\cite{2019PRBKaneko} which are established as the zero-gap semiconductors in paramagnetic metal phase as discussed in Sect.~\ref{sec:6_Iridates MIT}~\cite{2015NCommKondo,2016PRLNakayama}.
According to the band calculation based on local density approximation (LDA) and dynamical mean-field theory (DMFT) for $x=0.10$ [Fig.~\ref{5-4_ecio_kaneko}(e)], the QBT is located at 10 meV above the Fermi energy, which is shifted due to the hole doping.
Based on this band structure, the temperature dependence of Seebeck coefficient is calculated [Fig.~\ref{5-4_ecio_kaneko}(f)]; it reproduces the peak feature at a temperature corresponding to the half of Fermi energy, which is reconciled with the experiment [Fig. \ref{5-4_ecio_kaneko}(d)].
These results indicate that the hole doping causes the transition from the all-in all-out antiferromagnetic insulator to the paramagnetic semimetal characterized by the QBT.

\begin{figure}[h]
\begin{center}
\includegraphics[width=0.95\columnwidth]{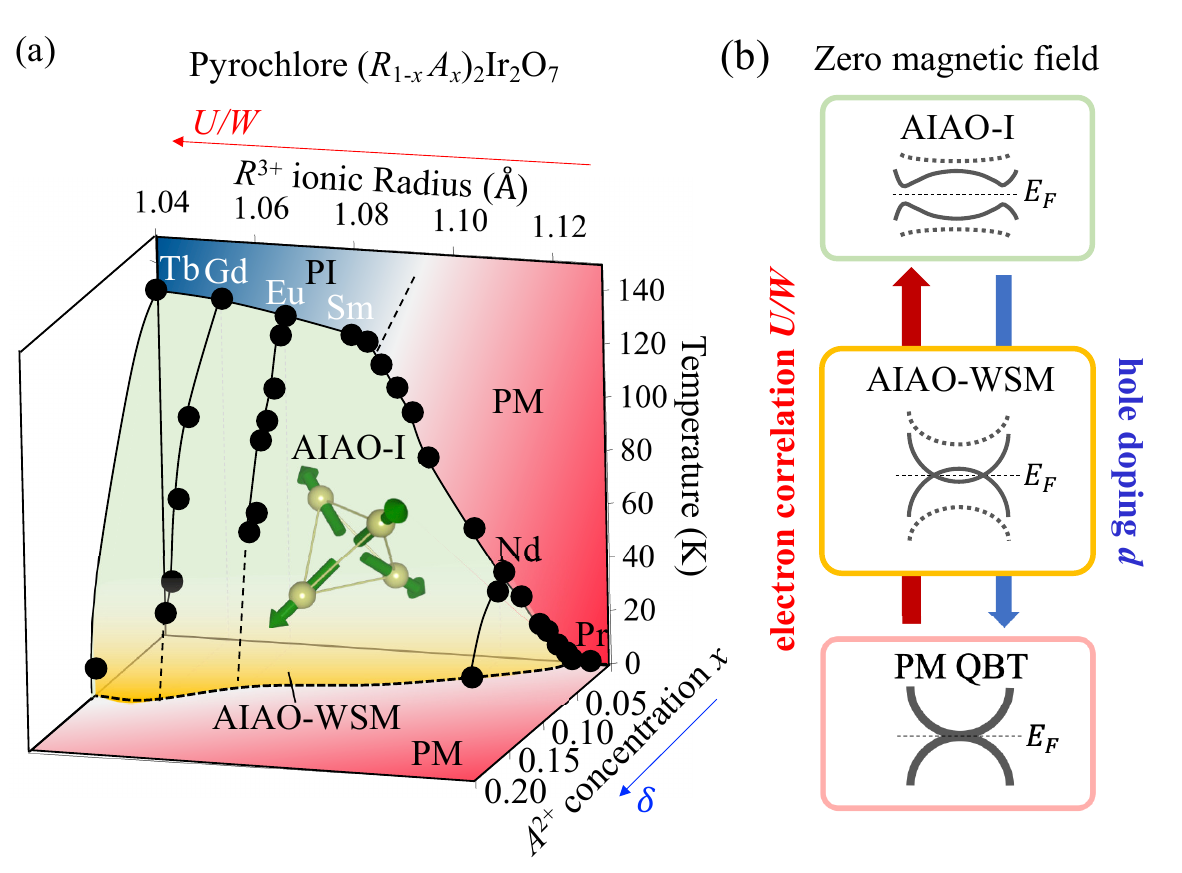}
\caption{
(a) Phase diagram of $R_2$Ir$_2$O$_7$ as a function of rare-earth ionic radius, divalent-ion concentration, and temperature.
Here PI stands for paramagnetic insulator, PM stands for paramagnetic metal, AIAO stands for all-in all-out state, and WSM stands for Weyl semimetal.
(b) Schematic picture of modulation of electronic band structures while changing electron correlation $U$ and hole doping $\delta $. Here QBT stands for quadratic band touching.
Reprinted with permission from \cite{2020PRBUeda}, Copyright (2020) by the American Physical Society.
}
\label{5-4_rcio_phasediagram}
\end{center}
\end{figure}

Figure~\ref{5-4_rcio_phasediagram}(a) exhibits the three-dimensional phase diagram with axes of temperature, $R^{3+}$ ionic radius (representing the effective change of one-electron bandwidth $U/W$), and $A^{2+}$ concentration (hole doping $\delta $).
It implies that the zero-gap semiconducting state is realized in a broad range of doping-induced paramagnetic metal phase, which enables us to explore the versatile magnetic topological states and functionalities by a choice of $R$ ions.
For instance, at zero field, the all-in all-out magnetic order transforms the zero-gap semiconductor into the Weyl semimetal state [Fig.~\ref{5-4_rcio_phasediagram}(b)].
Therefore, it is interesting to see what happens when the exchange field arises from the $R$ moments.

\begin{figure}[h]
\begin{center}
\includegraphics[width=0.95\columnwidth]{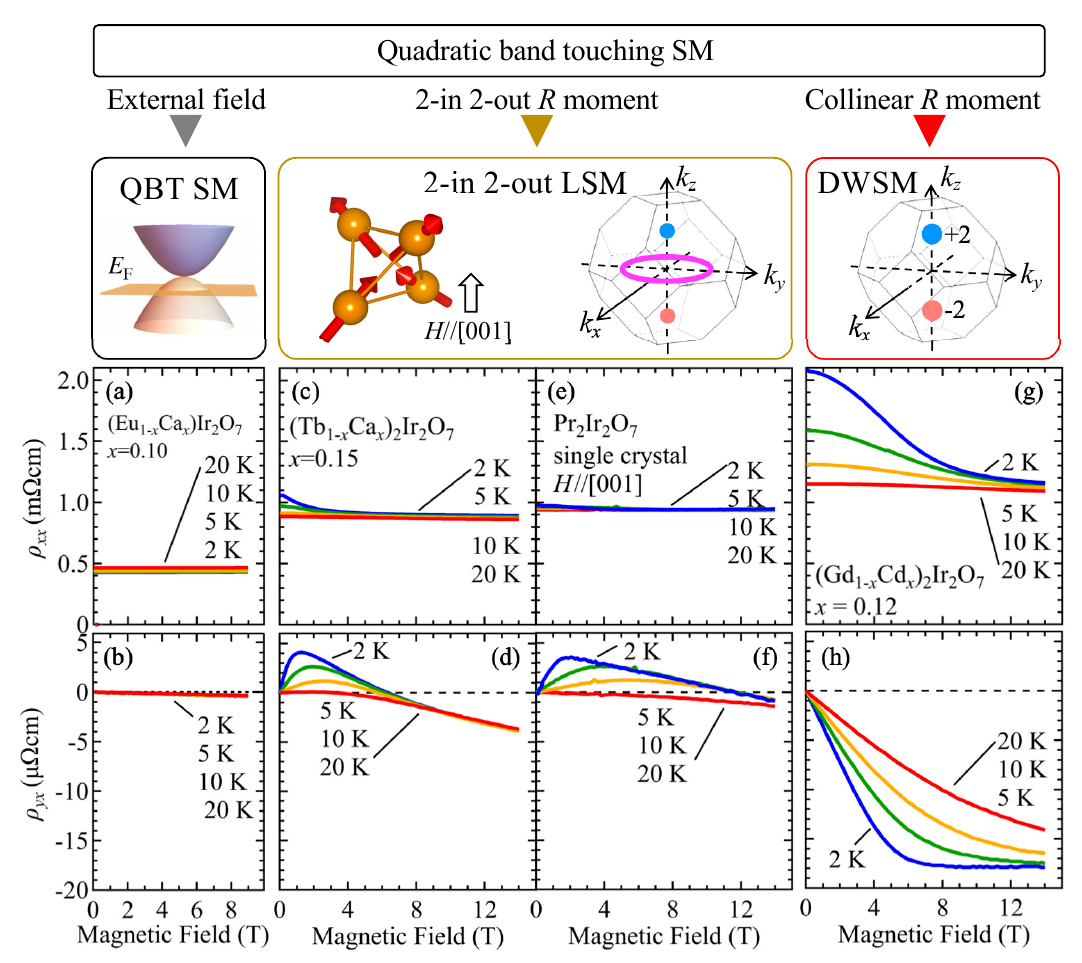}
\caption{
Magnetic-field dependence of (a), (c), (e), and (g) longitudinal resistivity and (b), (d), (f), and (h) Hall resistivity for (a) and
(b) (Eu$_{1-x}$Ca$_{x}$)$_{2}$Ir$_{2}$O$_{7}$ ($x=0.10$), (c) and (d) (Tb$_{1-x}$Ca$_{x}$)$_{2}$Ir$_{2}$O$_{7}$ ($x=0.15$), (e) and (f) Pr$_2$Ir$_2$O$_7$, and (g) and (h) (Gd$_{1-x}$Cd$_{x}$)$_{2}$Ir$_{2}$O$_{7}$ ($x=0.12$).
Schematic magnetic configurations of $R$ spins and expected electronic structures are displayed on top of the respective windows.
Reprinted with permission from \cite{2020PRBUeda}, Copyright (2020) by the American Physical Society.
}
\label{5-4_rcio_hall_rdep}
\end{center}
\end{figure}

The transport properties are examined for three different hole-doped $R$ compounds; nonmagnetic Eu, Heisenberg-like Gd, and Ising-like anisotropic Tb~\cite{2020PRBUeda}.
Figure~\ref{5-4_rcio_hall_rdep} shows the magnetic field dependence of resistivity and Hall resistivity in hole-doped metallic $R$ = Eu, Gd, and Tb compounds, respectively.
For the $R$ = Eu compound, both resistivity and Hall resistivity show only slight magnetic field dependence [Figs.~\ref{5-4_rcio_hall_rdep}(a) and (b)], presumably because the Ir magnetic moment is small (estimated as $\sim $0.2-0.5 $\mu _{\rm B}$/mol~\cite{2012JPSJTomiyasu,2012PRBShapiro}).
On the other hand, for the $R$ = Tb compound, the resistivity decreases as the field increases while the Hall resistivity shows a peak like structure at around 1 T [Fig.~\ref{5-4_rcio_hall_rdep}(d)].
This peak becomes more conspicuous as the temperature is lowered, indicating that the magnetic interaction between localized $4f$ moments and itinerant $5d$ electrons plays a vital role in transport properties.
Such a Hall response is quite similar to that of Pr$_2$Ir$_2$O$_7$ single crystals for $B//[001]$ [Fig.~\ref{5-4_rcio_hall_rdep}(f)], implying that the transport properties of $R$ = Tb polycrystals are mostly governed by the line-node semimetal state with the 2-in 2-out configuration [see the upper panel of Figs.~\ref{5-4_rcio_hall_rdep}(c) and (e)].
More importantly, large magnetotransport responses are observed in the isotropic $R$ = Gd compound [Figs.~\ref{5-4_rcio_hall_rdep}(g) and (h)].
In particular, the Hall resistivity is negative in the whole magnetic field range and its absolute value is about four times larger than those of $R$ = Tb or Pr.
Remarkably, despite polycrystals, the observed Hall angle reaches 1.5\%, which is larger than those of existing ferromagnetic oxides such as SrRuO$_3$~\cite{2003ScienceFang}, (La,Sr)CoO$_3$~\cite{2007PRLMiyasato}, Nd$_2$Mo$_2$O$_7$~\cite{2007PRLIguchi}.
According to Ref.~\cite{2018PRBOh}, a pair of double Weyl points, each of which hosts monopole charge $\pm 2$, can be induced by the uniform magnetic field [upper panel of Fig.~\ref{5-4_rcio_hall_rdep}(g)].
In the case of $R$ = Gd compounds, the $f$-$d$ coupling with the surrounding Gd spins can give rise to the large band splitting and thereby giant Hall effect.

\subsection{$A$ = Bi and Pb iridates}


Bi$_2$Ir$_2$O$_7$ single crystals are synthesized by a self-flux method containing Bi$_2$O$_3$ and IrO$_2$~\cite{2012JPCMQi}.
The resistivity is metallic and the specific heat shows no long-range order down to 50 mK.
However, the magnetization is distinct from the simple paramagnetic behavior.
It increases abruptly below 10 K as the temperature decreases. Further, the magnetic field dependence of magnetization is somewhat similar to the Brillouin function below 50 K like soft ferromagnets.
These may imply the strong electron correlation and magnetic instability in Bi$_2$Ir$_2$O$_7$, possibly due to the strong hybridization of Ir $5d$ and Bi $6s$ orbitals.

Pb$_2$Ir$_2$O$_{7-\delta }$ crystallizes $F\overline{4}3m$ space group, which is lower symmetry than the ordinal pyrochlore lattice ($Fd\overline{3}m$).
As explained in Sect.~\ref{sec:2_LatticeStructure}, pyrochlore oxides are formulated in a standard practice as $A_2$$B_2$O$_6$O' where 6 O sites and 1 O' site are distinguished in crystallographic positions.
In Pb$_2$Ir$_2$O$_{7-\delta }$, the O and O' sites in $Fd\overline{3}m$ structures are split into two sites, termed O1, O2 and O'1, O'2, respectively, resulting in the inversion symmetry breaking, as verified by the large second-harmonic generation signals~\cite{2013PRLHirata}.
The noncentrosymmetricity may lead to additional interesting physical properties.
For example, in the absence of the inversion center, antisymmetric spin-orbit couplings play an important role in the electronic band structures.
Thus, another topological feature may show up in Ir $5d$ bands.


\subsection{Summary}
\label{sec:6_summary}

A substantial amount of theoretical and experimental research has been conducted on $R_2$Ir$_2$O$_7$, establishing it as a leading platform at the intersection of strongly correlated physics and topological electronic properties.
Notably, compared to other $5d$ systems, $R_2$Ir$_2$O$_7$ offers an exceptionally broad range of tunable physical parameters, including bandwidth, filling, and $d$-$f$ interactions.
These parameters give rise to a diverse array of metal-insulator transitions driven by temperature, pressure, and magnetic field.
Furthermore, various topological semimetals emerge near these transitions, leading to unconventional phenomena, such as large magnetoresistance, anomalous Hall effects, and metallic states on magnetic domain walls.
Interestingly, the multiple topological phases appear to accumulate near the quantum critical point, which warrants further investigation through comprehensive measurements.
Furthermore, controlling magnetism via thin-film engineering presents promising opportunities.
This method could lead to the discovery of exotic magnetic phenomena, such as nonlinear magnetoelectric effects driven by magnetic multipoles~\cite{2024PRBOike} and quantum spin liquids emerging from low-dimensionality~\cite{2002PRLKawamura}.
These possibilities point to a broad and exciting direction for future research.

\section{Summary and outlook}
\label{sec:7_Perspective}

This article has highlighted the remarkable electronic, magnetic, and transport properties of pyrochlore oxides positioned at the metal-insulator phase boundary. These characteristics arise from strong electron correlation, magnetic frustration, spin-orbit coupling, and topology.

\subsection{Strong electron correlation}
The metal-insulator phenomena observed in pyrochlore oxides are attributed to the strong electron correlation inherent in transition-metal ($d$ electron) oxide systems.
These phenomena can be controlled through physical pressure, chemical substitution, and external magnetic fields.
For $B$=Mo, Ru, and Ir oxides with the pyrochlore structure (Sects.~\ref{sec:3_Molybdates}, \ref{sec:4_Ruthenates}, and \ref{sec:6_Iridates}), the ionic radius of the $A$-site ion affects the $B$-O-$B$ bond angles.
Consequently, its variation governs the magnitude of the $B$-site $d$ electron hopping amplitude $t$ (bandwidth control), the effective electron correlation $U/t$, and ultimately the Mott transition or Mott criticality.
This also modulates magnetic interactions, such as ferromagnetic double-exchange and antiferromagnetic/ferromagnetic super-exchange interactions for the magnetic moments of localized $d$ orbital electrons.
Bandwidth control can also be achieved through the application of physical pressure, which induces structural changes in the $B$-O-$B$ bonds analogous to the case of compositional changes at the $A$ site. This method provides a cleaner perturbation on $U/t$ without introducing chemical disorder effects.
Another powerful mechanism for driving the Mott transition is band-filling control or carrier doping. This is realized by modifying the $A$-site chemical composition, such as forming solid solutions of trivalent and divalent ions.
Using both bandwidth and band-filling controls, broad metal-insulator phase diagrams have been developed within a wide space of bandwidth and band-filling parameters. Being characteristic of $d$ electron correlated oxides with orbital degrees of freedom, these metal-insulator phenomena are deeply intertwined with $d$ electron magnetism.
External magnetic fields tend to align magnetic moments, favoring ferromagnetism mediated by double-exchange interactions while competing with antiferromagnetism mediated by super-exchange interactions.
In pyrochlore oxides, magnetic field effects occasionally lead to insulator-to-metal transitions or topological phase transitions beyond perturbative actions, as observed in Mo oxides (Sect.~\ref{sec:3_Molybdates}) and Ir oxides (Sect.~\ref{sec:6_Iridates}).

\subsection{Magnetic frustration}
In addition to electron correlation effects, a key characteristic of pyrochlore oxides is magnetic frustration.
The pyrochlore lattice is a prime example of strong geometrical frustration in magnetic interactions.
On one hand, magnetically ordered states with periodic 2-in 2-out and 3-in 1-out arrangements of magnetic $f$ moments on tetrahedral $A$ sites lead to the macroscopic emergence of scalar spin chirality.
This noncoplanar spin canting, when coupled to conduction $d$ electrons, produces an emergent magnetic field in real space, giving rise to unconventional phenomena such as the topological Hall effect.
On the other hand, the paramagnetic metallic state near Mott criticality is prone to hosting chiral spin liquids due to the local Dzyaloshinskii-Moriya interactions inherent to the pyrochlore lattice.
This can result in non-Fermi liquid behavior and field-induced topological Hall effects, as proposed for Mo-oxide and Ir-oxide pyrochlores, although conclusive experimental evidence is still lacking.
Abundant examples of exotic spin-related charge transport are observed in pyrochlore oxides, most notably in $R_2$Mo$_2$O$_7$ pyrochlores (Sect.~\ref{sec:3_Molybdates}).
However, achieving a microscopic and quantitative understanding of these phenomena remains challenging, partly due to the complexity of addressing the topological aspects of strongly correlated electron dynamics.
In particular, first-principles-based calculations of the electronic structure, alongside theoretical elucidation and prediction of transport and magnetic properties of compositionally specific pyrochlore oxides, are critically needed.
Accurately understanding the competition and interplay among strong correlation, magnetic frustration, and electronic topology in pyrochlore oxides represents one of the most demanding issues in contemporary condensed-matter physics.
Conversely, this complex yet fascinating entanglement in the metal-insulator pyrochlores may provide opportunities to discover new states of matter in the future.

\subsection{Spin-orbit coupling and topology}
Another significant characteristic of pyrochlore oxides is the strong relativistic spin-orbit coupling of the $4d$/$5d$ electrons on the pyrochlore $B$ sites, which plays a crucial role in the emergence of topological electronic states. A quintessential example is $R_2$Ir$_2$O$7$, with $j_{\mathrm{eff}} = 1/2$ conduction electrons derived from the Ir $5d$ electrons (Sect.~\ref{sec:6_Iridates}).
In this case, the spin-orbit coupling strength is comparable to the Coulombic repulsion energy ($U$). The antiferromagnetic Weyl semimetal was first theoretically proposed for this pyrochlore iridate, and the all-in all-out Weyl semimetal was subsequently confirmed experimentally in close proximity to the Mott transition.
Interestingly, the Ir $5d$ electrons can exchange-couple with the pyrochlore $A$-site $f$ moments, stabilizing versatile Weyl semimetal states across a wide phase space.
These states manifest not only in the all-in all-out configuration but also in the 2-in 2-out and 3-in 1-out arrangements of $f$ moments, and even in the forcibly ferromagnetic collinear configuration.
From the paramagnetic metallic side near the Mott transition, the foundational state from which magnetic Weyl semimetal states emerge is the quadratic band-touching semimetal state created by spin-orbit coupling.
Strong exchange coupling with $f$ moments, regardless of collinear or noncollinear configurations, can induce substantial exchange splitting in the band-touching semimetal.
This interaction leads to the formation of Weyl semimetal states characterized by pairs of single or double Weyl nodes in the band structure.
Magnetically tunable Weyl semimetal states, as observed in barely metallic pyrochlores, offer promising avenues for further research.
Noteworthy areas include exploring the magneto-thermoelectric (Nernst) effect and magneto-photogalvanic effect.
Additionally, the interface and domain-wall charge dynamics associated with the Fermi-arc-like surface states of Weyl semimetals present intriguing targets for future studies.
Partial evidence of these phenomena has already been observed in the metallic conduction along antiferromagnetic domain walls in nascent Weyl semimetals (Sect.~\ref{sec:6_Iridates DW}).

\subsection{Broken inversion symmetry}
Large nonlinear and nonreciprocal phenomena are widely observed in quantum materials when both inversion symmetry and time-reversal symmetry are broken~\cite{2018NCTokura}.
Spin-orbit coupling plays a pivotal role in enhancing these nonlinear and nonreciprocal effects through strong spin-momentum locking in the electronic bands under broken inversion symmetry.  
Although the pyrochlore lattice is originally centrosymmetric, there are ferroelectric pyrochlores like Cd$_2$Nb$_2$O$_7$~\cite{1955PRJona} and parity-broken conducting/superconducting pyrochlores such as Cd$_2$Re$_2$O$_7$.
These compounds have the potential to host novel electronic states, including previously unexplored multipoles and bond/spin-current orders driven by electron correlation (Sect.~\ref{sec:5_Cd2Re2O7}).
The creation of pyrochlore heterointerfaces also offers a promising approach to breaking inversion symmetry.
Notably, electric-current control of the magnetic state in the correlated (semi)metals of pyrochlore oxides with broken inversion symmetry—such as the emergence of spin-orbit torque—represents an exciting avenue for future research, particularly in the development of spintronic functionalities.

\subsection{Superconductivity}
Among the electronic phases yet to be thoroughly explored in pyrochlore oxides, the superconducting state stands out. Cd$_2$Re$_2$O$_7$ remains the sole known superconductor within the family of pyrochlore oxides to date (Sect.~\ref{sec:5_Cd2Re2O7}).
Nevertheless, the pyrochlore lattice holds promise, as evidenced by a well-known superconductor, the spinel oxide LiTi$_2$O$_4$, which exhibits a relatively high $T_{\mathrm{c}}$ ($\sim 12$ K)~\cite{1976Johnston}.
In LiTi$_2$O$_4$, the spinel Ti (with 1/2 $d$-electron per Ti) forms a network of a pyrochlore lattice.  The strong electron correlation and lattice frustration intrinsic to the pyrochlore structure—which inhibits long-range charge, spin, or orbital order—may contribute to the relatively high $T_{\mathrm{c}}$ observed in LiTi$_2$O$_4$.
It is plausible that analogous conditions could manifest in pyrochlore oxides, though no superconductivity has been reported so far.
Characterized by strong electron correlation, pronounced geometrical frustration, and robust spin-orbit coupling, pyrochlore oxides with $4d$/$5d$ electrons remain a fertile ground for exploring unique condensed-matter physics and phenomena.

\subsection{Concluding remarks}
In this article, we have explored the metal-insulator transition in the pyrochlore oxides. For those oxide compounds exhibiting Mott criticality, exotic or unique excited states and elementary excitations within spin-charge-orbital coupled systems are anticipated to play crucial roles not only in charge transport but also in nonlinear and non-equilibrium dynamics.
These include phenomena such as photo-excited dynamics and photo-induced Mott or topological transitions.
Pyrochlore oxides situated at the metal-insulator boundary provide an invaluable testbed for cutting-edge research in these areas. Collaboration between theoretical and experimental studies will be essential to advance understanding and achieve breakthroughs in this field.

\section*{Acknowledgements}

We are grateful to Yasujiro Taguchi, Satoshi Iguchi, Noriaki Hanasaki, Istv\'{a}n K\'{e}zsm\'{a}rki, Jun Fujioka, Maximilian Hirschberger, Hikaru Fukuda, Naoto Nagaosa, Bohm-Jung Yang, Hiroaki Ishizuka, Ryotaro Arita, Shiro Sakai, Yusuke Nomura, Hiroshi Shinaoka, Takashi Miyake, Shoji Ishibashi, Yusuke Tomita, Satoru Hayami, Yuki Yanagi, and Hiroaki Kusunose for collaborations that deepened our interest and understanding of the subject.
This work was supported by JSPS/MEXT Grant-in-Aid for Scientific Research (Grants No. 23H05431, 24H01649, and 25K00957) and CREST, JST (Grant No. JPMJCR16F1 and JPMJCR1874).



\section*{References}
\bibliographystyle{iopart-num}
\providecommand{\newblock}{}

\end{document}